\documentclass[twocolumn,prd,floatfix,preprintnumbers,a4paper,nofootinbib,superscriptaddress,groupedaddress]{revtex4-1}
\usepackage[utf8]{inputenc}
\usepackage{amsmath,amssymb}
\usepackage{amsfonts}
\usepackage{bm}
\usepackage{graphics}
\usepackage{float}
\usepackage{amsmath}
\usepackage{amssymb}
\usepackage[normalem]{ulem} 
\usepackage[mathscr]{euscript}
\usepackage{acronym}
\usepackage{float}
\usepackage[caption=false]{subfig}
\usepackage{lipsum}
\usepackage{mathtools}
\usepackage{multirow}
\usepackage{graphicx}
\usepackage{mathtools}
\usepackage{multirow}
\usepackage{graphicx}
\usepackage[usenames,dvipsnames,table,xcdraw]{xcolor}
\usepackage[colorlinks, pdfborder={0 0 0}]{hyperref} 
\usepackage[nameinlink,capitalise]{cleveref}

\graphicspath{{fig}}
\newcommand{\Msun}{\ensuremath{M_\odot}}

\newcommand{\todo}[1]{\textcolor{black}{#1}}

\newcommand{\xj}[1]{\todo{#1}}

\graphicspath{{fig}}

\begin{document}

\title{Spectroscopy of binary black hole ringdown using overtones and angular modes}

\author{Xisco Jim\'enez Forteza$^{1}$,
Swetha Bhagwat$^2$,
Paolo Pani$^{2}$,
Valeria Ferrari$^2$}\email{francisco.jimenez.forteza@aei.mpg.de}

\affiliation{$^1$ Max Planck Institute for Gravitational Physics (Albert Einstein Institute), Callinstra{\ss}e 38, 30167 Hannover, Germany}
\affiliation{$^2$ Dipartimento di Fisica, ``Sapienza'' Universit\`a di Roma \& Sezione INFN Roma1, Piazzale Aldo Moro 
5, 00185, Roma, Italy}

\begin{abstract}

The black hole uniqueness and the no-hair theorems imply that the quasinormal spectrum of any astrophysical black hole is determined solely by its mass and spin. The countably infinite number of quasinormal modes of a Kerr black hole are thus related to each other and any deviations from these relations provide a strong hint for physics beyond the general theory of relativity. To test the no-hair theorem using ringdown signals, it is necessary to detect at least two quasinormal modes. In particular, one can detect the fundamental mode along with a subdominant overtone or with another angular mode, depending on the mass ratio and the spins of the progenitor binary. Also in the light of the recent discovery of GW190412, studying how the mass ratio affects the prospect of black hole spectroscopy using overtones or angular modes is pertinent, and this is the major focus of our study. First, we provide ready-to-use fits for the amplitudes and phases of both the angular modes and overtones as a function of mass ratio $q\in[0,10]$. Using these fits we estimate the minimum signal-to-noise ratio for detectability, resolvability, and measurability of subdominant modes/tones. We find that performing black hole spectroscopy with angular modes is preferable when the binary mass ratio is larger than $q\approx 1.2$ (provided that the source is not located at a particularly disfavored inclination angle). For nonspinning, equal-mass binary black holes, the overtones seem to be the only viable option to perform a spectroscopy test of the no-hair theorem. However this would require a large ringdown signal-to-noise ratio ($\approx 100$ for a $5\%$ accuracy test with two overtones) and the inclusion of more than one overtone to reduce modeling errors, making black hole spectroscopy with overtones impractical in the near future.
\end{abstract}

\maketitle

\section{Introduction}


In the general theory of relativity~(GR), the postmerger remnant of a binary black hole~(BBH) coalescence settles to a Kerr black hole~(BH) at a sufficiently late time after the merger. To reach this stationary state, the perturbed BH remnant emits gravitational waves~(GWs) during a process known as the ringdown~(RD). The GW corresponding to the RD phase is described by the linear perturbation theory of a Kerr  BH~\cite{Vishveshwara:1970zz,Teukolsky1,Teukolsky2,Teukolsky3,Chandrasekhar,10.2307/2397876,Kokkotas:1999bd,Berti:2009kk}. The strain takes the following form:
\begin{align}
\label{eq:rdmodel}
    h(t) = \Sigma_{lmn} {\cal A}_{lmn} e^{- \iota \omega_{lmn} t} e^{- (t-t_0)/ \tau_{lmn}}\, _{s}\mathcal{Y}_{lm}\,.
\end{align}
Here $\omega_{lmn}$ and  $\tau_{lmn}$ are the quasinormal-mode~(QNM) frequencies and damping times, respectively\xj{. For a given choice of the $(lmn)$ indices, there exist two families of solutions, those with $\omega_{lmn}>0$ and those with $\omega_{lmn}<0$ corresponding to the co-rotating and counterrotating modes~\cite{berti:2005ys,Leaver:1985ax,Leaver:1985ax,Berti:2009kk,Kokkotas:1999bd,Ferrari:1984zz}. The} $(l,m)$ indices describe the angular decomposition of the modes, $_s\mathcal{Y}_{lm}$ are the spin-weighted \xj{$s=-2$} spheroidal harmonics, $n$ accounts for the $n$-tone excitations of a given $(l,m)$ mode, with $n=0$ being the fundamental ``tone'', \footnote{In this paper, we call the overtones simply as the ``tones'' and the angular modes as ``modes''. We shall often refer to a \emph{2-tone} or to a \emph{2-mode} RD model, when considering Eq.~\eqref{eq:rdmodel} with two different tones ($n=0,1$) or two different angular modes, respectively.}, and $t_0$ is the starting time of the RD, i.e., a suitable time where the linear perturbation theory is expected to describe the dynamics accurately~\cite{bhagwat:2017tkm,Bhagwat:2019dtm}. The amplitude ${\cal A}_{lmn}=A_{lmn}e^{\iota \phi_{lmn}}$ is a complex number that depends on the perturbation conditions set up during the inspiral-plunge-merger phase of the BBH evolution and is determined by the mass ratio and the spins of the BBH system.

As a consequence of the BH uniqueness and the no-hair theorems~\cite{Figueras:2009ci,Carter:1971zc,Hawking:1973uf,Robinson:2004zz, Mazur:1982db,Gibbons:2002av,Israel:1967wq}, the QNM frequencies $\omega_{lmn}$ and the damping times $\tau_{lmn}$ define a countably infinite set of modes uniquely related to each other by the final BH mass~($M_{f}$) and final spin~($a_{f}$) only.

Customarily, there are at least two approaches to test the BH no-hair theorem with RD signals (see \cite{Barack:2018yly,Berti:2018vdi} for reviews). The first consists of performing an \emph{inspiral-merger-RD consistency test}~\cite{theligoscientific:2016src} that checks whether the final mass and spin of the remnant BH estimated using the inspiral signal only is consistent with the final mass and spin measured from its RD phase only or from the merger-RD phase.
This approach is a null-hypothesis test and assumes that GR is the correct theory of gravity: an inconsistency between the different measurements of $\{ M_{f}, a_{f} \}$ \xj{larger than the statistical errors} would provide evidence for new physics beyond GR describing the BBH coalescence.
The second and more direct approach is performing \emph{BH spectroscopy}, where one aims at extracting several QNM parameters from the RD signal. These are then used both to measure $\{ M_{f}, a_{f} \}$ and to check whether the QNM spectrum is consistent with the Kerr hypothesis~\cite{Berti:2009kk,Berti:2016lat}. This is a more stringent test but requires a higher \xj{signal-to-noise ratio~(SNR) $\rho$}  in the RD, as it relies on \emph{independent measurements} of at least two QNM modes.
While measuring only the fundamental $l=m=2$ QNM frequency and damping time provides an estimate of $M_f$ and $a_f$, subdominat mode parameters (frequencies and damping times of other modes/tones) are needed to perform a consistency test of the QNM spectrum~\cite{Gossan:2011ha}.
The premise of this work is to investigate the prospects of performing a BH spectroscopy with BBH RD. 

As a rule of thumb, for a given value of $\{M_{f}, a_{f} \}$, the frequency of a mode scales approximately as ${\omega_{lmn}\approx \frac{l}{2} \omega_{22n}}$. Thus, the relative difference between the frequencies of two angular modes \xj{ ($l=2\neq l'$) and for a moderate final spin $a_f \lesssim 0.7$ } is approximately $\gtrsim 50\%$. \xj{However, the damping times are typically comparable for similar values of $l$ and for the same overtone number, at least for a moderately spinning BH~\cite{Kokkotas:1999bd,Berti:2009kk}}. 
Contrarily, the frequencies difference between two overtones [with same angular indices $(l,m)$ but $n\neq n'$] is typically less than a few percent. For instance, for a GW150914-like BBH event~\cite{theligoscientific:2016agk}, $1-{\omega_{221}/\omega_{220}\sim 2\%}$.
Several earlier studies of BH RD spectroscopy focused solely on the angular modes~\cite{TGR,greg1,greg2,Baibhav,Bhagwat:2016ntk,2019arXiv191013203B,Gossan:2011ha} and typically neglected the overtones (with some notable exceptions~\cite{London:2014cma,Taracchini:2013rva,Baibhav,Baibhav:2017jhs}).
However, more recently, RD tests with overtones were explored in more details~\cite{Giesler:2019uxc,Bhagwat:2019dtm,Ota:2019bzl}, and an attempt to perform such tests with GW150914 data was demonstrated in Ref.~\cite{Isi:2019aib}.

Loosely speaking, the accuracy to which a particular set of modes or tones in a BBH RD allows for a BH spectroscopy depends on the power contained in them. The relative amplitudes to which different RD modes/tones are excited depend on the initial perturbation conditions that are set up during the plunge-merger phase. This, in turn, depends on the mass ratio and the spins of the progenitor BBH. Asymmetric RD modes (such as $l=m=3$ and $l=2$, $m=1$) are excited when the merging BHs are either of unequal masses or have misaligned spins. However, the BBH GW events detected by the LIGO-Virgo observations show a high concentration of comparable mass binaries with low spins~\cite{LIGOScientific:2018mvr} that prevents extracting higher angular QNMs from their RD.

The recent discovery of a BBH system with a mass ratio significantly different from unity, GW190412~\cite{LIGOScientific:2020stg}, provides further motivation to detect asymmetrical RD angular modes from stellar mass BBHs. This state of affairs might become common in the third and future observational runs, where several events with unequal masses and possibly nonvanishing spins are expected.

Thus, it is timely to study the role of overtones and angular modes in the RD signal as a function of the BBH mass ratio. A recent analysis in this direction was performed in Ref.~\cite{Bhagwat:2019dtm} and, more recently, Ref.~\cite{Ota:2019bzl}, quantified the amplitude ratio of different tones/modes relative to the dominant $l=2$, $m=2$, $n=0$ mode as a function of the mass ratio for nonspinning binaries.

It is unlikely that a large number of QNMs will be detected in a single event  with the current ground-based GW observatories~\cite{Cabero:2019zyt}, although the third-generation facilities like Einstein Telescope and Cosmic Explorer ~\cite{hild:2010id,evans:2016mbw,essick:2017wyl} and the space-based mission LISA~\cite{audley:2017drz} are expected to detect multiple modes. Therefore, the current and the near-future RD-based GR tests shall rely either on a \emph{2-mode analysis} or on a \emph{2-tone analysis}. This motivates the question we try to address in this work, namely what is the optimal set of QNMs to perform 2-mode/tone BH spectroscopy. In particular, we quantify the prospects of observing the fundamental $l=2$, $m=2$, $n=0$ mode along with either:
\begin{itemize}
    \item[a)]  its first overtone, i.e., the $l=2$, $m=2$, $n=1$ mode, or
    \item[b)]  another angular mode, either the $(l=m=3,n=0)$ or the $l=2$, $m=1$, $n=0$ mode \,.
\end{itemize} 
The optimal set of QNMs to analyze a BBH RD signal depends on the mass ratio and initial spins of the progenitor BBH. We quantify the notion of an optimal set of QNMs by estimating the minimum SNR required to perform BH spectroscopy with a 2-mode/tone analysis.

In order to address the above question, we introduce three specific criteria: the \emph{detectability}, the \emph{resolvability}, and the \emph{measurability} of the QNMs.
The detectability criterion demands that the amplitude ratio between the subdominant mode/tone and the dominant mode should be nonzero \xj{at the $1\sigma$ confidence level}. Furthermore, the resolvability criterion requires that the subdominant mode/tone QNM parameters should be distinguishable from the corresponding parameters of the fundamental mode~\cite{Berti:2009kk}. Finally, the measurability criterion quantifies the measurement uncertainties in the inference of these parameters and requires that their uncertainty must be smaller than a given threshold. 

We estimate the minimum SNR for which each of the three criteria (and combinations thereof) are met using a Fisher information matrix framework. Thus, our estimate of the minimum SNR provides an optimistic lower bound to what is required to perform a Bayesian analysis using GW data.
Compared to Ref.~\cite{Ota:2019bzl}, our analysis has three notable differences: (i)~we assess the role of the starting time of the RD differently; (ii)~we use the combination of the three criteria mentioned above -~detectability, resolvability, and measurability~- as the discriminator between an overtone-based and an angular-mode-based BH spectroscopy, as compared to the criteria used in Ref.~\cite{Ota:2019bzl}; (iii)~in our RD model we use the quality factor ${Q_{lmn}\equiv \pi f_{lmn} \tau_{lmn}}$ in place of the damping time $\tau_{lmn}$. As discussed below, this choice leads to some differences for what concerns the resolvability criterion, but it does not affect our final conclusions.

For this study, we fit for the amplitude $A_{lmn}$ and phase $\phi_{lmn}$ in Eq.~\eqref{eq:rdmodel} to a set of numerical relativity~(NR) simulations corresponding to nonspinning binaries with a mass ratio $ q\in [1,10]$ from the simulating eXtreme spacetimes~(SXS) catalog~\cite{sxscatalog}.
In Secs.~\ref{sec:fit} and~\ref{sec:parfits}
we provide ready-to-use fits for the amplitude ratio and the phase difference of the different modes/tones for various choices of the RD starting time. Then, in Sec.~\ref{sec:crit}, we address the issue of choosing a set of subdominant mode/tone $(l,m,n)$ to perform BH spectroscopy.
Finally, in Sec.~\ref{sec:discussion}, we discuss the qualitative aspects of our results.

\begin{table}[]{%
\begin{tabular}{|c|c|c|c|c|} 
\hline
Sim No. & SXS ID   & $q$         & $a_{f}$       & $M_f$   \\ \hline
\hline
1 & \text{SXS:0066} & 1.00 & 0.686 & 0.952 \\
2 & \text{SXS:0070} & 1.00 & 0.686 & 0.952 \\
3 &\text{SXS:1143} & 1.25 & 0.680 & 0.953 \\
4 &\text{SXS:0007} & 1.50 & 0.664 & 0.955 \\
5 &\text{SXS:1354} & 1.83 & 0.638 & 0.959 \\
6 &\text{SXS:0169} & 2.00 & 0.623 & 0.961 \\
7 &\text{SXS:0201} & 2.32 & 0.596 & 0.965 \\
8 &\text{SXS:0259} & 2.50 & 0.581 & 0.967 \\
9 &\text{SXS:0191} & 2.51 & 0.580 & 0.967 \\
10 & \text{SXS:0030} & 3.00 & 0.541 & 0.971 \\
11 &\text{SXS:1221} & 3.00 & 0.541 & 0.971 \\
12 &\text{SXS:0294} & 3.50 & 0.504 & 0.975 \\
13 &\text{SXS:1220} & 4.00 & 0.472 & 0.978 \\
14 &\text{SXS:0182} & 4.00 & 0.472 & 0.978 \\
15 &\text{SXS:0107} & 5.00 & 0.417 & 0.982 \\
16 &\text{SXS:0296} & 5.50 & 0.393 & 0.984 \\
17 &\text{SXS:0181} & 6.00 & 0.372 & 0.985 \\
18 &\text{SXS:0166} & 6.00 & 0.372 & 0.985 \\
19 &\text{SXS:0298} & 7.00 & 0.337 & 0.988 \\
20 &\text{SXS:0186} & 8.27 & 0.300 & 0.990 \\
21 &\text{SXS:0301} & 9.00 & 0.282 & 0.991 \\
22 & \text{SXS:1107} & 10.00 & 0.261 & 0.992 \\
 \hline
\end{tabular}%
}
\caption{Set of NR simulations used in this work obtained from the SXS public catalog \cite{sxscatalog,boyle:2019kee}. Here, $q=m_1/m_2 \geq 1$ is the mass ratio while $a_f$ and $M_f$ are the final dimensionless spin and the final mass (in units of the binary total mass), respectively.}
\label{tab:NR_set}
\end{table}

\section{Fitting NR-RD waveforms}\label{sec:fit}
In this section we discuss our procedure to fit the NR waveforms listed in Table~\ref{tab:NR_set} with the RD model given in Eq.~\eqref{eq:rdmodel} for different modes/tones and discuss the choice of the starting time $t_0$.

\subsection{Setup}

We fit the set of simulations listed in Table \ref{tab:NR_set} following the prescription described in~\cite{Bhagwat:2019dtm,Giesler:2019uxc}. For each of the SXS-BBH simulations, we select the NR-RD waveform that corresponds to that with the best resolution (see Appendix~\ref{app:NRerror}). 
For each simulation the SXS catalog provides a mode decomposition, $h_{lm}$, in a basis of \emph{spherical} harmonics, which are related to the spheroidal harmonics in Eq.~\eqref{eq:rdmodel} as discussed in Appendix~\ref{app:spheroidal}.
The reference time $t=0$ is defined as the time at which the amplitude of $l=m=2$ mode peaks.

The data set consists of BBH systems with nearly zero initial dimensionless spins $a_{{1,2}}$ and a mass ratio $q=m_1/m_2 \in \left[1,10\right]$. 
The motivation for neglecting the binary component spins is twofold. First, all binaries detected in the first two observational runs of LIGO/Virgo are compatible with small or negligible spins\footnote{Nonetheless, in the light of the very recent BBH detection GW190412 during the third observational run~\cite{LIGOScientific:2020stg}, including the binary spins is a natural and urgent extension of our analysis.}, except possibly for one event~\cite{LIGOScientific:2018mvr}. Second, neglecting the spins reduces the parameter space significantly, since all dimensionless quantities (e.g., mode amplitude ratios and phases) depend solely on $q$.  

In Table~\ref{tab:NR_set}, we also list the final mass $M_f$ and spin $a_f$ of the BH remnant as provided in the metadata of the SXS catalog; the details of their computation can be found in Ref.~\cite{boyle:2019kee}. Loosely speaking, for a nonspinning BBH system, the final BH spin monotonically decreases as a function of $q$ as a consequence of the conservation of angular momentum~\cite{davis:1971gg,buonanno:2007sv,hofmann:2016yih,healy:2014yta,hughes:2002ei}. Our study spans a range of final spin with ${a_f\approx 0.69}$ for  $q=1$ to $a_f\approx 0.26$ when $q\approx 10$.

The fits are performed by fixing the frequencies $\omega_{lmn}$ and the damping times $\tau_{lmn}$ to the their corresponding values predicted by GR~\cite{Berti:2009kk}, choosing a starting time $t_0$, and then fitting for the complex mode/tone amplitudes $\mathcal{A}_{lmn}$. 
A complex least-square fit is used and the best-fit parameters correspond to the one that minimizes the $\chi^2$ value,
\begin{equation}
\chi^2=\sum_{i} |\bar{h}(\vec{\lambda})_i-h_i|^2,
\end{equation}
where  $\vec{\lambda}=\left\lbrace A_{lmn},\phi_{lmn}\right\rbrace$ are the fit parameters for a given $t_0$.
To quantify the deviations of the fits with respect to the NR waveform,  we compute the mismatch $\mathcal{M}$ which is defined as 
\begin{equation}
    \mathcal{M} = 1 - \frac{\langle h_{\rm NR}|h_x\rangle}{\sqrt{\langle h_{\rm NR}|h_{\rm NR}\rangle \langle h_x|h_x\rangle}}\,,
    \label{eq:mismatch}
\end{equation}
where 
\begin{equation}
    \langle f|g\rangle = \int_{t_i}^{t_f} f(t) g(t) dt\,.
\end{equation}
Here $h_{\rm NR}$ is the NR-RD waveform with the highest resolution~\cite{sxscatalog,boyle:2019kee} and $h_{x}$ stands, respectively, for the fit model ($h_{x}=h_M$) or for the  NR waveform with the next-to-the-highest resolution (when $h_x=h_{\rm LNR}$) when computing the NR error estimates. The integration domain ranges from $t_i=t_0$ and $t_f=60M$ for $h_{M}$, and from $t_i=\delta t^p_{lm}$ (where $\delta t^p_{lm}$ is the peak time of the strain $h_{lm}$, see Sec.~\ref{subsec:ang} below) to $t_f=60M$ for $h_{\rm LNR}$.

\begin{figure}[t]
    \includegraphics[width=0.49\textwidth]{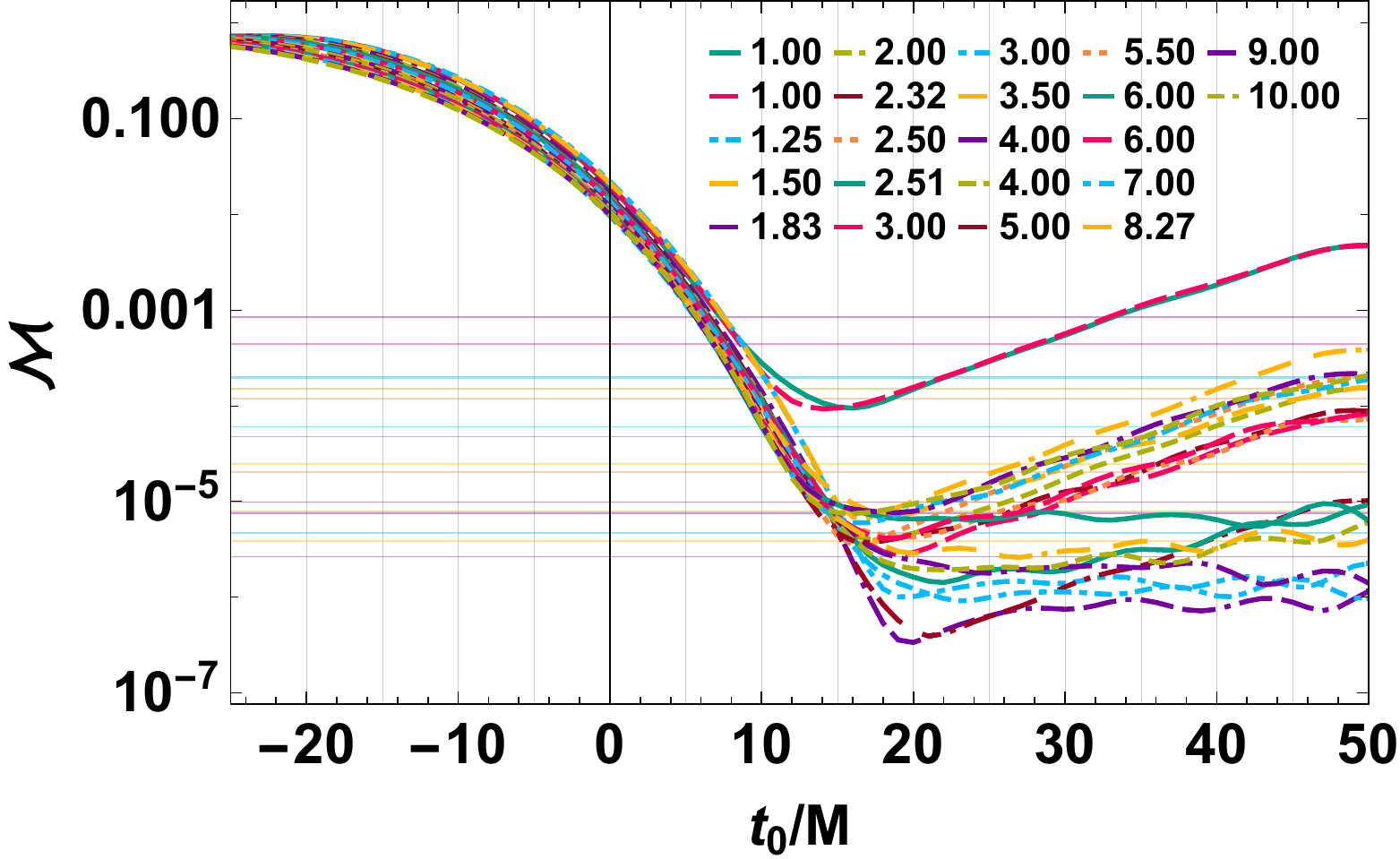}
    \caption{Mismatch computed between a 2-tone model fit and its respective NR-RD waveform as a function of the starting time $t_0$ for $l=m=2$. Each curve corresponds to an entry in Table~\ref{tab:NR_set}, while the horizontal lines show the NR error estimates using the same color scheme (see Fig.~\ref{fig:goodness} below). Note that, for all cases, the error lines cross the solid curves at ${t_0/M\approx 5-15}$. The reference time $t=0$ is defined by the peak of the $l=m=2$ strain mode.The plot starts at $t_0=-25M$ (i.e. before the peak of the strain) in order to explore the premerger behavior.} 
    \label{fig:matchplot}
\end{figure}

\subsection{Role of the starting time}\label{sec:fit_overtones}

\subsubsection{Overtones}

In Fig.~\ref{fig:matchplot}, we show the mismatch $\mathcal{M}$ between the $l=m=2$ mode of the NR-RD waveforms from Table~\ref{tab:NR_set} and a \emph{2-tone} RD model --~i.e., Eq.~\eqref{eq:rdmodel} with the fundamental mode $l=m=2, n=0$ and its first overtone~-- as a function of $t_0$. The thin horizontal lines in the plot correspond to the NR error estimates for each of the NR-RD simulations (see Appendix~\ref{app:NRerror}) and are marked using the same color scheme as their corresponding mismatch curves. These horizontal lines correspond to the mismatch between the two highest resolutions of the NR waveforms available in the SXS catalog. 

The mismatch has a similar behavior for all mass ratios considered in this study: it decreases monotonically up to a minimum at $t_0 \in \left[15M,20M\right]$ and then gradually rises up as the NR simulation saturates with numerical noise.
One expects that the accuracy of the RD model given by Eq.~\eqref{eq:rdmodel} increases at late times in the BBH postmerger and, therefore, $\mathcal{M}$ should monotonically decrease. However, the mismatch curves reach a minima due to the numerical noise floor, which is exacerbated by the exponential damping of the RD waveform.

Furthermore, we note that when ${t_0/M\approx 5-15}$ (depending on the value of $q$) or higher, the mismatch between the highest-resolution NR simulation and the analytical RD model is smaller than the corresponding NR error estimate. This happens because, at late times, the waveform is accurately described by a superposition of the QNM excitations. \xj{Notice that the error estimate is computed here from $t_0/M=0$. Therefore it can be considered as a conservative estimate of the error for $t_0/M>0$}.
This suggests that $t_0/M\in\left[5,15\right]$ would provide a reasonable range for the choice for the starting time for the $2$-tone RD model. Ideally, to ensure that the system is in a quasilinear regime, one would pick the largest possible value of $t_0$. However, the fast exponential decay of the $n=1$ overtone implies that its amplitude is very small for large values of $t_0$. Therefore, we opt for an agnostic strategy\footnote{This is different from Ref.~\cite{Ota:2019bzl}, in which $t_0$ has been identified with the value that minimizes $\mathcal{M}$ using either the strain or the time derivative of the phase of the waveform. In such a case both methods identify the range ${t_0/M \in \left[5, 18\right]}$ for ${q \in\left[1,10\right]}$. We did not follow this prescription since the minimum of ${\cal M}$ is a quantity solely determined by the accuracy of the NR waveforms, and not by the faithfulness of the RD model, as explained in the main text. Nonetheless, our prescription identifies a range compatible with that of Ref.~\cite{Ota:2019bzl}, namely $t_0/M\in\left[10,15\right]$.} and in the following sections, we present the results for $t_0/M=\{0,5,10,15\}$. We emphasize that a 2-tone RD model is insufficient to describe a BBH RD accurately from $t=0$; higher overtones should be included to ensure accuracy~\cite{Isi:2019aib,Giesler:2019uxc,Bhagwat:2019dtm}. However, including a large number of overtones is impractical from the point of view of parameter estimation in the context of BH spectroscopy with current detectors~\cite{Cabero:2019zyt}.

\subsubsection{Angular modes} \label{subsec:ang}

The choice of the starting time is much less of a problem for angular modes~\cite{bhagwat:2017tkm} as the damping time of subdominant angular modes (with $n=0$) is comparable to that of the fundamental $l=m=2$ mode. For the case of a \emph{2-mode} RD model (with either $l=m=3$ or $l=2$, $m=1$ along with the fundamental $l=m=2$ mode), we choose two values of the starting time: $t_0=10M$ and $t_0=15M$. Unlike for the case of a $2$-tones model, the choice of the starting time does not affect the amplitude-ratio and phase-difference fits significantly, as we will discuss later in detail.

\begin{figure}[t]
\includegraphics[width=0.49\textwidth]{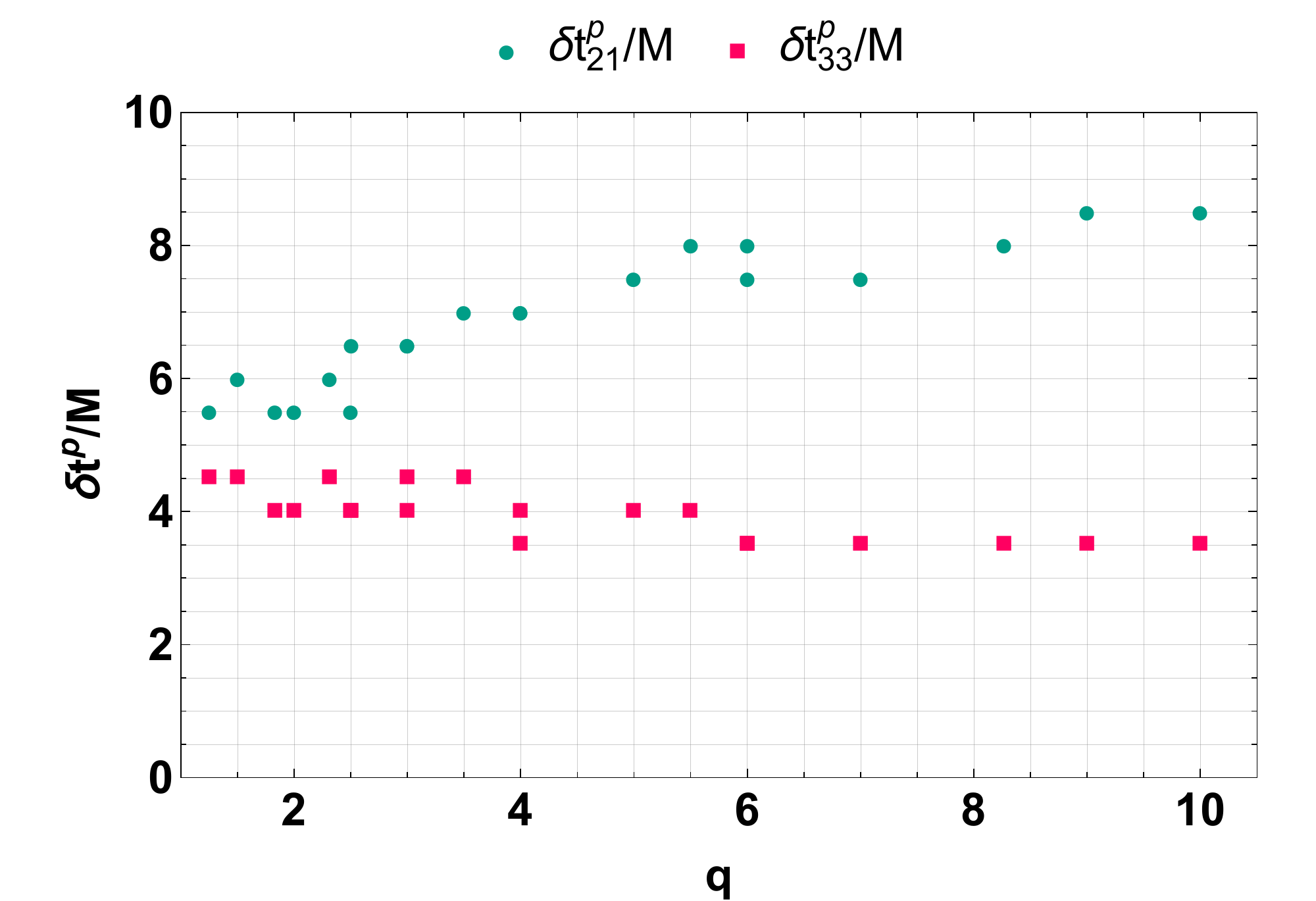}
    \caption{Difference between the time at which the (22) mode strain peaks (conventionally chosen at $t=0$) and the peak time of the higher harmonics, $\delta t^{p}_{lm}$. The (21) peak time $\delta t^{p}_{21}$ raises up as $q$ increases and eventually reaches $\delta t^{p}_{21}\approx 9M$. On the other hand, $\delta t^{p}_{33}$ remains approximately constant, $\delta t^{p}_{33}\approx 4M$. In both cases, $\delta t^{p}_{lm}$ is smaller than the starting time chosen for the ($lm$) modes fit, $t_0=10M$ or $t_0=15M$.
    }
    \label{fig:dts}
\end{figure}

Furthermore, the amplitudes of different modes do not peak at the same time. We denote the peak amplitude of the $h_{lm}$ mode as $\delta t^{p}_{lm}$ (with $\delta t^{p}_{22}=0$ conventionally). In Fig.~\ref{fig:dts}, we show the time shifts $\delta t^{p}_{lm}$ as a function of $q$ for our set of waveforms. Note that both $l=2$, $m=1$ and ${l=m=3}$ modes peak at a later time compared to the $l=m=2$ mode (i.e. $\delta t^{p}_{lm}>0$ for these cases), for all values of $q$. The time shifts $\delta t^{p}_{21}$ corresponding to the $l=2$, $m=1$ mode (green circles) approximately increases linearly with $q$ up to $\delta t^{p}_{21}\approx9M$ at $q=10$, while $\delta t^{p}_{33}$ (red squares) tends to be approximately constant ($\delta t^{p}_{33}\approx4M$) for all $q$'s. Thus, our choices $t_0=10M$ and $t_0=15M$ ensure that $t_0>\delta t^{p}_{lm}$, i.e., that we start the RD analysis after all the modes peak.

\subsection{Amplitude and phase fits}\label{sec:parfits}
Using the procedure described in the previous section, we can compute the fits for the amplitude and phases of each mode/tone. Since ${\cal A}_{lmn}=A_{lmn}e^{\iota \phi_{lmn}}$ in Eq.~\eqref{eq:rdmodel}, the total phase for the $(lmn)$ mode is
\begin{equation}
    \Phi_{lmn}(t)=\omega_{lmn}t-\phi_{lmn}\,.
\end{equation}
The phase $\phi_{lmn}$ depends on the initial orbital phase of the BBH system, which is generically different for different NR simulations. Thus, to make a meaningful comparison among different simulations, we align the phases by adding an extra constant phase to the NR waveform for each mode, such that $\Phi_{lmn}(10M)=0$. The choice of the reference time $t=10M$ is arbitrary and does not affect our results.

We choose to the fit the amplitude ratio,
\begin{equation}
  A_{R,lmn}=\frac{A_{lmn}}{A_{220}}\,, \label{eq:ratio}
\end{equation}
and the phase difference,
\begin{equation}
 \Delta \phi_{lmn}=\phi_{220}-\phi_{lmn}\,, \label{eq:phasediff}
\end{equation}
relative to the $l=m=2$, $n=0$ mode.

We find that these quantities can be conveniently expressed by the following closed-form approximations
\begin{align}
\label{eq:amp_an}
A_{R,lmn}&=a_{0,lmn} +\frac{a_{1,lmn}}{q}+\frac{a_{2,lmn}}{q^2}+\frac{a_{3,lmn}}{q^3}\,,\\
\label{eq:phse_an}
\Delta \phi_{lmn}&=b_{0,lmn} - \frac{b_{1,lmn}}{b_{2,lmn} + q^2}\,,
\end{align}
where the fit parameters $a_{i,lmn}$ and $b_{i,lmn}$ are obtained for different choices of $t_0$. 
Their values are listed in Tables~\ref{tab:fitovertones}, \ref{tab:fitmodes210}, and \ref{tab:fitmodes330}, for the 2-tone model ($l=m=2$, $n=1$) and for both the 2-mode models ($l=m=3$ or $l=2$, $m=1$, both with $n=0$), respectively. 
For the subdominant angular modes, the parameter $a_{0,lmn}$ is fixed in terms of the others by requiring that ${A_{R,210}=A_{R,330}=0}$ at equal mass ratio, ${q=1}$, as imposed by symmetry arguments.

These fits are discussed in the following sections:

\begin{table}[!]{%
\begin{tabular}{|c|c|c|c|c|}
\hline
$t_0$   & $0 M$        & $5 M$       & $10 M$      & $15 M$      \\ \hline\hline
$a_{0,221}$ & 0.901327   & 0.641972    & 0.373974  & 0.170129   \\ \hline
$a_{1,221}$ & 0.705107   & 0.297579  & 0.074412  & 0.10284  \\  \hline
$a_{2,221}$ & -0.386356   & 0.218103  & 0.416288  & 0.0918048 \\ \hline
$a_{3,221}$ & 0.045801 & -0.25445 & -0.322963 & -0.0765307 \\ \hline \hline
$b_{0,221}$ & 0.426307 & 0.396928 & 0.381152 & 0.415696 \\ \hline
$b_{1,221}$ & -5.59052 & -4.84138  & -6.3857  & -6.70373  \\ \hline
$b_{2,221}$ & 56.7092  & 18.6836  & 14.9772  & 13.2363  \\ \hline
\end{tabular}%
}
\caption{Fitting coefficients of Eqs.~\eqref{eq:amp_an} and \eqref{eq:phse_an} for the 2-tone model (i.e., $l=m=2$ fundamental mode plus its first overtone) for different values of the starting time $t_0$.
}
\label{tab:fitovertones}
\end{table}

\begin{table}[!]{%
\begin{tabular}{|c|c|c|}
\hline
$t_0$   & $10 M$      & $15 M$      \\ \hline\hline
$a_{0,210}$    & 0.473846   & 0.479966   \\ \hline
$a_{1,210}$    & -1.22756  & -1.23848   \\  \hline
$a_{2,210}$    & 1.61047   & 1.61757    \\ \hline
$a_{3,210}$    & -0.85676   & -0.859064   \\ \hline \hline
$b_{0,210}$    & 1.8082    & 1.82032    \\ \hline
$b_{1,210}$    & 9.9702    & 8.79577     \\ \hline
$b_{2,210}$    & 10.3096    & 9.30836     \\ \hline
\end{tabular}%
}
\caption{Same as Table~\ref{tab:fitovertones} for for the 2-mode RD model with $l=m=2$, and $l=2$, $m=1$ fundamental modes. }
\label{tab:fitmodes210}
\end{table}

\begin{table}[!]{%
\begin{tabular}{|c|c|c|}
\hline
$t_0$   & $10 M$      & $15 M$      \\ \hline\hline
$a_{0,330}$    & 0.439698   & 0.437926   \\ \hline
$a_{1,330}$    & -0.611581  & -0.651738   \\  \hline
$a_{2,330}$    & 0.199865   & 0.301015    \\ \hline
$a_{3,330}$    & -0.0279826   & -0.0872038   \\ \hline \hline
$b_{0,330}$    & 2.66306    & 2.68764    \\ \hline
$b_{1,330}$    & -6.81421    & -6.39255     \\ \hline
$b_{2,330}$ & 6.65011       & 6.03077     \\ \hline
\end{tabular}%
}
\caption{Same as Table~\ref{tab:fitovertones} for for the 2-mode RD model with $l=m=2$, and $l=m=3$ fundamental modes.}
\label{tab:fitmodes330}
\end{table}

\subsubsection{Assessing the accuracy of the RD model by varying the starting time}
\label{sec:shifts}

The RD waveform model~\eqref{eq:rdmodel} has the following symmetry:
\begin{equation}
    \left\{\begin{array}{l}
    t_0\to t_0+\Delta t\\
    A_{lmn}\to A_{lmn}e^{-\frac{\Delta t}{\tau_{lnm}}} \\
    \phi_{lnm}\to\phi_{lnm}
    \end{array}\,.\right. \label{eq:symmetry}
\end{equation}
Thus, if Eq.~\eqref{eq:rdmodel} is a faithful model for the real signal, shifting the starting time and rescaling the amplitudes as presented in the above formula should not affect our best-fit parameter values. Conversely, the dependence of the fit parameters on the starting time after the above rescaling would imply that the Eq.~\eqref{eq:rdmodel} (for a certain number of modes/tones) does not reproduce the NR RD accurately.

We expect this symmetry to be broken for small values of $t_0$ for the following reasons: (i) we included only one overtone ($n=1$), whereas higher overtones with $n > 1$ should be included, especially for small values of $t_0$~\cite{Isi:2019aib}; and (ii) there could be possible nonlinearities in the source frame that cannot be accounted for by a superposition of the QNMs in the asymptotic frame, especially near the peak at $t\approx0$~\cite{London:2014cma,London:2017bcn,Baibhav:2017jhs,Bhagwat:2019dtm,Okounkova:2020vwu,PhysRevD.97.084028}.
These inaccuracies in the 2-tone RD model lead to systematic errors when performing BH spectroscopy. In Ref.~\cite{Bhagwat:2019dtm}, we study these effects and their impact on the QNM spectrum and find that the inferred QNM frequencies can be significantly biased if $t_0$ is chosen very close to the time of peak amplitude. In principle, one can alleviate this problem by choosing a sufficiently large value of $t_0$, but this also reduces the power contained in the highly damped $n=1$ overtone.

Finally, note that the symmetry~\eqref{eq:symmetry} can be broken when either the amplitude is not rescaled appropriately or when the phase $\phi_{lmn}$ does not remain constant upon shifting $t_0$.

\begin{figure}[]
\includegraphics[width=0.49\textwidth]{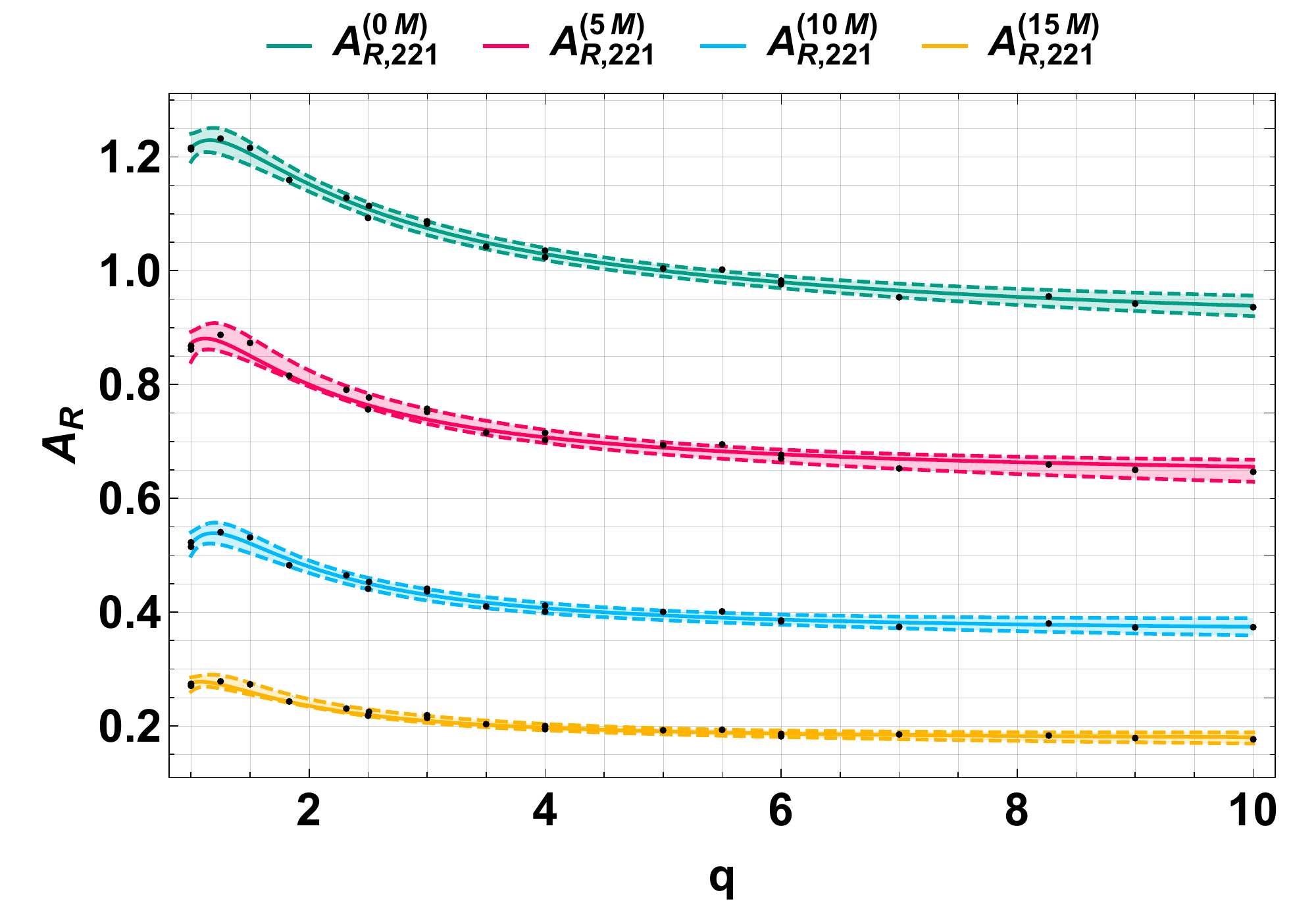}
    \caption{Amplitude ratio fits $A_{R,221}$ for the $n=1$ overtone computed for the $2$-tone RD model using Eq.~\eqref{eq:amp_an} and the fitting coefficients in Table~\ref{tab:fitovertones} for four different starting times, $t_0/M=\left\lbrace 0,5,10,15\right\rbrace$. Solid curves denote the best fit for each case while shaded regions denote the $90\%$ credible range. The agreement between the data (dots) and the best fit is good for all mass ratios and choices of $t_0$. 
    The amplitude among different choices of $t_0$ is not rescaled as in Eq.~\eqref{eq:symmetry}; see Fig.~\ref{fig:fit_and_starttime_221} for the corresponding rescaled quantity.
    } 
    \label{fig:fit_and_starttime_amp}
\end{figure}

\subsubsection{Fits for overtones}

In Fig.~\ref{fig:fit_and_starttime_amp}, we show the amplitude ratio $A_{R,221}$ [see Eq.~\eqref{eq:ratio} and Table~\ref{tab:fitovertones}] for the 2-tone RD models corresponding to four different starting times $t_0/M=\{0,5,10,15\}$. The fits reproduce the NR amplitude ratio and phase difference for all the simulations considered in this study with a $90 \%$ credibility. For a given choice of $t_0$, the amplitude ratio for the overtones decreases as the mass ratio increases until the slope flattens out and $A_{R,221}\to a_{0,221}$ as $q\gg1$.

The ratio between the test-particle limit, $a_{0,221}=A_{R,221}^{q\rightarrow \infty}$, and the equal mass-ratio case $A_{R,221}^{q=1}$ is $\left\lbrace 0.71, 0.71, 0.69, 0.59 \right\rbrace$ for the four values of $t_0$ considered in the plot. This shows that -~independent of the choice of $t_0$~- the relative amplitude does not vary significantly with the mass ratio and can be used to improve the coherent mode stacking algorithm such as the one outlined in Refs.~\cite{Yang:2017zxs,Ota:2019bzl}. This is  especially important for third-generation GW interferometers, where one expects to detect ${\cal O}(100-1000)$ BBH RDs. Improvements to current stacking algorithms using this empirical observation will be explored in a forthcoming work.

\begin{figure*}[]
    \subfloat{\includegraphics[width=0.48\textwidth]{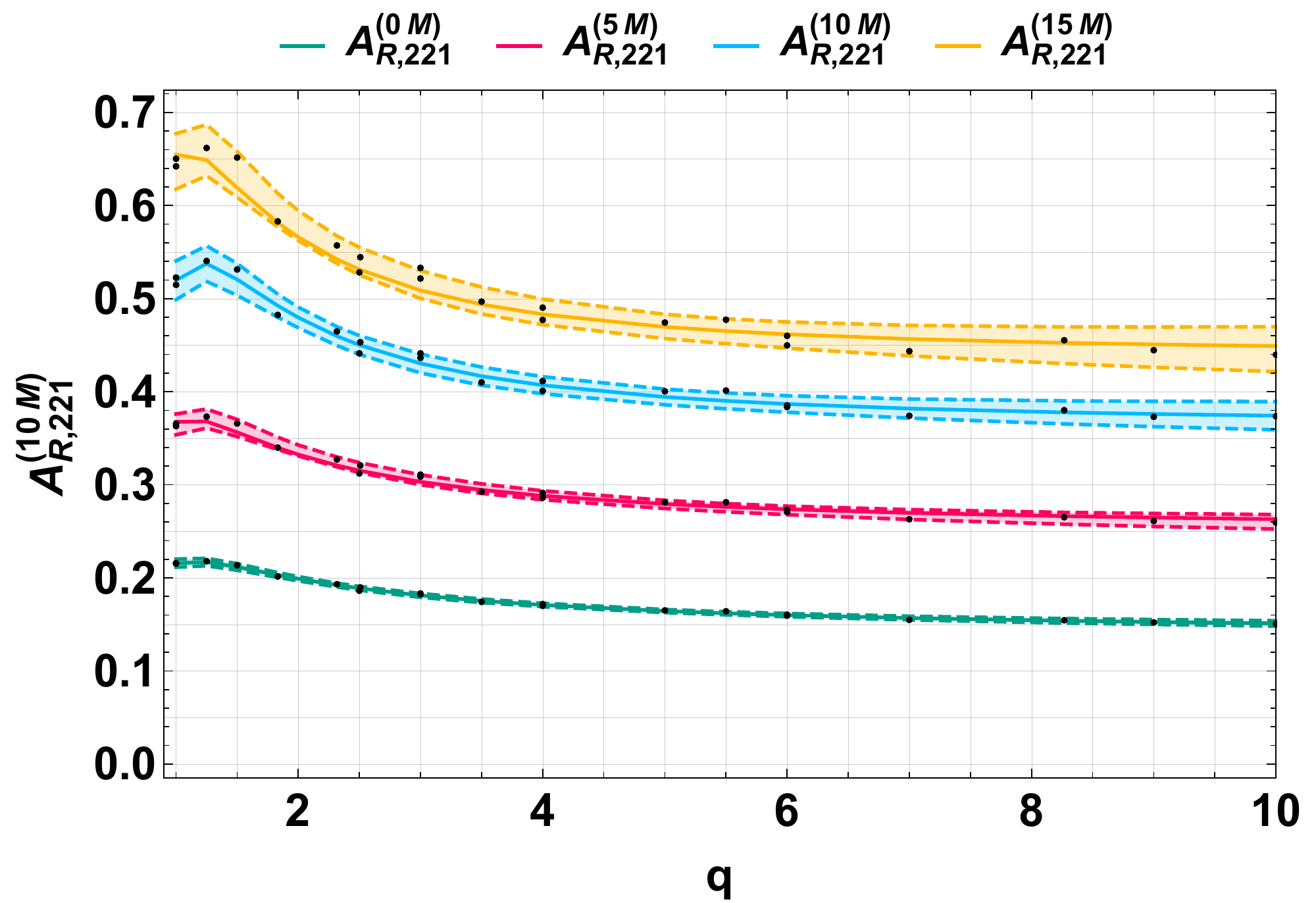}} 
    \subfloat{\includegraphics[width=0.48\textwidth]{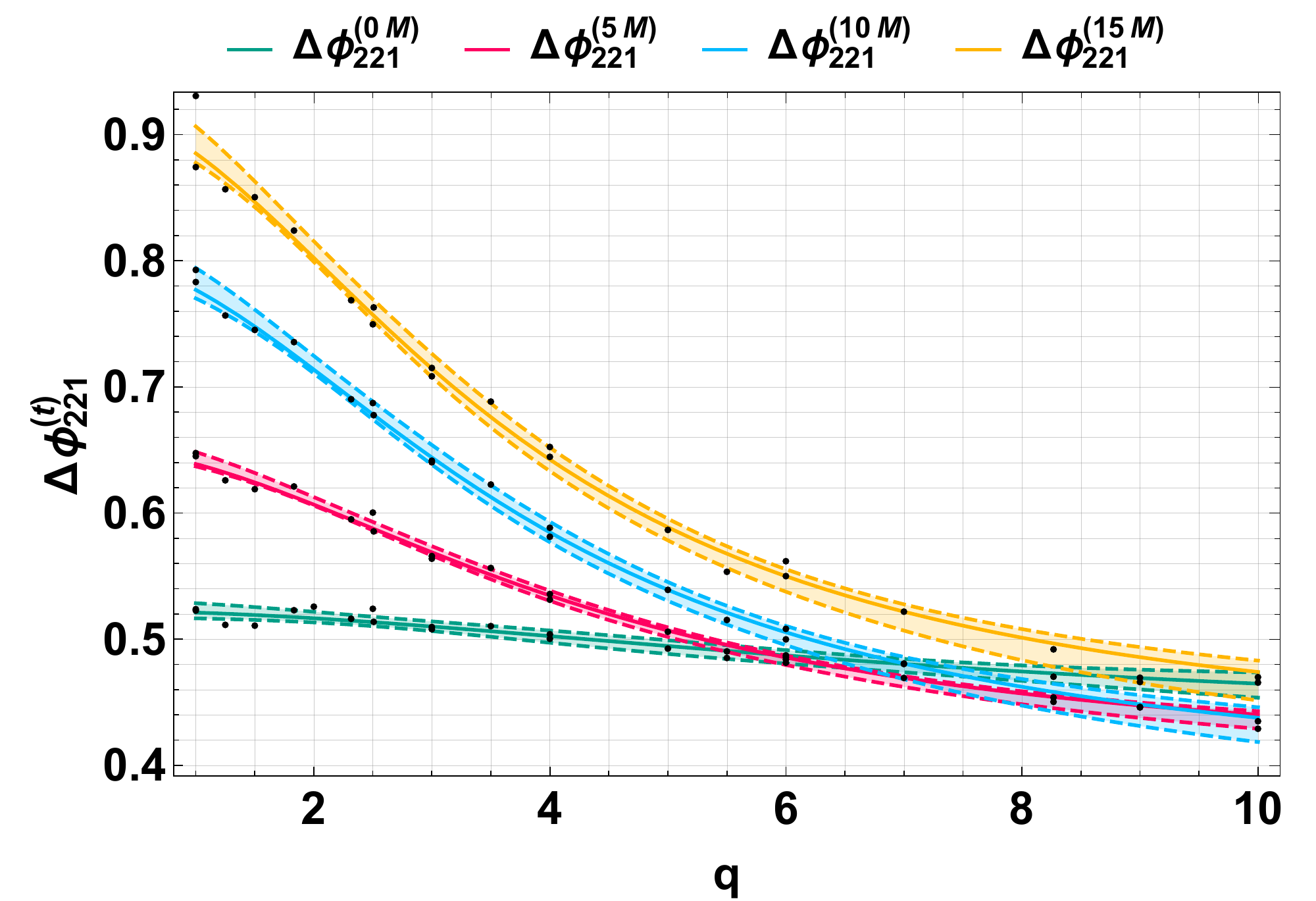}}\\
    \caption{Fits for the rescaled amplitude ratio  $A_{R,221}$  (left panel) and the phase difference (right panel) as a function of $q$. The rescaling~\eqref{eq:symmetry} is performed using $t_0=10M$ as reference starting time. The bands denote the $90\%$ credible intervals for each fit. The symmetry~\eqref{eq:symmetry} is broken for $t_0=0M,5M$ by a large amount and to a minor extend for $t_0=15M$. Although the variations of the phase difference $\Delta \phi_{lmn}$ for $t_0=10M,15M$ are larger than the corresponding $90\%$ credible regions, they are within the phase uncertainty of the NR waveforms; see Appendix~\ref{app:NRerror}}
    \label{fig:fit_and_starttime_221}
\end{figure*}

Note that the amplitude ratios shown in Fig.~\ref{fig:fit_and_starttime_amp} are not rescaled by Eq.~\eqref{eq:symmetry} for different choices of $t_0$. The rescaled amplitude ratios and phase differences are presented in the left and right panels of Fig.~\ref{fig:fit_and_starttime_221}, respectively.
To rescale the ratio in Eq.~\eqref{eq:ratio}, we use Eq.~\eqref{eq:symmetry} for both the numerator and the denominator, i.e.,
\begin{equation}
\label{eq:amp_an_resc}
A_{R,lmn}=\frac{A_{lmn}}{A_{220}}\to A_{R,lmn} e^{-\Delta t(1/\tau_{lmn}-1/\tau_{220})}\,,
\end{equation}
and we rescale the amplitude ratios with respect to that at the reference starting time $t_0=10M$.

As shown in the left panel of Fig.~\ref{fig:fit_and_starttime_221}, the rescaled ratios at $t_0=0$ and $t_0=5M$ (green and red curves) are not compatible with the $90\%$ confidence intervals for all mass ratios and they break the symmetry~\eqref{eq:symmetry} by a large amount. The same is true for $t_0=10M$ and $t_0=15M$ (blue and yellow curves) although, in this case, the differences are smaller for all $q's$ considered. 

The situation is similar for the phase difference $\Delta \phi_{221}$.
From Eq.~\eqref{eq:symmetry}, we should expect that $\Delta \phi_{221}$ be the same independently of $t_0$. Instead, we observe that, for fixed $q$, $\Delta \phi_{221}$ depends on $t_0$ especially when $t_0=0M,5M$ (green and red curves) and mostly for low mass ratios. The differences between the $t_0=10M$ and the $t_0=15M$ curves (yellow) are at most about $10\%$ ($\approx 0.1~{\rm rad}$ absolute error) for low mass ratios, and are within the NR errors on the phase. In Appendix~\ref{app:NRerror} we estimate the latter to be approximately $\pm0.04\,{\rm rad}$.

Using the symmetry criterion in Eq.~\eqref{eq:symmetry}, we argue that one need to wait for at least $t_0\geq 10M$ after the peak amplitude to start a reasonable accurate RD analysis with a $2$-tone RD model.

\begin{figure}[h!]
    \subfloat{\includegraphics[width=0.48\textwidth]{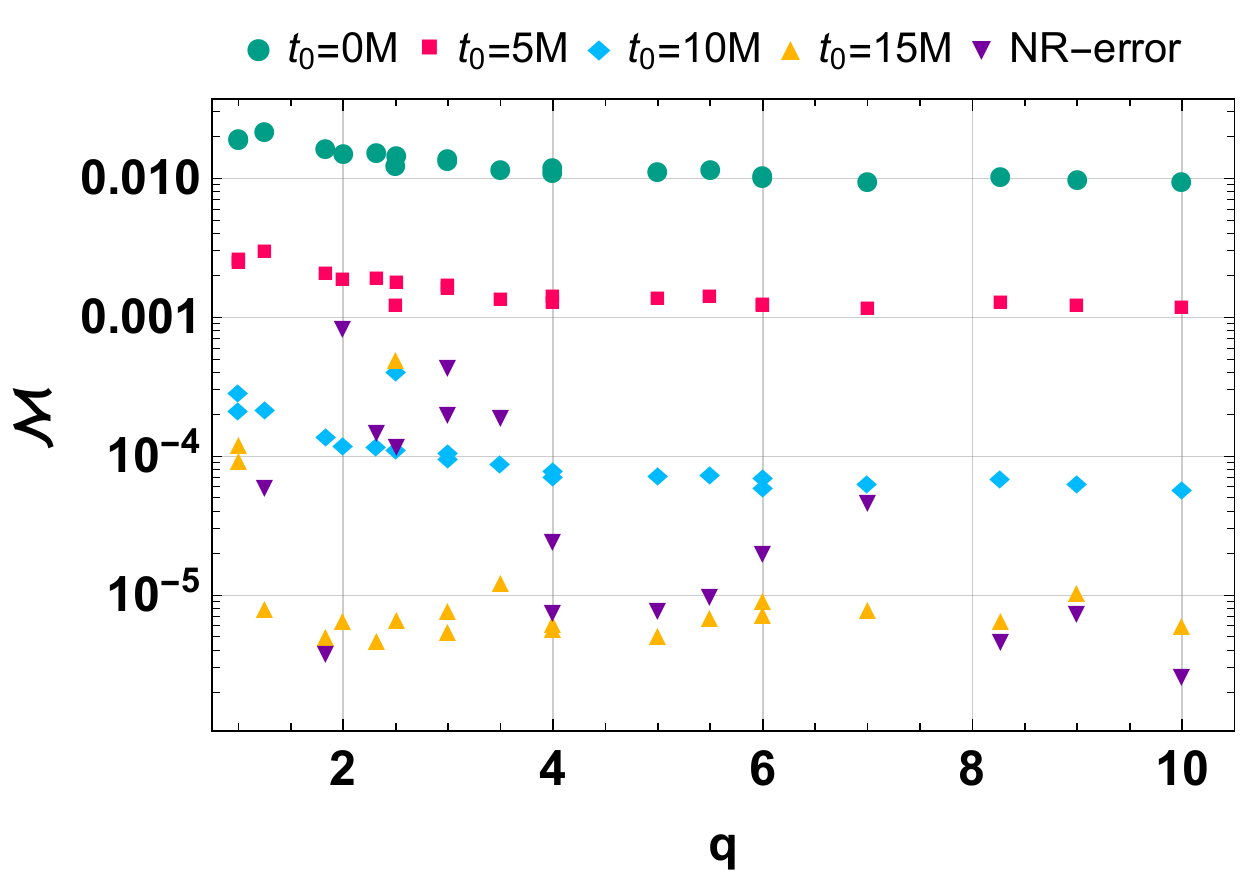}}
    \caption{Measure of the accuracy of the fits for the 2-tone RD model in terms of mismatch between the fits and the NR waveform for $t_0/M=\{0,5,10,15\}$ (green to yellow markers). The mean values for the mismatch are $\left\lbrace 0.0134,0.0016,0.00012,0.00004\right\rbrace$ for $t_0/M=\{0,5,10,15\}$, respectively. As a reference, we also show the NR error (purple triangles) as in Fig.~\ref{fig:matchplot} (see Appendix~\ref{app:NRerror} for details). \xj{Since we compute the NR error from $t_0$, the purple triangles marking the NR errors must be compared with the green dots corresponding to $t_0=0$. Note that the NR errors lie orders of magnitude below the fit corresponding to $t_0=0$, showing that the fit does not completely capture all the features of the NR simulation.} 
    }
    \label{fig:goodness}
\end{figure}

In Fig.~\ref{fig:goodness}, we show the mismatch  between the fits and the NR simulations computed using Eq.~\eqref{eq:mismatch} for $t_0/M=\{0,5,10,15\}$ (green to yellow markers). We also show the NR error estimates similar to Fig.~\ref{fig:matchplot} (purple triangles).
Furthermore, from Fig.~\ref{fig:goodness}, we observe that the NR errors can be important at $t=15M$ for some of the simulations considered in this study. Loosely speaking, the mismatch decreases as $t_0$ increases, confirming that a 2-tone model provides a more accurate fit at late times. 
The mismatch reduces approximately by an order of magnitude for every $5 M$ increase in  $t_0$.

\subsubsection{Fits for angular modes}

Fixing $t_0= 10M$ and $t_0= 15M$ as a reference starting time, we fit for the amplitude ratio and the phase difference of a 2-mode RD model~\footnote{Recall that $t_0\geq10M$ ensures that the peak of the subdominant angular modes is contained in the data.}.
The SXS catalog provides the waveform decomposed into spherical harmonic modes and we use this to perform fits for each angular modes independently. 
Specifically, we consider a signal with a secondary mode with either $l=m=3$ or $l=2$, $m=1$, and $n=0$ in both cases. 

It is worth noticing that the phase difference and the amplitude ratio presented in this section are inherent to the BBH system and do not include extrinsic factors such as the inclination angle. To perform an analysis on the astrophysical BBH events, one needs to add the effect of the inclination angle as it modifies the amplitude ratio and phase difference between the different angular modes. See Appendix~\ref{app:spheroidal} for details.

\begin{figure*}[t]
   \subfloat{\includegraphics[width=0.48\textwidth]{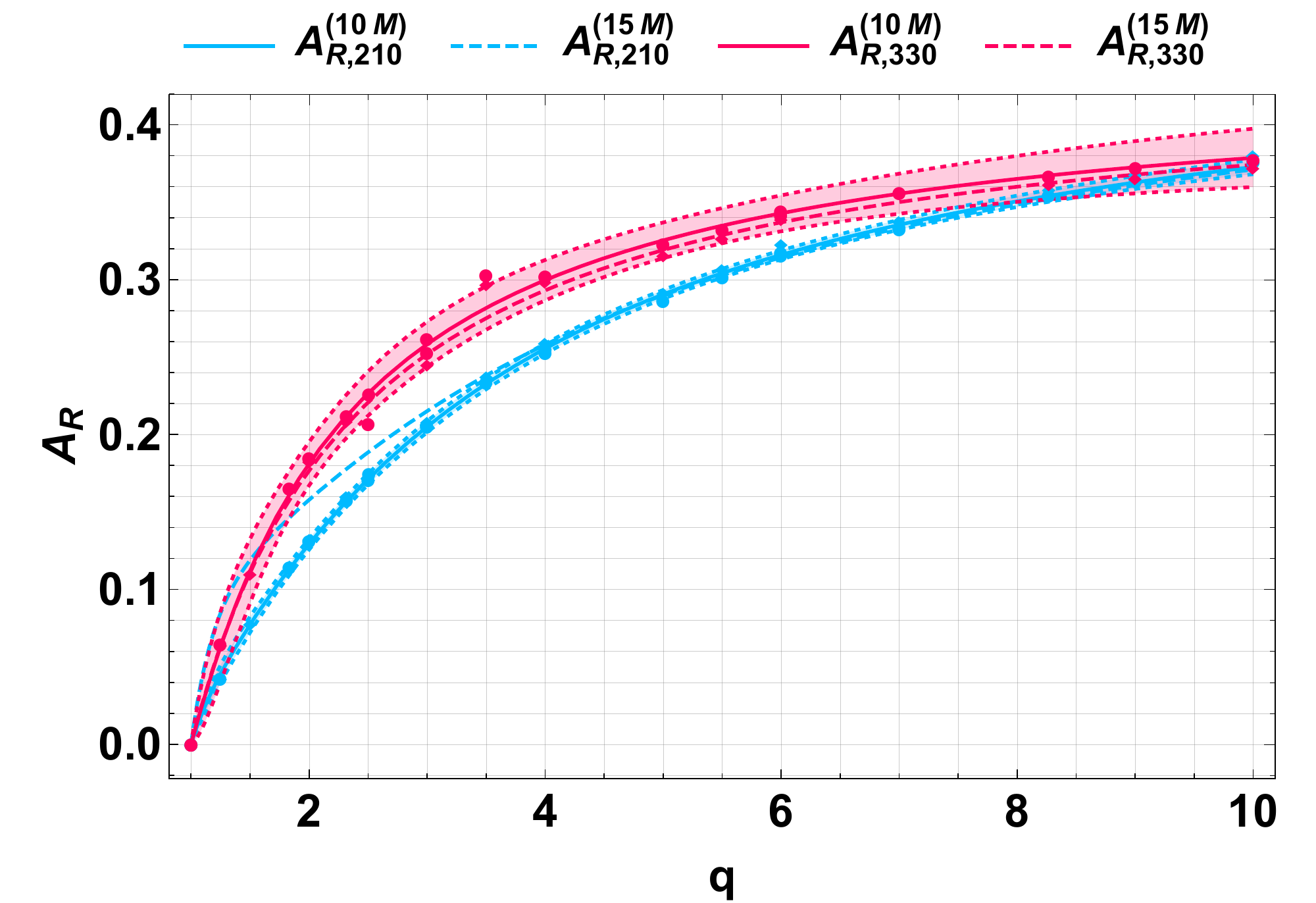}} 
    \subfloat{\includegraphics[width=0.48\textwidth]{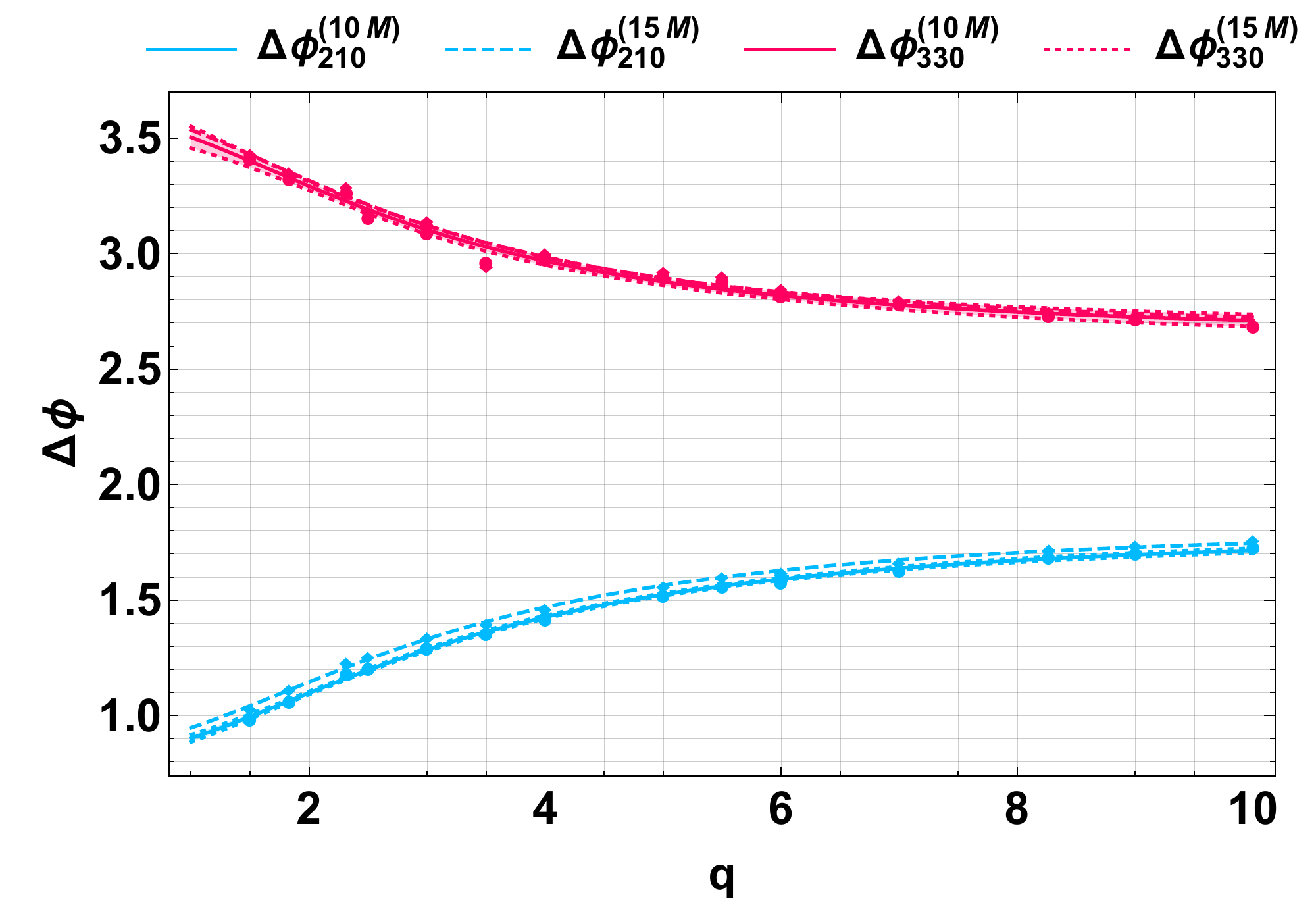}} \\
    \caption{Fits for the rescaled amplitude ratio and phase difference as a function of $q$ using the fundamental $l=m=2$ mode and different (fundamental) angular modes (either $l=m=3$ or $l=2$, $m=1$). Unlike for the overtone case shown in Fig.~\ref{fig:fit_and_starttime_221}, in this case the symmetry~\eqref{eq:symmetry} is preserved between $t_0=10M$ (solid) and $t_0=15M$ (dashed). The bands show the $90\%$ credible intervals of the $t_0=10M$ fit. The NR data are shown with solid circles and diamonds at $t_0=10M$ and $t_0=15M$ respectively.
    } 
    \label{fig:fit_and_starttime_lmn}
\end{figure*}

In Fig.~\ref{fig:fit_and_starttime_lmn}, we show the fits for the rescaled amplitude ratio (left panel) and the phase difference (right panel) corresponding to $t_0=10M,15M$. To illustrate the fit accuracy, we also provide an overlay of the data points that have been fitted. The fits are reasonably accurate (within their own error bars) for both choices of starting time $t_0$. A few data points are scattered with respect to the best-fit curve; we attribute this to the inaccuracy of some NR waveforms.
Similarly, we notice that $A_{R,210}^{(15M)}$ happens to lose accuracy at low mass ratio regime, due to errors of the corresponding NR simulations (see Fig.~\ref{fig:goodness}). Note also that for subdominant angular modes the deviation of the rescaling relative to the symmetry~\eqref{eq:symmetry} is much smaller than the one observed in Fig.~\ref{fig:fit_and_starttime_221} for overtones. This happens because the damping factors $\tau_{lmn}$ for different angular modes $(l,m)$ of equal tone index $n$ are comparable. 
We also note that for $t_0= \{10M,15M\}$, the values for $A_{R,210}$ and $A_{R,330}$ increase monotonically with $q$ until they asymptote to $A_{R,210}=a_{0,210}$ and $A_{R,330}=a_{0,330}$ in the high-$q$ limit. For these modes, the amplitude ratio is constrained approximately in the range $A_{R,lmn} \in \left[0,0.5\right]$.
For the phase difference $\Delta \phi_{lmn}$ we observe two opposite trends for the modes:  $\Delta \phi_{330}$ monotonically decreases with the massratio $q$ whereas the opposite is true for $\Delta \phi_{210}$.
The phase $\Delta \phi_{lm0}$ is constrained within $[0.5 b_{210},b_{210}]$ and $[b_{330},1.3 b_{330}]$ for the two modes considered here. This result is independent of the arbitrary phase alignment at $t_0=10M$ that is used in this study.

\subsubsection{Test-particle limit}
In the test-particle limit (${q\to\infty}$), our fitting formulas~\eqref{eq:amp_an} and~\eqref{eq:phse_an} reduce to ${A_{R,lmn}\to a_{0,lmn}}$ and ${\Delta \phi_{lmn}\to b_{0,lmn}}$. For ${t_0=10M}$, we obtain 
${a_{0,lmn}=\left\lbrace 0.374 \pm 0.07,\, 0.474 \pm 0.1 ,\,0.440 \pm 0.03 \right\rbrace}$ for the ${(l=m=2,n=1),\, (l=2, m=1, n=0)}$ and $(l=m=3, n=0)$ modes respectively, where the uncertainties mark the $90\%$ credible intervals. Our fits are consistent with previous work on the test-particle limit of QNM excitation factors. For instance, we obtain results similar to Ref.~\cite{Berti:2007zu}, wherein the authors estimated the test-particle limit amplitude ratio $\approx 0.43$ for the ${l=m=3}$ mode by fitting a set of nonspinning NR simulations. A different approach was taken in Ref.~\cite{Barausse:2011kb}, where the authors provide an independent estimates of the test-particle limit for amplitudes of the ${l=m=3}$ and ${l=2,m=1}$ modes computed at the peak of the strain, ${t_0=0}$, by solving numerically the Teukolsky equation. Their results at $t_0=10M$ translate\footnote{To translate the results, we need to apply Eq.~\eqref{eq:amp_an_resc} with $\Delta t=10M$.} to $\left\lbrace 0.36, 0.40 \right\rbrace$ for the amplitude ratio of the ${l=m=3}$ and ${l=2,m=1}$ modes, respectively. Although these values are compatible  with our fit errors, the small difference could arise from the numerical errors of the NR waveforms analyzed in this study, especially in the high mass-ratio regime.
Furthermore, in Ref.~\cite{Ota:2019bzl} the amplitude ratio in the test-particle limit is estimated to be approximately $0.3-0.4$ for ${(l=3, m=3, n=0)}$, ${(l=2, m=1, n=0)}$, and ${(l=2, m=2, n=1)}$, also in agreement with our fits.

For what concerns the phase difference, to the best of our knowledge no similar estimates have been made for in the test particle limit that can be used to benchmark our results. For ${t_0=10M}$, we obtain the following phase difference in the test-particle limit: ${b_{0,lmn}=\left\lbrace 0.381 \pm 0.018, 1.808 \pm 0.080 , 2.663\pm 0.103 \right\rbrace}$ for the $(l=m=2,n=1), (l=2, m=1, n=0), (l=m=3, n=0)$ modes, respectively.
The trend descends monotonically for $\Delta \phi_{221}$ and $\Delta \phi_{330}$, whereas it increases for $\Delta \phi_{210}$. Since these values depend on the phase alignment (see Sec.~\ref{sec:parfits}), they may be freely shifted by a constant phase term. However, the quantity physically relevant ${\Delta \phi_{lmn}({q=1)}-b_{0,lmn}}$ remains constant independently of the phase alignment. Thus, at $t_0=10M$, we obtain ${\Delta \phi_{lmn}({q=1)}-b_{0,lmn}}=\left\lbrace 0.400,-0.882 , 0.891 \right\rbrace$ for the ${(l=m=2,n=1), (l=2, m=1, n=0)}$, ${(l=m=3, n=0)}$ modes, respectively.


\section{Quantifying the prospects for BH spectroscopy with overtones or angular modes}

RD signals from BBH mergers serve as a powerful probe to test the nature of the remnant compact object as well as the behavior of gravity around it. Experimental validation of certain predictions of GR can be performed to some extent even when explicit spectroscopy of BBH RD is not feasible. For example: i) an inspiral-merger-RD consistency test is based on the comparison of $\{M_f, a_f\}$ of the final BH, estimated from the inspiral phase, to those estimated directly from the merger-RD phase; ii) even when the signal is not strong enough to measure the subdominant QNMs confidently, an inference can be made that disfavors large deviations from GR by using methods such as the Bayes factor, or by checking whether there is strong support of the posterior distribution of the subdominant QNM parameters in unexpected frequency ranges. These tests are, however, not as stringent as performing an explicit BH spectroscopy and confirming that a \emph{measurement} of the QNM mode frequencies and damping times is consistent with the prediction of GR. BH spectroscopy can then be used to validate the predictions of GR such as the no-hair theorem, the BH uniqueness theorem and, indirectly, the area-increase theorem. 

We note that in this section all the mode amplitudes are computed starting the RD analysis at $t=t_0$ (see Tables~\ref{tab:fitovertones}, \ref{tab:fitmodes210}, and \ref{tab:fitmodes330}) without performing the rescaling discussed above.

\subsection{Detectability, resolvability, and measurability criteria for BH spectroscopy}
\label{sec:crit}

The possibility to perform BH spectroscopy relies on some necessary criteria that a given RD signal should satisfy. To quantify this issue, for a 2-mode or 2-tone RD model we define the following criteria:
\begin{enumerate}
    \item \emph{Detectability criterion}:
     \begin{equation}
         \sigma_{A_{R}} < A_{R} \label{eq:det}
     \end{equation}
where $A_{R}$ is the amplitude ratio between the dominant and the subdominant mode, and $\sigma_{A_{R}}$ is the uncertainty in the recovery of the parameter $A_{R}$. We name this as the detectability criterion as it is a necessary (albeit not sufficient) condition to claim the presence of the subdominant mode in the RD signal at a $1 \sigma$ level\footnote{\xj{See also Appendix B of Ref.~\cite{Berti:2007zu}, where a different criterion for detectability (therein called "amplitude resolvability") of angular modes was adopted. Here we use the name "detectability" to distinguish this criterion from the resolvability of the frequency and damping time discussed later on.}}.
\item \emph{Resolvability criterion}: 
\begin{subequations}
     \begin{align}
         \max[\sigma_{f_{220}}, \sigma_{f_{\rm sub}}]&<|f_{220} - f_{\rm sub}|\,, \label{eq:resf}\\
         \max[\sigma_{Q_{220}}, \sigma_{Q_{\rm sub}}]&<|Q_{220} - Q_{\rm sub}|\,, \label{eq:resQ}
     \end{align}\label{eq:res}
\end{subequations}
where $\sigma_{X}$ is the uncertainty in recovering a quantity $X$, $f_{220}$ and $f_{\rm sub}$ (respectively, $Q_{220}$ and $Q_{\rm sub}$) are the QNM frequencies (respectively, quality factors, with $Q_{lmn}=\pi f_{lmn} \tau_{lmn}$) of the dominant mode and subdominant mode, respectively. 
Henceforth, we call this the resolvability criterion as it ensures that a putative measurement of the subdominant mode can be resolved from that of the dominant mode at a $1 \sigma$ level. Note that this is the traditional Rayleigh resolvability criterion that was introduced in the context of BBH RD in Refs.~\cite{berti:2005ys,Berti:2007zu}. The resolvablity criterion requires that the RD must satisfy either Eq.~\eqref{eq:resf} or Eq.~\eqref{eq:resQ}, depending on which quantity is measured (see below). We expect that for angular modes it is much easier to satisfy Eq.~\eqref{eq:resf}, whereas for overtones it is much easier to satisfy Eq.~\eqref{eq:resQ}, since their frequency is close to that of the fundamental mode.  
 \item \emph{Measurability criterion:}
      \begin{subequations}
 \begin{align}
     \left\{  \frac{\sigma_{f_{220}}}{f_{220}}, \frac{\sigma_{Q_{220}}}{Q_{220}}, \frac{\sigma_{f_{\rm sub}}}{f_{\rm sub}} \right\} &\leq T \,, \label{eq:measf}\\
     \left\{  \frac{\sigma_{f_{220}}}{f_{220}}, \frac{\sigma_{Q_{220}}}{Q_{220}}, \frac{\sigma_{Q_{\rm sub}}}{Q_{\rm sub}} \right\} &\leq T\,,\label{eq:measQ}
 \end{align}
      \end{subequations}
 where $T$ is a given threshold. In other words, we require that at least three QNM parameters are measured within a relative accuracy $T$. For concreteness, below we shall consider $T= \{1, 5, 10 \}\%$.  The combined requirement of measurability and resolvability imposes Eqs.~\eqref{eq:measf} and~\eqref{eq:resf} \emph{or} Eqs.~\eqref{eq:measQ} and~\eqref{eq:resQ}.
\end{enumerate}

We define the minimum SNR that allows for a spectroscopic analysis of the BH as the one for which \emph{all of the above three conditions} [i.e., either Eqs.~\eqref{eq:det},~\eqref{eq:resf}, and~\eqref{eq:measf} or Eqs.~\eqref{eq:det},~\eqref{eq:resQ}, and~\eqref{eq:measQ}] are satisfied.

We estimate the errors on the RD waveform parameter using a Fisher matrix framework, which is valid in the high SNR limit and when the statistical properties of the noise can be assumed to be Gaussian.  Therefore, our estimates of the minimum SNR required for performing BH spectroscopy are \emph{optimistic} lower bounds. The details of the Fisher-matrix framework used in this study are similar to that described in Sec.~III of Ref.~\cite{Bhagwat:2019dtm}, to which we refer for technical aspects.
In particular, we note that the analysis does not depend significantly of the noise curve of the detector, since the latter is approximately flat in the frequency range of interest~\cite{berti:2005gp}. 
Since the amplitude ratio for different angular modes depends on the inclination angle of the source, we have averaged out the GW strain on the location angles $\theta$ and $\phi$, as discussed in Appendix~\ref{app:spheroidal}.
\subsection{Progenitor mass ratio and its effect on BH spectroscopy}

The amplitudes to which different RD modes are excited depend on the properties of the progenitor system, in particular the mass ratio and spins of the two coalescing BHs. As a rule of thumb, the more the asymmetry in the BBH system i.e., the higher the mass ratio and spins, the larger is the excitation of angular modes. However, the dependence of overtone excitation on the progenitor mass ratio is less intuitive. We note from Fig.~\ref{fig:fit_and_starttime_221} that the amplitude ratio for the $n=1$ overtone slightly decreases as the mass ratio of the BBH increases, while the opposite is true for angular modes (see Fig.~\ref{fig:fit_and_starttime_lmn}).

The $l=m=2,\, n=0$ mode is always the dominant one when the progenitor BBH system undergoes inspiral, plunge and merger; these are the main signals of interest for GW RD tests using LIGO/Virgo. As in the previous section, we consider the following three subdominant modes: a) $l=m=2,\, n=1$;  b) $l=m=3,\, n=0$; c) $l=2,\, m=1,\, n=0$.

As discussed in our earlier work~\cite{Bhagwat:2019dtm}, picking the optimal value of starting time depends on the interplay between the systematic error of RD modeling and statistical uncertainty due to SNR. For overtones, here we adopt a more agnostic approach and perform the analysis using four different choices of the starting time, $t_0/M=\{0,5,10,15\}$. As discussed in the previous section, the fact that the symmetry~\eqref{eq:symmetry} is broken for $t_0=0, 5M$, and to a minor degree at $t_0 =15M$, shows that a simple 2-tone RD model is not accurate at early times, and one should include higher-order overtones to accurately capture features close to the peak amplitude of the waveform~\cite{Isi:2019aib,Giesler:2019uxc}. We include $t_0=0,5M$ for the sake of completeness and because using two tones is the minimum requirement to perform BH spectroscopy (measuring \emph{three} modes/tones will require even higher SNR). Nonetheless, as discussed below, our qualitative results do not significantly depend on the choice of $t_0$. 

On the other hand, from Fig.~\ref{fig:fit_and_starttime_lmn}, we note that the amplitude ratio between the angular modes does not change significantly with the starting time of RD, and therefore, for these, we assume that the RD starts $15 M$ after the peak amplitude of the waveform (note that any other choices in the range $t_0\in[10,15]M$ would not change our analysis significantly).

Although the total mass of the progenitor BBH system is a simple rescaling factor for all dimensionful quantities, for concreteness, we set $m_1+m_2=70 \Msun$ and rescale all quantities accordingly.

We can now proceed to study the minimum SNR required to satisfy the detectability, resolvability and measurability criteria for different values of $q$.

\begin{figure}[th]
    \includegraphics[width=0.49\textwidth]{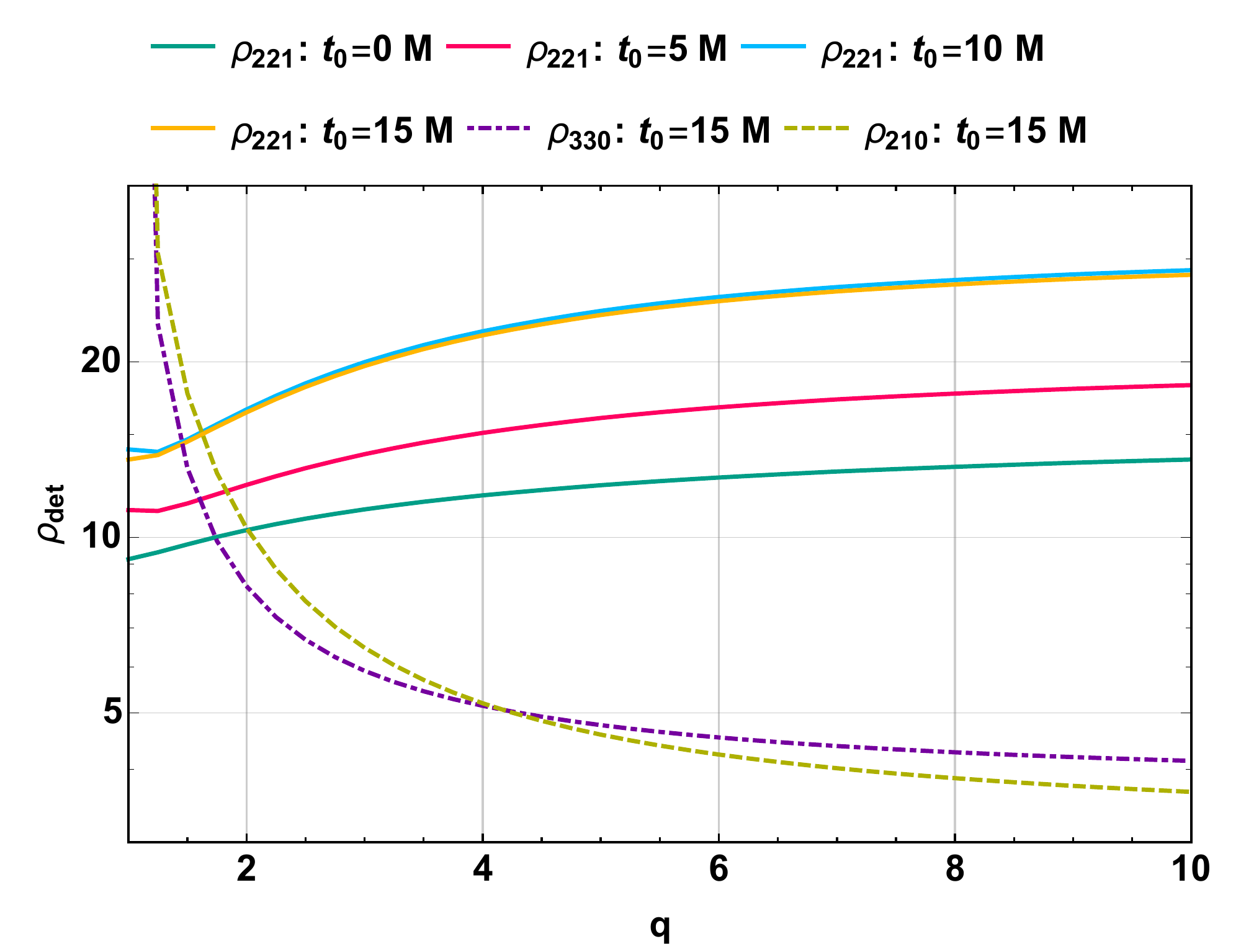}
     \caption{The minimum SNR required to detect the subdominant mode [see Eq.~\eqref{eq:det}]. The dotted purple curve corresponds to the angular mode $l=m=3, n=0$ and the dashed green curve corresponds to $l=2, m=1, n=0$ mode, with $t_0=15M$ as starting time. The continuous teal, pink, blue and orange curves show the minimum SNR for detectability of the overtone for different starting times, $t_0/M=\{0,5,10,15\}$. 
     }
     \label{fig:det}
 \end{figure}

\begin{figure*}[]
\subfloat{\includegraphics[width=0.99\textwidth]{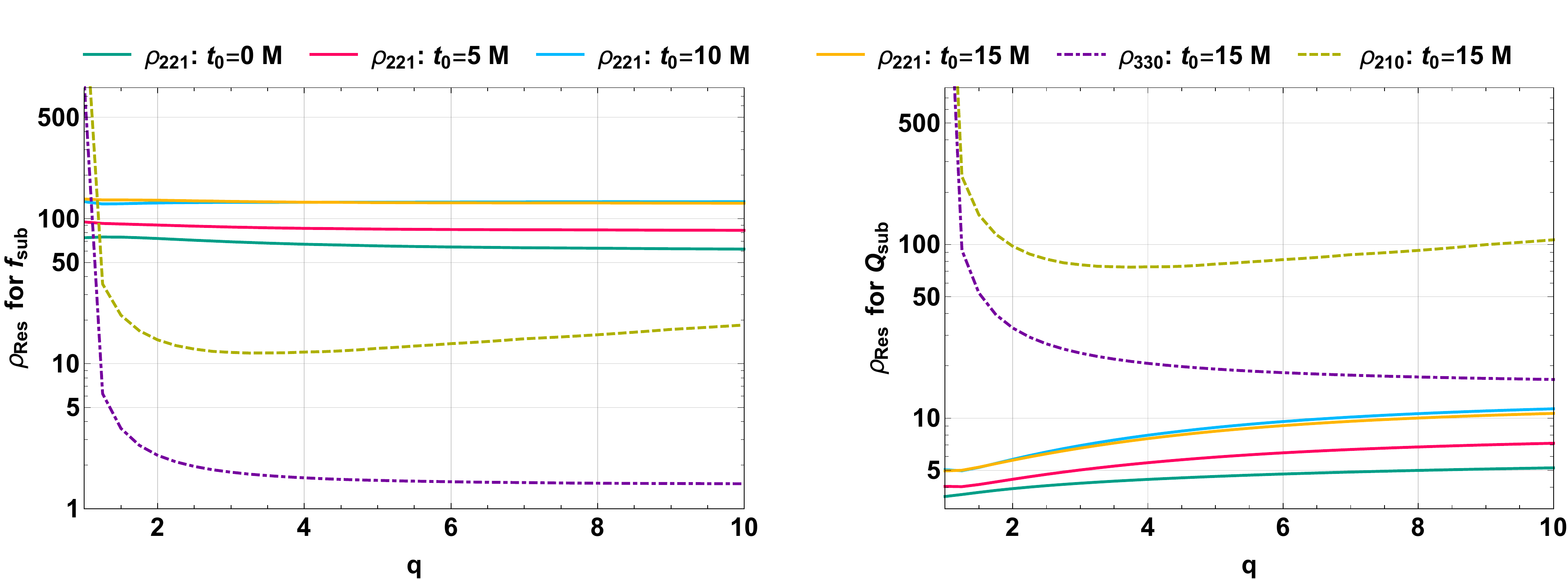}}  
     \caption{The resolvability criterion [see Eq.\eqref{eq:res}]. We show the minimum SNR required to resolve the subdominant mode frequency $f_{\rm sub}$ (left panel) and quality factor $Q_{\rm sub}$ (right panel) from the dominant mode. The color scheme is identical to that used in Fig.~\ref{fig:det}. Notice that $Q_{\rm sub}$ for the overtones and $f_{\rm sub}$ for the angular modes are easily resolved.
     }  
     \label{fig:res}
 \end{figure*}

\begin{figure*}[th!]
\subfloat{\includegraphics[width=0.99\textwidth]{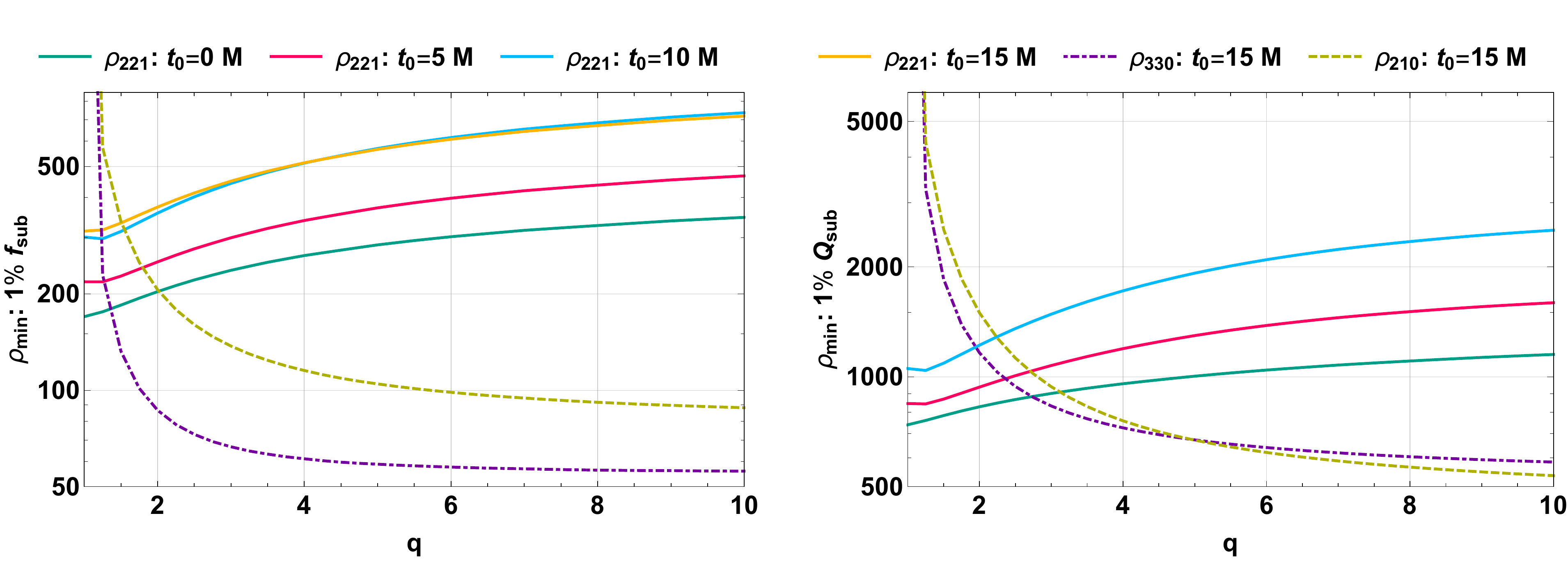}} \\
     \subfloat{\includegraphics[width=0.99\textwidth]{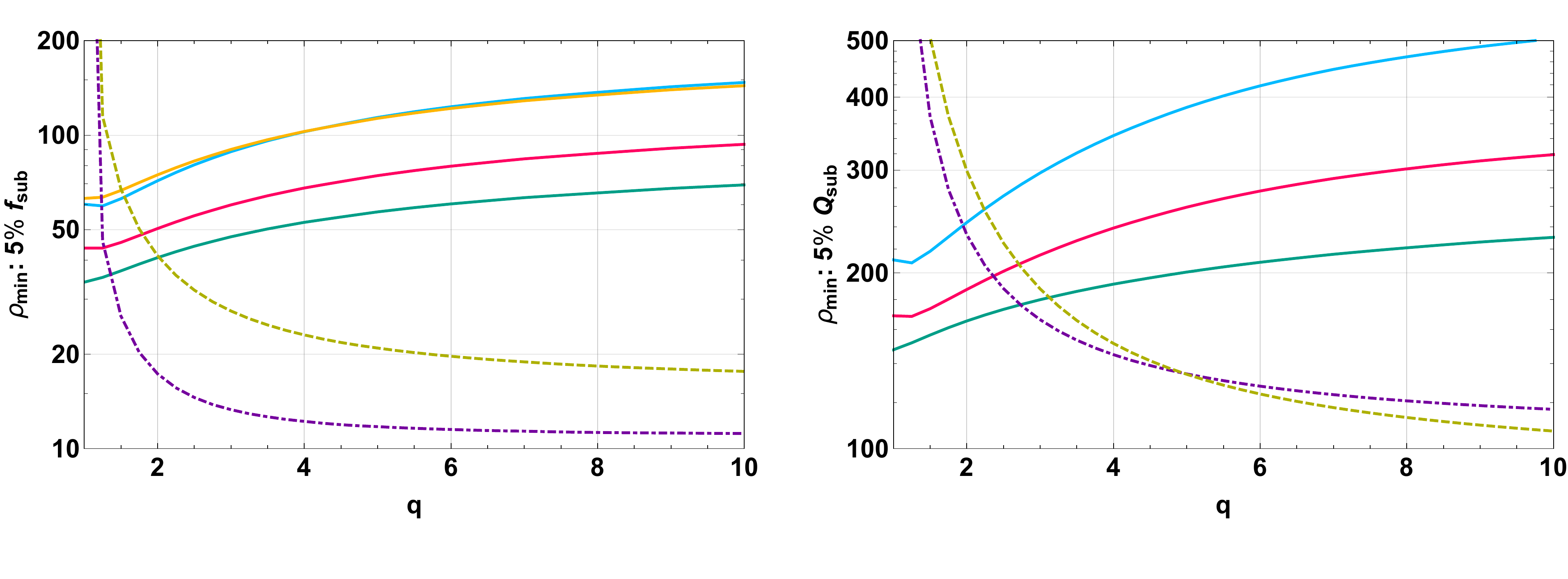}} \\
     \subfloat{\includegraphics[width=0.99\textwidth]{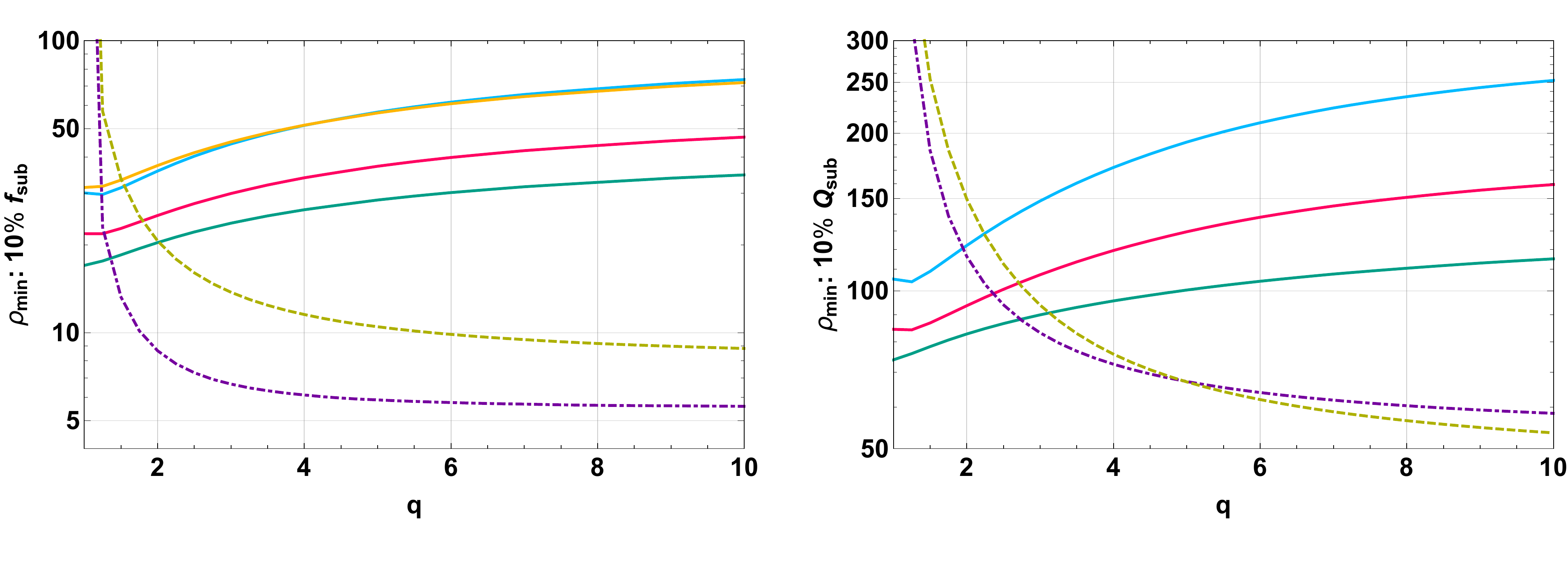}}
     \caption{Measurability of the subdominant mode. Top to bottom panels show the minimum RD SNR required to measure the QNM parameters with $10 \%, 5 \%$ and $1 \%$ precision, respectively. We use the same color scheme as in Fig.~\ref{fig:res}. The left and right panels correspond to the recovery of the frequency and of the quality factor, respectively. Notice that for a given RD SNR, the subdominant mode frequency can be measured with much less uncertainty compared to the quality factor.
     The orange curve corresponding to $l=m=2$, $n=1$ for $t_0=15M$ does not appear in the right panels since it is above the scale of the plots. 
     }  
     \label{fig:res-mes}
 \end{figure*}

\begin{figure*}[]
     \subfloat{\includegraphics[width=0.99\textwidth]{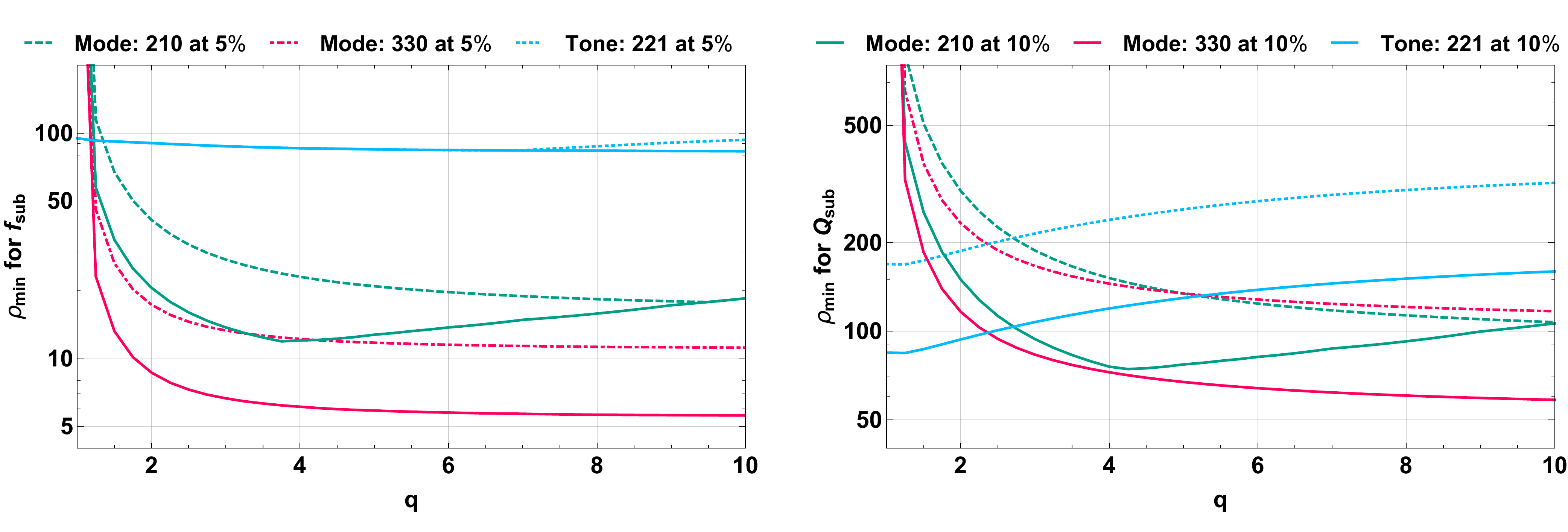}} 
     \caption{Minimum SNR for BH spectroscopy with a secondary mode, combining the detectability, resolvability, and measurability criteria [Sec.~\ref{sec:crit}]. Left panel refers to measuring and resolving the triad $\{f_{220},Q_{220},f_{\rm sub}\}$, whereas the right panel refers to $\{f_{220},Q_{220},Q_{\rm sub}\}$. The solid teal, pink and blue curves correspond to measurement with $10 \%$ precision of  $l=2, m=1, n=0$, $l=m=3, n=0$ and $l=m=2, n=1$ subdominant mode frequency, respectively. The dot-dashed curves correspond to $5 \%$ precision.  In this figure, for the overtones, we choose $t_0=5M$. For unequal mass-ratio BBHs, $l=m=3$ is the most promising subdominant mode to perform BH spectroscopy and for near-equal mass ratio systems, overtone performs better. This is a conservative estimate, since the choice of small values of $t_0$ leads to higher SNR in the overtones. 
     }
     \label{fig:SNRmin-final}
 \end{figure*}

First, we investigate the minimum SNR required to satisfy the three criteria listed in Sec.~\ref{sec:crit} individually and present the results in Figs.~\ref{fig:det}, \ref{fig:res} and \ref{fig:res-mes}. Then, in Fig.~\ref{fig:SNRmin-final} (which is one of the main results of this work), we combine the three criteria and provide the minimum SNR required in the RD to allow for BH spectroscopy using angular modes and overtones. 
We note that the results of this section are based on the amplitude ratios and phase difference obtained for the list of NR simulations in Table~\ref{tab:NR_set}. In Appendix~\ref{app:contour} we provide contour plots for the minimum SNR needed to satisfy the three criteria above as a function of $A_R$ and $\Delta\phi$, i.e., for generic initial configurations.

In Fig.~\ref{fig:det} we present the minimum SNR required in the RD to satisfy the detectability criterion as a function of the mass ratio of the BBH system. We find that the SNR required for detectability of the subdominant angular mode rapidly decreases with increasing mass ratio while that for overtones slowly increases. For a nearly equal-mass system, an extremely high SNR is required for detecting the subdominant angular mode (asymptoting to infinity for an equal mass system) while an SNR between $8$ to $15$ is sufficient for the case of overtone, depending on the starting time of RD assumed in the analysis. From this plot we see that subdominant angular modes become a more promising candidate for detection when $q\gtrsim 1.5$.

Next, in the left panel of Fig.~\ref{fig:res} we present the minimum SNR required in the RD to satisfy the resolvability criterion\footnote{Our results for the resolvability of angular modes agree (both numerically and analytically) with those of Ref.~\cite{berti:2005ys} (the latter differ by a factor of $2\pi$ with the results of Ref.~\cite{Berti:2007zu}). However, note that, differently from Refs.~\cite{berti:2005ys,Berti:2007zu} we also include $\Delta\phi$ as one of the parameters of the Fisher matrix. Compared to the resolvability criterion used in Ref.~\cite{Ota:2019bzl}, our results for the overtones and angular modes are in agreement, after taking into account two notable differences: (i)~we use $Q_{\rm sub}$ instead of $\tau_{\rm sub}$ in the RD model. Since $dQ_{\rm sub}/Q_{\rm sub}\sim df_{\rm sub}/f_{\rm sub}+ d\tau_{\rm sub}/\tau_{\rm sub}$, it is easier to resolve $Q_{\rm sub}$ rather than $\tau_{\rm sub}$ (we thank Cecilia Chirenti and Iara Ota for this comment). Assessing whether using $Q_{lmn}$ or $\tau_{lmn}$ in a proper RD analysis is more convenient is an interesting question that we postpone for future work; (ii)~we require resolvability of \emph{either} frequencies \emph{or} quality factors, whereas Ref.~\cite{Ota:2019bzl} required resolvability of both frequencies and damping times.\label{footnote}} for the frequency, Eq.~\eqref{eq:resf}. We find that, except for nearly equal-mass BBHs, it is always easier to resolve the subdominant angular mode frequency from the dominant mode frequency. The angular mode $l=m=3$ generally has a QNM frequency that is separated from the dominant mode frequency by tens of Hz (for total mass $m_1+m_2=70\Msun$) and can be resolved even at a low SNR, $\rho\sim 3$. For angular modes, therefore, frequency resolvability is not a limiting factor for BH spectroscopy unlike for the case of overtones. 

Furthermore, in the right panel of Fig.~\ref{fig:res} we present minimum SNR required to resolve the quality factor of the subdominant mode $Q_{\rm sub}$ from that of the dominant mode, Eq.~\eqref{eq:resQ}. We find that $Q_{\rm sub}$ can be resolved for $\rho\approx 10$ or smaller for the case of overtones for all mass ratios considered in this study. Conversely, for nearly equal mass-ratio systems, resolving $Q_{\rm sub}$ for angular modes requires very high SNR: $\rho\sim 20-100$ for $q\gtrsim 2$.

In Fig.~\ref{fig:res-mes} we present the minimum SNR required to measure the subdominant mode frequency (left panel) and quality factor (right panel) to a precision of $10 \%$, $5 \%$, and $1\%$ (top to bottom). The uncertainty in the measurements of QNMs using overtones slowly increases at higher mass ratios, whereas the uncertainty in the measurements of QNMs using angular mode decreases rapidly as the mass ratio of the progenitor BHs of the BBH system increases. 
We find that, for a mass ratio greater than $1.5$, the  $l=m=3, n=0$ subdominant angular mode allows for the most precise measurement among the subdominant modes considered in this study. For a near equal mass system, since angular modes can be measured poorly, the overtones provide a much better measurement precision.  

Furthermore, comparing the SNR values in the left panel with that in the right panel, we infer that, for a given SNR, the frequency of a given mode can be estimated with higher accuracy than its quality factor, as expected. 

Finally, in Fig.~\ref{fig:SNRmin-final}, we combine the three above criteria (Sec.~\ref{sec:crit}) to establish the minimum SNR required to perform BH spectroscopy. 
Solid (dashed) lines provide an estimate of the minimum SNR required to perform  BH spectroscopy to a level of $10 \%$ (\emph{$5 \%$}) accuracy. To compare the performance of a set of modes/tones, one needs to compare the solid (dashed) curves to solid (dashed) curves in this figure.
Among the three criteria discussed in Sec.~\ref{sec:crit}, which is the most stringent one depends on the value of $q$. The transition from one criteria to another corresponds to the derivative discontinuity seen in Fig.~\ref{fig:SNRmin-final}. For example, for the $l=2, m=1, n=0$ mode the most stringent criterion at $q\approx1$ is detectability, whereas for $q\gtrsim 3.7$ is resolvability, and for intermediate values of $q$ the most stringent criterion is measurability.

To summarize, we find that for all values of $q \gtrsim 1.2$, the $l=m=3$ mode allows for the best spectroscopic analysis of BBH RD, with a minimum required SNR $\approx 10$ or less when $q\gtrsim 1.75$ for a $10\%$ accuracy test using the combination ${\left\lbrace f_{220}, Q_{220}, f_{sub}\right\rbrace}$ and slightly larger for a $5\%$ accuracy test. However, the minimum SNR required for BH spectroscopy with angular modes quickly deteriorates as $q\to 1$ and becomes larger than that required with overtones in this limit. For nearly equal mass systems ($q\lesssim 1.2$), overtones allow us to perform a spectroscopic analysis of the RD at a $10 \%$ accuracy level, but only for high SNR RD signals ($\rho_{\rm RD}\gtrsim 100$).
\section{Discussion}
\label{sec:discussion}
In this work we have investigated the prospects for a spectroscopic analysis of the BBH RD using the angular modes as well as the overtones. Specifically, we have investigated the minimum SNR required to detect, resolve, and measure the QNM frequencies and quality factors of a single subdominant mode/tone for three specific choices: the $l=m=3$ and $l=2, m=1$ subdominant angular modes (both with $n=0$), and the first overtone ($l=m=2, n=1$) of the dominant angular mode.

We presented the fits for the amplitude ratio $A_{R,lmn}$ and phase difference $\Delta \phi_{lmn}$ of the subdominant mode using the NR RD waveforms corresponding to nonspinning BBHs with different mass ratios and for different choices of the RD starting time $t_0$. 
We found that choosing $t_0 \gtrsim 10 M$ is appropriate to reduce the impact of the systematic errors in the RD modeling (either due to a limited number of overtones or to possible nonlinearities). However, this choice of the starting time is impractical for a 2-tone RD model, as the overtones are quickly damped. Therefore, for the cases of a subdominant overtone the analysis has been repeated for a starting time of $\{0 M, 5 M, 10 M, 15 M \}$; an appropriate value needs to be chosen using techniques such as that outlined in Ref.~\cite{Bhagwat:2019dtm} (see also Appendix~\ref{app:Bias}). Our amplitude fits are consistent with the results in Ref.~\cite{Ota:2019bzl}.

The amplitude ratio $A_{R,lmn}$ has been fit to a fourth-degree polynomial while the phase difference $\Delta \phi_{lmn}$ has been fit to a Lorenzian function.
In the range $q \in [1,10]$, the amplitude ratio $A_{R,lmn}$ is approximately bounded within $~[\,a_{0,221},2.5\, a_{0,221}]$, $~[0, \,{a_{0,330}}]$, $~[0, \,{a_{0,210}}]$, where $a_{0,lmn}$ (Tables~\ref{tab:fitovertones}, \ref{tab:fitmodes210}, and \ref{tab:fitmodes330}) is the asymptotic value at $q\to\infty$. Similarly, for the phase difference we obtain that ${|\Delta \phi_{lmn}({q=1)}-b_{0,lmn}|}\lesssim 0.9$ for the all three modes considered.
The fact the amplitude ratio and phase difference do not change significantly across the entire range of $q$ is an interesting empirical feature that could be used to tailor a more efficient
mode stacking algorithm for RD tests of GR~\cite{Yang:2017zxs,Ota:2019bzl}. We plan to explore this issue in a future work.

One of the main results of this work is the estimate of the minimum SNR required for BH spectroscopy with 2-modes/tones as a function of the mass ratio $q$ of the progenitor BBH. For this, we introduced a \textit{detectability} criterion -~i.e., whether the error in the secondary mode amplitude ratio is smaller than $100\%$~-, a \textit{resolvability} criterion -~i.e., whether the secondary QNM can be resolved from the dominant $l=m=2$, $n=0$ mode~-, and a \textit{measurability} criterion -~i.e., whether we can constrain at least three out of the four QNM parameters of a 2-mode/tone model within a threshold accuracy. 
By combining these three criteria we found that BH spectroscopy with angular modes (especially, $l=m=3$) is the most promising channel for nonspinning BBHs with $q\gtrsim 1.2$. Our conclusions differ from those of~\cite{Ota:2019bzl} for unequal-mass binaries due to the different set of criteria we require to perform BH spectroscopy (see Sec.~\ref{sec:crit} and footnote~\ref{footnote}), but our results are otherwise in good agreement.

In particular, BH spectroscopy could be successfully performed  with the $l=m=3$ mode and at $10\%$ accuracy level if a signal with $\rho_{\rm RD}\approx 8$ and $q \sim 2$ is observed. Analogously, the minimum RD SNR required for these tests is $\rho_{\rm RD}\approx 20$ if the $l=2$, $m=1$ mode is used.
On the other hand, the minimum required SNR for BH spectroscopy with angular modes diverges in the $q\to1$ limit since in this case $l=m=3$ and $l=2$, $m=1$ modes are not excited. For nonspinning binaries with $q\lesssim1.2$, BH spectroscopy with overtones becomes more convenient, although it requires very loud signals ($\rho_{\rm RD}\gtrsim 100$).
Such loud signals might be detectable with third-generation ground-based detectors~\cite{hild:2010id,evans:2016mbw,essick:2017wyl} (such as the Einstein Telescope and Cosmic Explorer) and with the future space mission LISA~\cite{audley:2017drz}. However, in these cases (especially for LISA), binaries with a broad distribution of mass ratios are expected so the actual benefit of overtone BH spectroscopy might be limited. We will explore this issue in a future work.
Future instruments will also detect RD signals at higher SNR, thus increasing the chances to detect more than 2-modes/tones. Besides allowing for independent BH spectroscopy tests, including more overtones will reduce the model systematics and will allow to start the RD analysis earlier after the merger.

Another possible extension of our analysis is to include spinning binaries (along the lines of Refs.~\cite{Baibhav:2017jhs,Baibhav:2018rfk}) and quantify the role of the binary component spins in the excitation and detectability of a secondary QNM (see also Appendix~\ref{app:contour}. This is particularly relevant in the light of GW190412, which has a nonnegligible effective spin parameter~\cite{LIGOScientific:2020stg}.
We note that, in our notation, $q_{\text{\tiny GW190412}}\approx 3.8$. Although the RD SNR for this system was low, our results show that a putative spectroscopic analysis of GW190412 would have been much more accurate if performed with angular modes rather than with overtones.

\section*{Acknowledgements}
We are indebted to Cecilia Chirenti and Iara Ota for useful suggestions on the draft and for comparing some of their results with ours,  to Emanuele Berti and Vitor Cardoso for useful correspondence on Refs.~\cite{berti:2005ys,Berti:2007zu} and to Lionel London for useful discussions on the symmetry tests.
We acknowledges financial support provided under the European Union's H2020 ERC, Starting Grant Agreement No.~DarkGRA--757480, and under the MIUR PRIN and FARE programmes (GW-NEXT, CUP:~B84I20000100001).
We also acknowledge networking support by the COST Action CA16104 and support from the Amaldi Research Center funded by the MIUR program ``Dipartimento di Eccellenza'' (CUP: B81I18001170001).


\appendix

\section{NR error estimate}
\label{app:NRerror}

Truncation errors in NR waveforms result from the discreteness of the grid that defines the domain of the simulations~\cite{boyle:2019kee,Hinder:2013oqa}. This error gets reduced as one increases the sampling of the numerical domain, that is, by increasing the resolution of the numerical grid. To estimate the NR error of a single NR simulation we need to compare the numerical outputs that result from changing the grid resolution.

\begin{figure}[th]
    \subfloat{\includegraphics[width=0.49\textwidth]{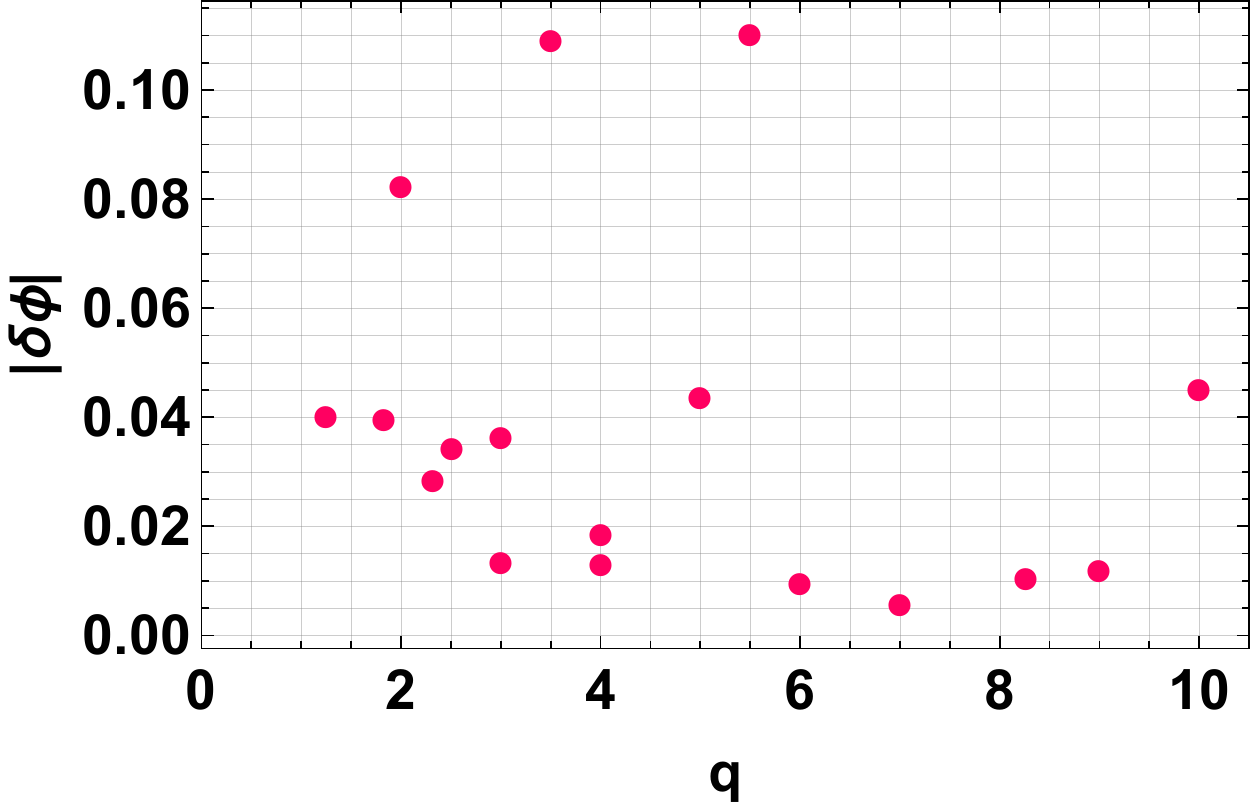}} 
    \caption{\xj{Deviation accumulated on the phase $|\delta\phi|$ between two waveforms with the two highest resolutions ($N:N-1$)
    for each of the simulations in the catalog~\cite{sxscatalog,boyle:2019kee}. The mean values are both approximately $\langle |\delta\phi|\rangle=0.04\,{\rm rad}$.}}
    \label{fig:phasedev}
\end{figure}

The SXS catalog~\cite{sxscatalog} provides the data for different resolution levels, labeled as \emph{Lev-N}, where the resolution increases with the value of N. The NR error estimate for GW data is usually given in terms of phase and amplitude deviations that one finds between different resolution levels.
This is computed after aligning all the waveforms at different resolutions in time (or frequency) and phase for a given mass ratio, since in general there exists a time and phase offset between the levels. 
We choose to align each resolution level $N$ such that $\Phi^{(N)}=0$ at $t=\delta t^p_{lm}$. In Fig.~\ref{fig:phasedev} we compute the accumulated (up to $t=60M$) deviation on the phase, $\delta\phi$, \xj{between the two highest resolution levels available in the SXS catalog for each NR simulation which we use as a rough upper limit on the simulation errors. Note that the simulations i.e., the full inspiral-merger-rindown signal, are always convergent~\cite{boyle:2019kee}}. We take the mean value as the averaged estimates of the NR error of our phase difference $\Delta \phi_{lmn}$. The mean value of $|\delta\phi|$ is about $0.04\,{\rm rad}$, which can approximately account for the differences of $\sim 0.1\, {\rm rad}$ observed in Fig.~\ref{fig:fit_and_starttime_221} at $t_0=10M$ and $t_0=15M$. 

\begin{figure}[b!]
    \subfloat{\includegraphics[width=0.49\textwidth]{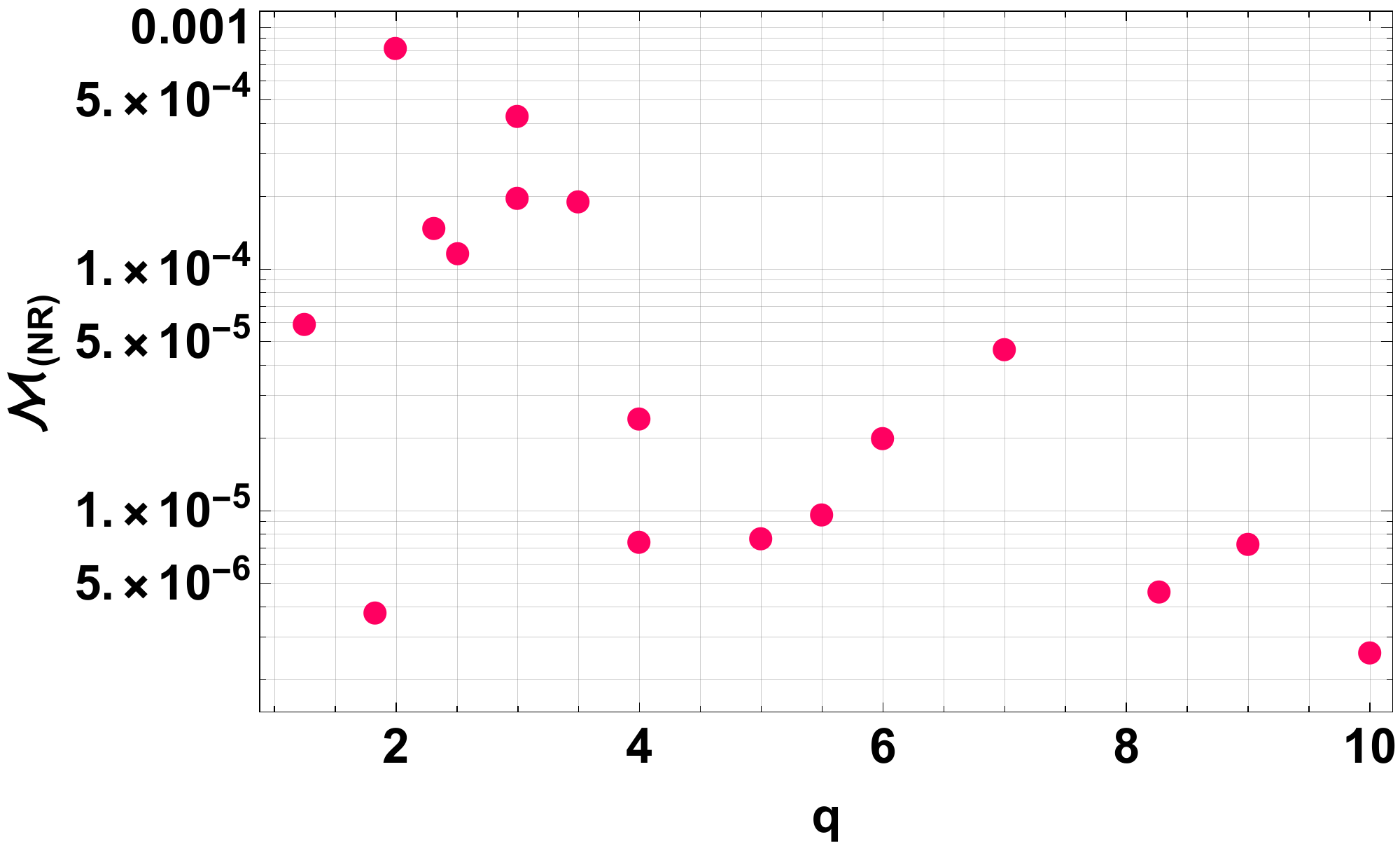}} 
    \caption{\xj{
    Mismatch for each SXS NR simulation computed between the two best resolved waveforms $N:N-1$. The median values for the set of points are ${2\times10^{-4}}$.}}
    \label{fig:nrmatch}
\end{figure}

Another possibility is to estimate the NR error by means of the mismatch Eq.~\eqref{eq:mismatch} between two waveforms with two different resolutions. This is the approach followed in this work for the RD signals. For each simulation listed in Table~\ref{tab:NR_set}, we compute Eq.~\eqref{eq:mismatch} with $t_i=0$ and $t_{f}=60M$ between \xj{two waveforms with the highest ($N$) and second highest ($N-1$)}. This analysis is shown in Fig.~\ref{fig:nrmatch}. \xj{The median value of the mismatch is ${2\times10^{-4}}$ and depends on the specific setup of the SXS simulation.} 

The red markers in Fig.~\ref{fig:nrmatch} correspond to the horizontal grid lines in Fig.~\ref{fig:matchplot} and to the purple points of Fig.~\ref{fig:goodness}.

\section{Spherical versus spheroidal harmonic decomposition and dependence on the inclination angle} \label{app:spheroidal}

The GW signal $h(t)$ can be decomposed onto a set of basis functions.
In the RD regime, \xj{the angular sector of the Teukolsky equations is described by spin-weighted spheroidal harmonics.
} Indeed, each of the $h_{lmn}(t)$ modes in Eq.~\eqref{eq:rdmodel} are weighted by the spin-two spheroidal harmonics \xj{${}_{-2}\mathcal{Y}_{lmn}(\theta,\phi)$}. Due to the angular dependence of the spheroidal harmonics, the relative amplitude and phase of different modes depend on the inclination angle $\theta$ of the source.
However, the spacetime during the inspiral-merger phase is better described by a decomposition in spin-two spherical harmonics \xj{${}_{-2}Y_{lm}(\theta,\phi)$}. Due to the dominant quadrupolar nature of GW emission during inspiral, a BBH merger signal is dominated by the $l=2$, $m=\pm 2$ modes.

The spherical and spheroidal basis functions are related by~\cite{Press1973,London:2018nxs}
\begin{equation} 
\xj{{}_{-2}\mathcal{Y}_{lmn}= {}_{-2}Y_{lm} + \sum_{l'\neq l}\mathfrak{H}_{ll'mn}(a,a^2){}_{-2}Y_{l'm}+ {\cal O}(a^3), \label{sphe}}
\end{equation}
where the specific form of $\mathfrak{H}_{ll'mn}(a,a^2)$ can be found in Appendix~A of~\cite{Berti:2014fga}. In the above equation, $Y_{lm}$ is the dominant term and $\mathfrak{H}_{ll'mn}=0$ for vanishing dimensionless spacetime angular momentum, $a=0$. Thus, ${}_2\mathcal{Y}_{lmn}={}_2Y_{lm}$ for Schwarzschild BHs. 
In the case of a spinning remnant, the spheroidal harmonics differ from the spherical ones as given in Eq.~\eqref{sphe}. However, since the overlap between spherical harmonics with different index $l$ is small, in practice one can neglect the difference at least for moderately spinning \xj{remnants~\cite{berti:2005gp,cook2020aspects}}.

In general, one must account for the dependence on the inclination angle through these angular functions to understand qualitatively the contribution of each $h_{lmn}$ to the final RD signal $h(t)$. The harmonics relevant to us are
\begin{align}
_{-2}Y_{22}&=
\frac{1}{2} \sqrt{\frac{5}{\pi }} \cos ^4\left(\frac{\theta }{2}\right) e^{\iota\, 2\phi}\,,\\
_{-2}Y_{33} &= -\sqrt{\frac{21}{2 \pi }} \sin \left(\frac{\theta }{2}\right) \cos ^5\left(\frac{\theta }{2}\right)e^{\iota\, 3\phi}\,,\\
_{-2}Y_{21} &= \frac{1}{4} \sqrt{\frac{5}{\pi }} \sin (\theta ) (1+\cos\theta)e^{\iota\, \phi}\,,
\end{align}
where $\theta$ is the inclination angle and $\phi$ is the orbital plane (or azimuthal) angle. We can obtain the  $m<0$ modes by means of the parity transformation \xj{${_{-2}Y_{l-m}(\theta,\phi)=(-1)^{l+m}{_{-2}Y_{lm}^{*}}(\pi-\theta,\pi+\phi)}$}.

\begin{figure}[th!]
    \subfloat{\includegraphics[width=0.47\textwidth]{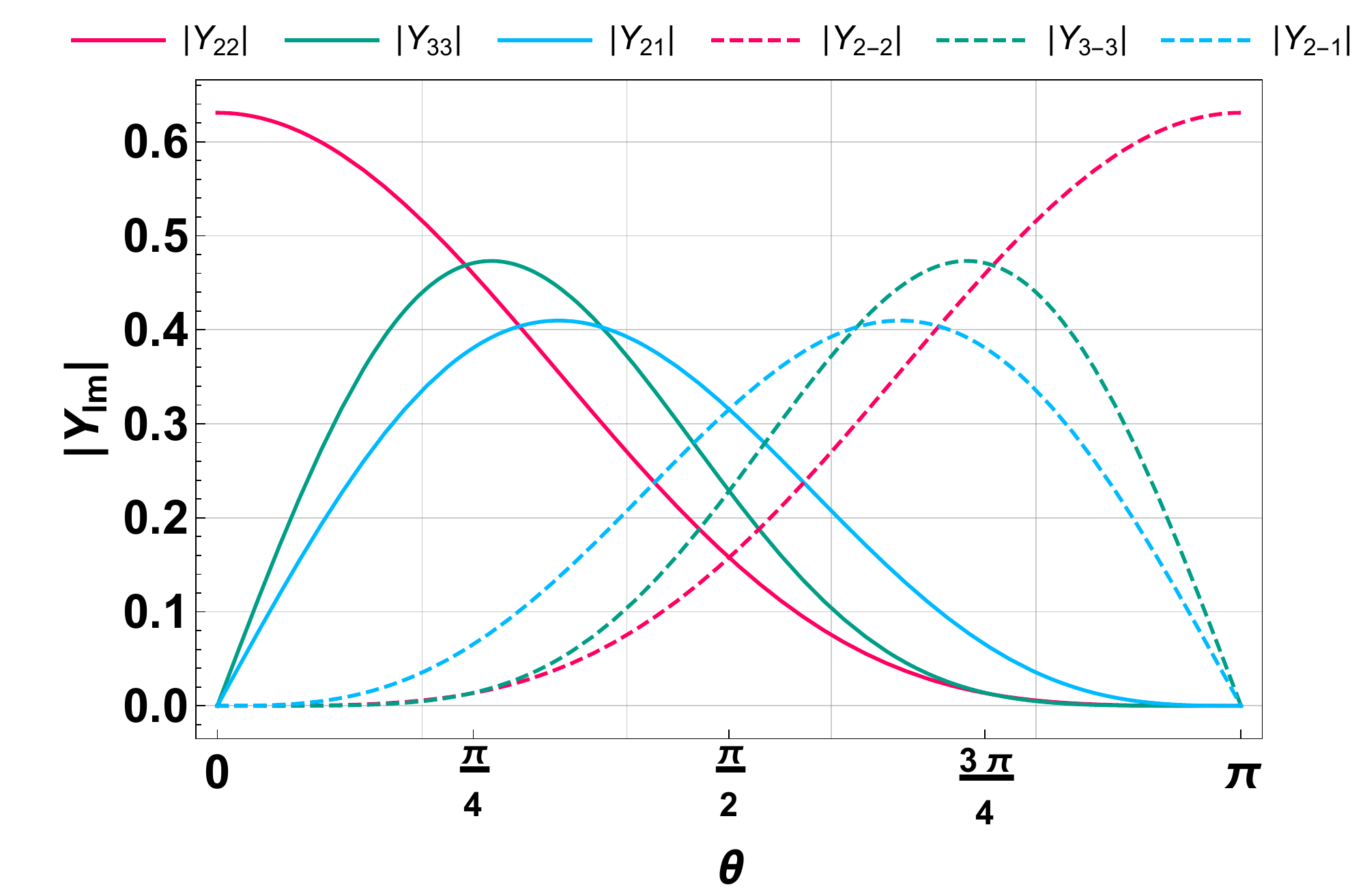}} 
    \caption{Absolute value of the \xj{${}_{-2}Y_{lm}$} harmonics for the $(2,\pm 2)$, $(2,\pm 1)$, $(2,\pm 2)$ modes. At face-on orientation ($\theta=0$) the \xj{${}_{-2}Y_{22}$} harmonic is the only nonzero contribution and dominates until $\theta\approx \pi/4$, where the contributions of the \xj{${}_{-2}Y_{21}$} and \xj{${}_{-2}Y_{33}$} harmonics become increasingly important. At edge-on orientation ($\theta=\pi/2$) both \xj{${}_{-2}Y_{21}$} and \xj{${}_{-2}Y_{33}$} are larger than \xj{${}_{-2}Y_{22}$}.}
    \label{fig:harmonics}
\end{figure}

In Fig.~\ref{fig:harmonics} we show that for nearly face-on/face-off systems (i.e., with inclination angle $\theta\approx 0,\pi$) the $l=3$, $m=\pm 3$ and  $l=2$, $m=\pm 1$ harmonics are negligible and $h(t)$ is accurately modeled with only the dominant ${l=2, m=\pm 2}$ mode for a face-on/face-of orientation, respectively. 
However, for an edge-on orientation ($\theta=\pi/2$), both \xj{${}_{-2}Y_{21}$} and \xj{${}_{-2}Y_{33}$} are larger than \xj{${}_{-2}Y_{22}$}. In general, unless the source is located at a particularly disfavored angle ($\theta\sim0,\pi$), the modes $h_{21}$ and $h_{33}$ can in principle (depending on their relative amplitude) carry an important fraction of the amplitude of $h(t)$ for these inclination.

The dependence on the inclination angle $\theta$ implies that the detectability, resolvability, and measurability criteria discussed in the main text will also depend on $\theta$. For simplicity, we have averaged out the GW strain on the $\theta$ and $\phi$ angles, which is equivalent to replacing ${_s\mathcal{Y}_{lm} \rightarrow 1/\sqrt{4\pi}}$ in  Eq.~\eqref{eq:rdmodel}~\cite{berti:2005ys}.

\section{Bias on the subdominant QNM frequencies due to the choice of the RD starting time}\label{app:Bias}

To gain an insight into how the choice of starting time $t_0$ leads to a bias in the estimated frequency for different angular modes, here we perform an analysis similar to the one we presented in Ref.~\cite{Bhagwat:2019dtm} for the 2-tone model (c.f.  Ref.~\cite{Bhagwat:2019dtm} for details of the analysis). 

In brief, for each mode we assume a modified RD waveform of the form
\begin{align}
 h_{lmn} (t) = {A}_{lmn} e^{- \iota \, t_{lmn} (1 + \frac{\alpha_{lmn}}{100}) } e^{-(t-t_0)/\tau_{lmn}} 
\label{modifiedRD}
\end{align}
where $\alpha_{lmn}$ is the relative (percent) deviation of the frequency from the GR-QNM spectrum, $\omega_{lmn}$. Since we studied the overtone case in Ref.~\cite{Bhagwat:2019dtm}, here we focus on $n=0$ only.

\begin{figure*}[h!]
    \subfloat{\includegraphics[width=0.48\textwidth]{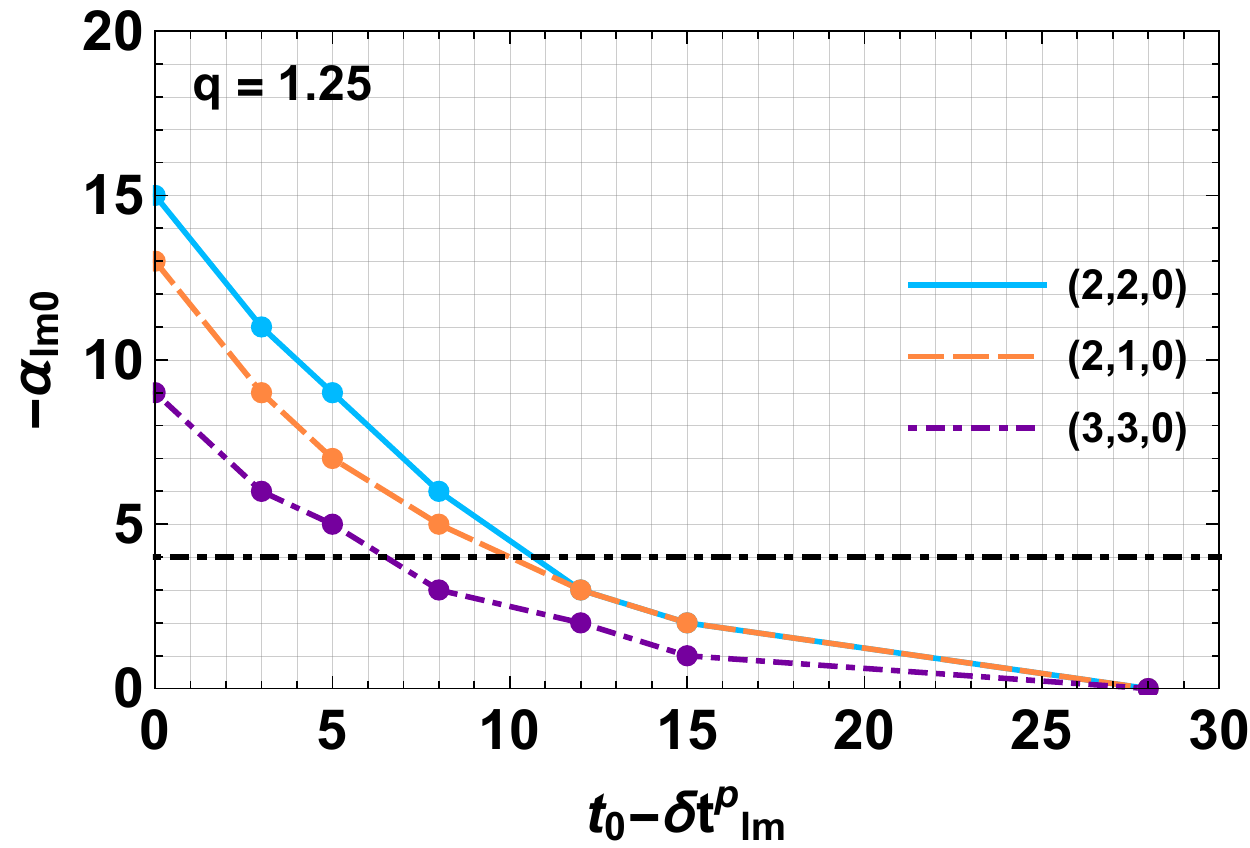}} 
    \subfloat{\includegraphics[width=0.48\textwidth]{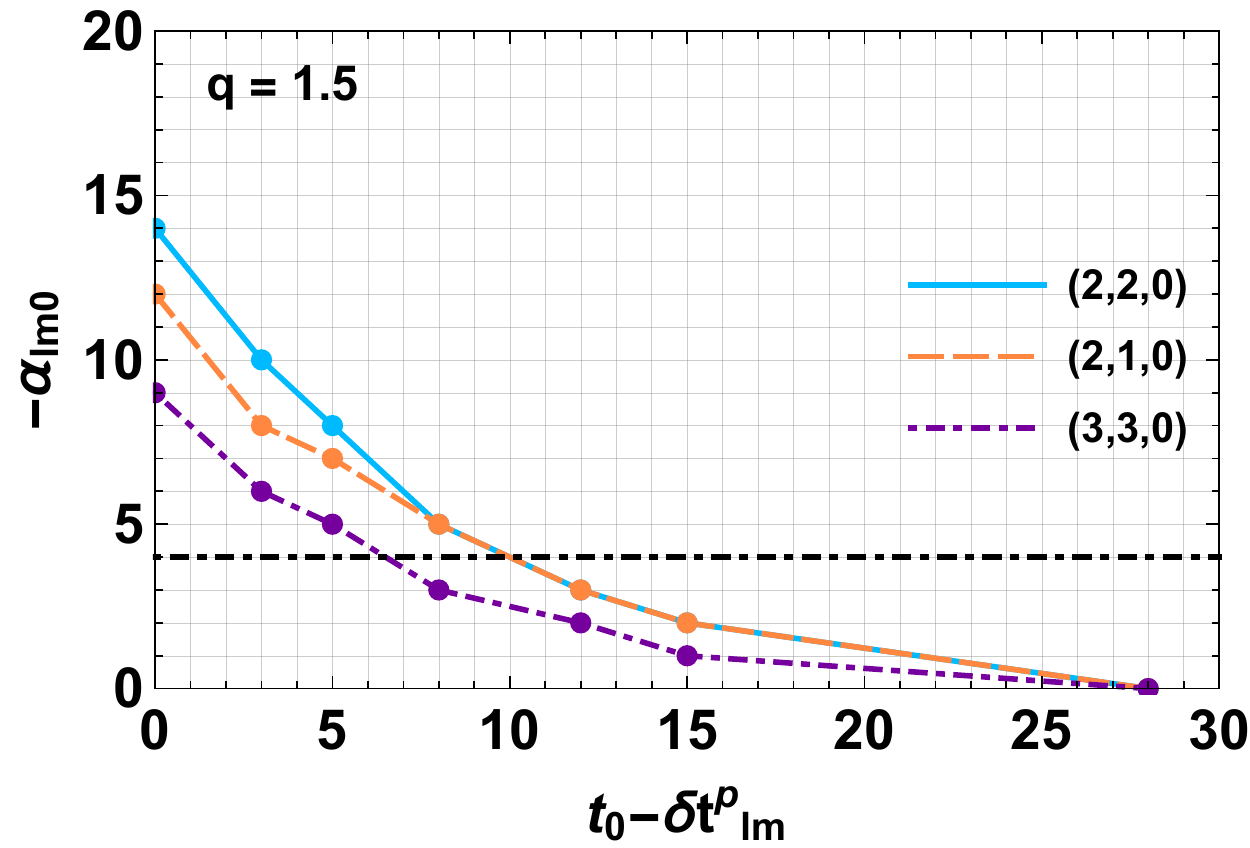}} \\
     \subfloat{\includegraphics[width=0.48\textwidth]{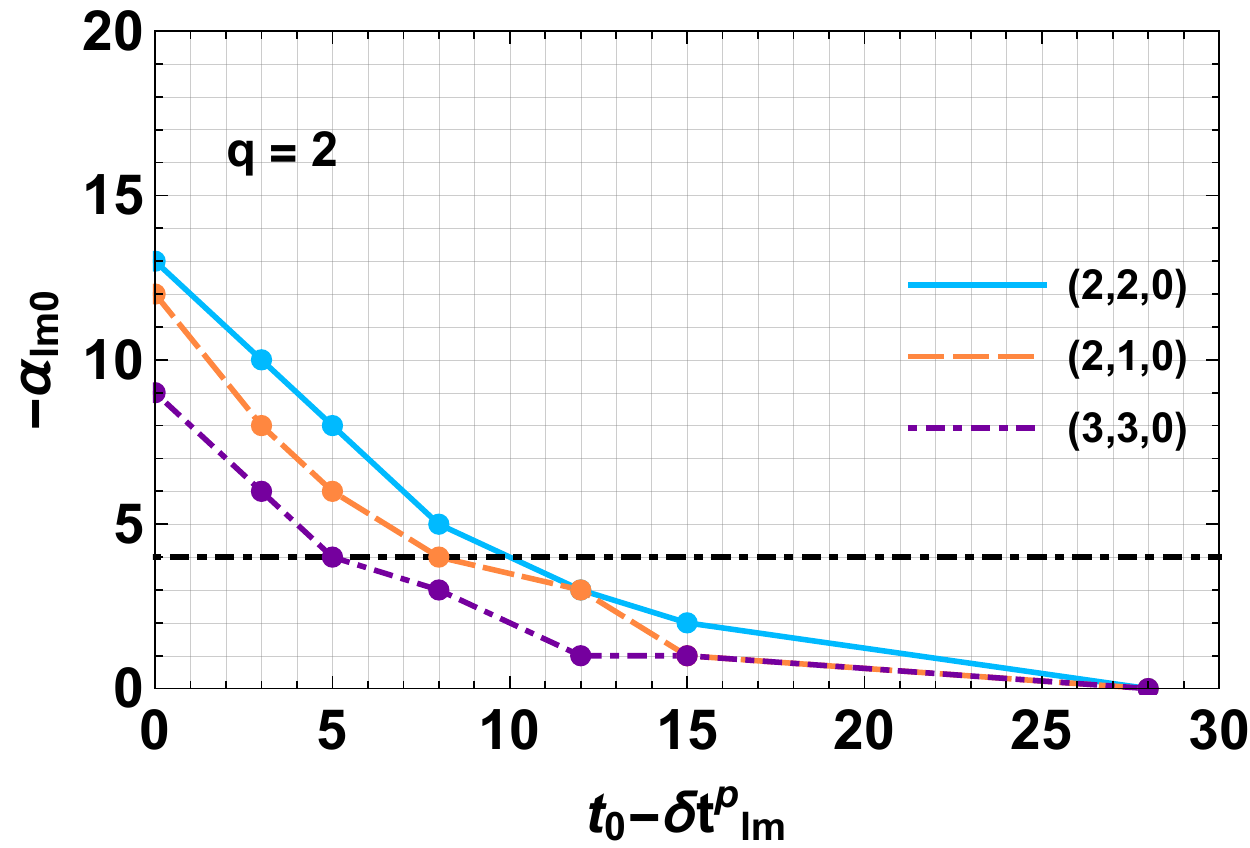}} 
    \subfloat{\includegraphics[width=0.48\textwidth]{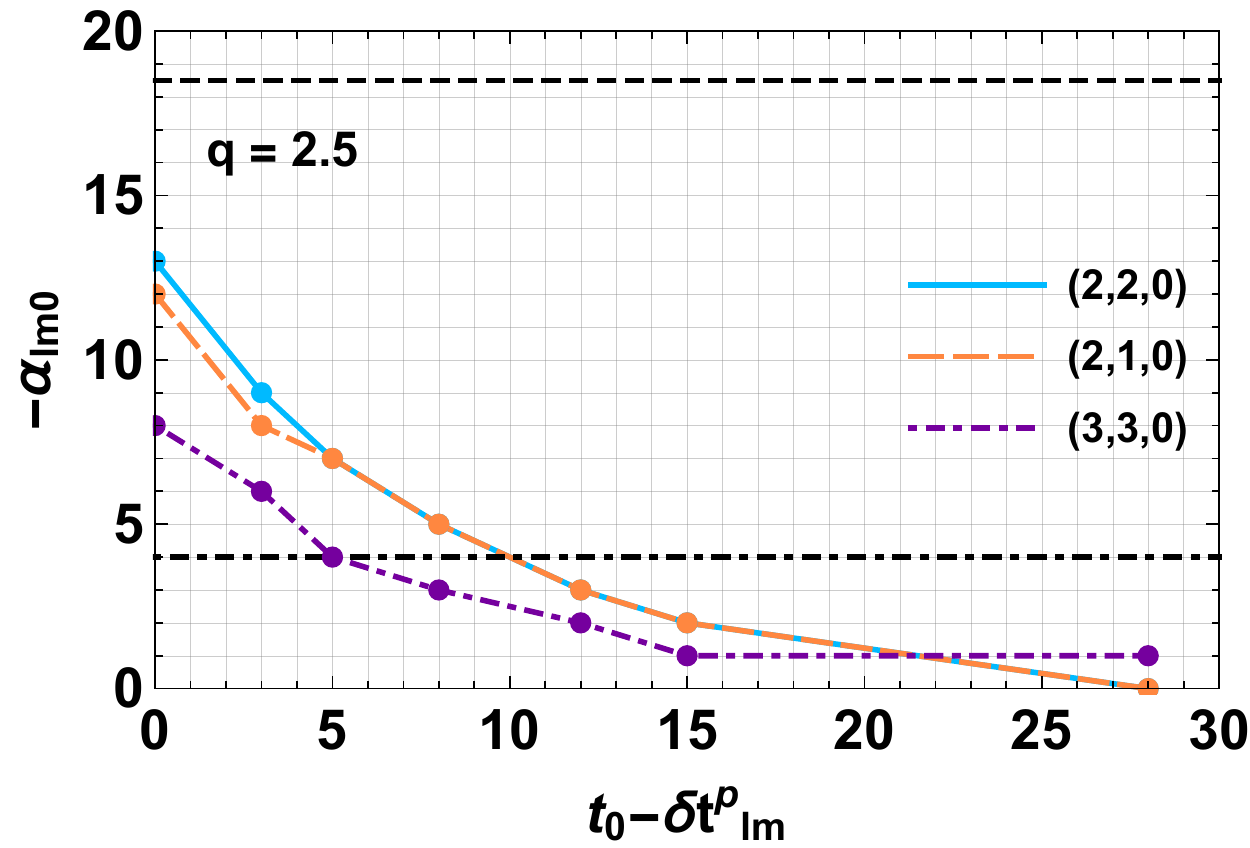}} \\
      \subfloat{\includegraphics[width=0.48\textwidth]{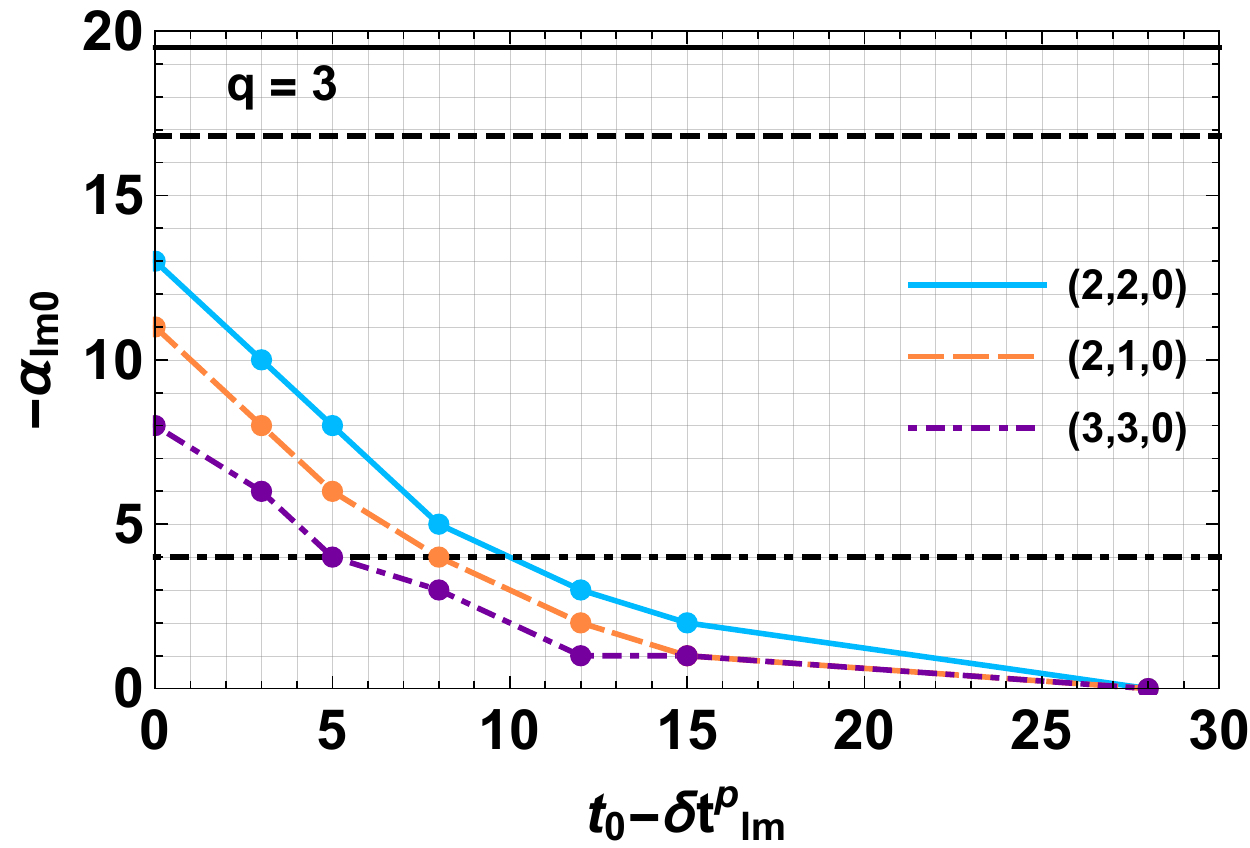}} 	
    \subfloat{\includegraphics[width=0.48\textwidth]{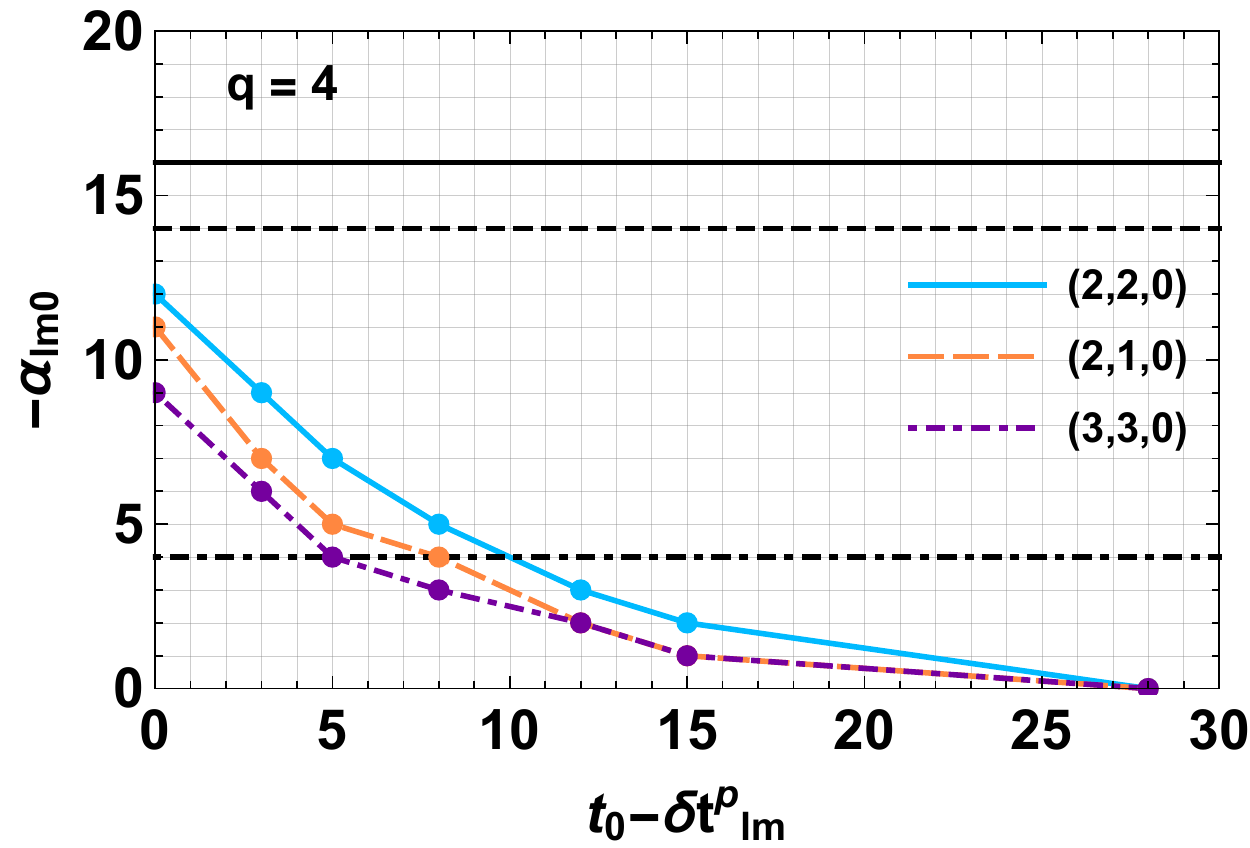}}
    \caption{Bias $-\alpha_{lm0}$ on the recovery of the QNM frequency of different (fundamental) angular modes. The $x$-axis $t_0-\delta t_{lm}^{p}$ denotes the starting time of each mode with respect to the time at which the amplitude of the waveform peaks. 
    The blue, the orange, and the purple curves correspond to the bias in $l=m=2$, $l=m=3$ and $l=2, m=1, n=0$ angular modes, respectively, neglecting overtones for each angular mode.  The dot-dashed, dashed, and solid horizontal lines correspond to the Fisher matrix spread expected in the recover of $f_{22}$, $f_{33}$, and $f_{21}$ respectively, assuming $\rho_{\rm RD}=15$ (the horizontal lines scales as $1/\rho_{\rm RD}$). Notice that the dominant $l=m=2$ mode is the one with the largest bias for any mass ratio.} 
    \label{fig:bias}
\end{figure*}

Since the overlap between different angular modes is not significant, we fit each angular mode to the corresponding angular decomposition of the NR simulation, and quantify the bias by evaluating the best fit values of $|\alpha_{lmn}|$. Ideally, should the above RD model match the NR-RD perfectly, we expect $\alpha_{lmn} =0$; i.e., we expect a zero bias in the recovered QNM frequencies. A nonvanishing value of $|\alpha_{lmn}|$ quantifies the systematic errors introduced by the choice in the starting time. 
Possible nonlinearity and the inclusion of a finite number of tones (here we consider only the fundamental $n=0$ tone for each angular mode) may explain this bias. In Fig.~\ref{fig:bias}, we show $-\alpha_{lm0}$ as a function of $t_0-\delta t_{lm}^{p}$ for various BBH mass ratios. 
From Fig.~\ref{fig:bias}, we see that the $l=m=2$ mode is most sensitive to the choice of the starting time in the sense that the blue curve always corresponds to the largest absolute value among the three curves for any given choice of the starting time. \xj{The $l=m=2, n=0$ mode damping time is slightly larger than for the other two modes considered in this work, which makes this mode longer lived and thus increasing the bias produced}. In other words, an earlier starting time leads to a larger amount of bias in the dominant mode compared to the subdominant mode. Therefore, if one desires that the systematic error due to the choice of starting time be limited to a certain percent, it is  sufficient that this criterion be imposed on just $l=m=2$ mode. 

\section{\xj{On the corotating and counterrotating modes}} 
\label{app:counter-rot}
 The complex QNM spectrum for a Kerr BH are the poles of the Teukolsky equation decomposed on a spin-weighted spheroidal harmonic basis~\cite{teukolsky:1974yv,Leaver:1985ax,Berti:2009kk,Kokkotas:1999bd,Ferrari:1984zz}. For a Kerr BH, the rotation leads to a Zeeman-like splitting of the QNM frequencies - similar to the energy levels of an atom in a static external magnetic field~\cite{Berti:2009kk}. The splitting depends on the type of perturbation as well as on the $(l,m,n)$ indices. Given a set of $(l,m,n)$, the solution to Teukolsky equation is a superposition of two damped sinusoids corresponding to the two poles of the equation - one with positive real frequency $\omega^+_{lmn}$, also known as the corotating mode, and another with a negative real frequency $\omega^-_{lmn}$, known as the counterrotating mode. Further, note that $\omega^+_{lmn} \neq |\omega^-_{lmn}|$ and $\tau^+_{lmn}=\tau^-_{lmn}=\tau_{lmn}$.
 To illustrate this point, We consider the remnant corresponding to the SXS:1143 and SXS:1107 systems that have spins $a_f= \{0.680,0.261 \}$ and masses $M_f= \{0.953,0.992\}$, respectively. 
\begin{table}[h!]{%
\begin{tabular}{|c|c|c|c|c|}
\hline
SXS ID   & $M\omega^+_{220}$      & $M\omega^-_{220}$  & $M\omega^+_{221}$      & $M\omega^-_{221}$      \\ \hline\hline
SXS:1143    & 0.5498   & -0.3266  & 0.5371 & -0.2872\\ \hline
SXS:1107    & 0.4158  & -0.3479   & 0.3937& -0.3163\\  \hline
\end{tabular}%
}
\caption{Examples of the $n=0,1$ frequencies of the corotating and counterrotating modes for the remnant of two simulations considered in this work.}
\label{tab:rotcount}
\end{table}

The NR waveform $h(t)$ is decomposed into a sum of modes $h_{lm}$ as given in Eq.~\eqref{eq:rdmodel}. Both the rotating and counterrotating frequencies of the given $(lm)$ mode are contained in the decomposition. We use the fitting techniques described in the main text to compute the amplitude excitation associated with the counterrotating frequency assuming that the signal is a superposition of the corotating and counterrotating damped sinusoids, namely
\begin{align}
\label{eq:rdowncc}
h_{lmn} &= {\cal A}^+_{lmn} e^{- \iota \omega^+_{lmn} t} e^{- (t-t_0)/ \tau_{lmn}} \\&+  {\cal A}^-_{lmn} e^{- \iota \omega^-_{lmn} t} e^{- (t-t_0)/ \tau_{lmn}}\,\,,
\end{align}
where the first term on the right side accounts for the corotating modes (those that have been used in the main text) while the second term accounts for the counterrotating modes, which have been neglect in the main analysis (i.e. we assumed ${\cal A}^-_{lmn}=0$). In this section, we check the validity of this approximation by fitting ${\cal A}^-_{lmn}$ for the two NR waveforms listed in Table~\ref{tab:rotcount}.

\begin{figure*}
 \subfloat{\includegraphics[width=0.48\textwidth]{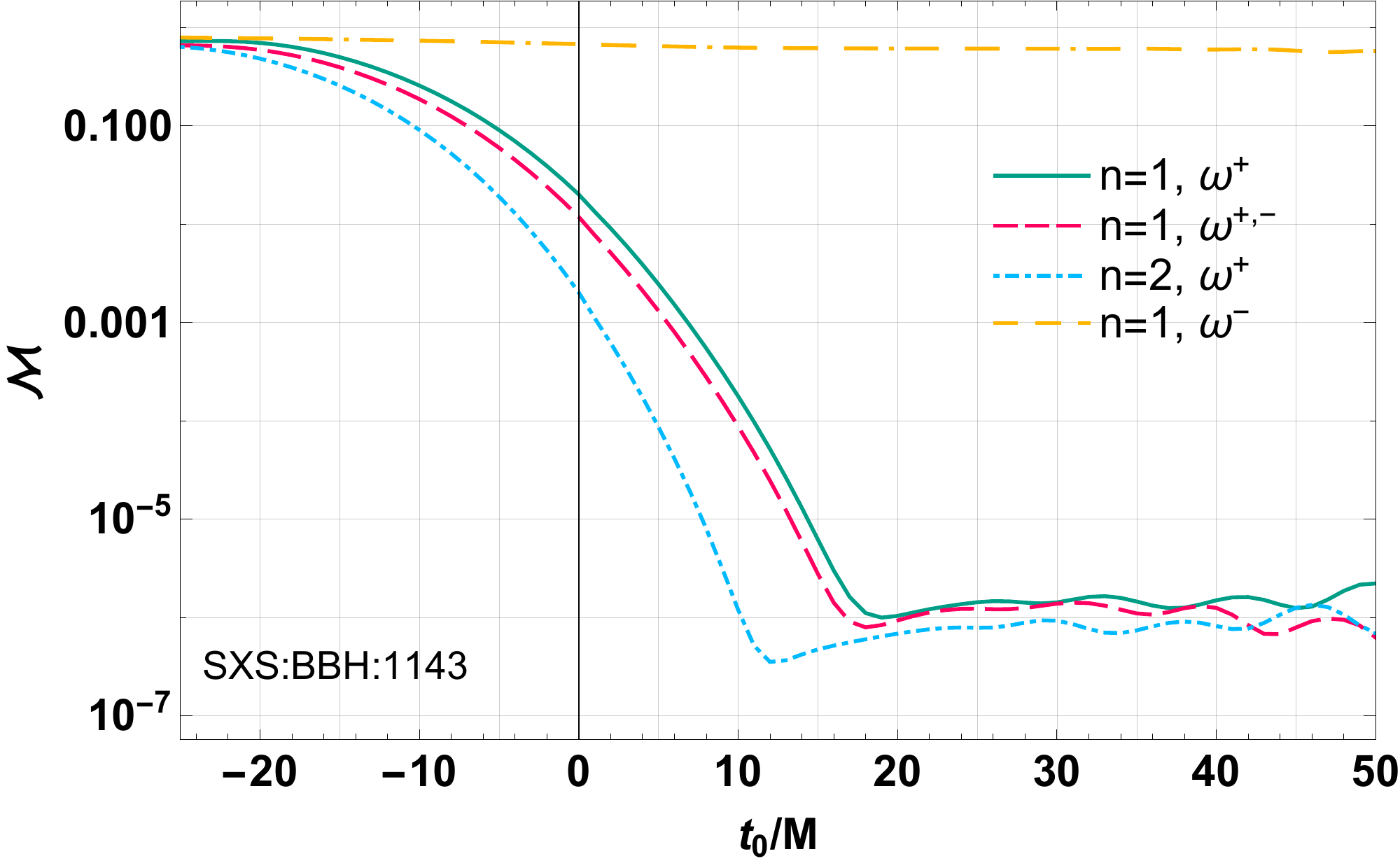}}
  \subfloat{\includegraphics[width=0.48\textwidth]{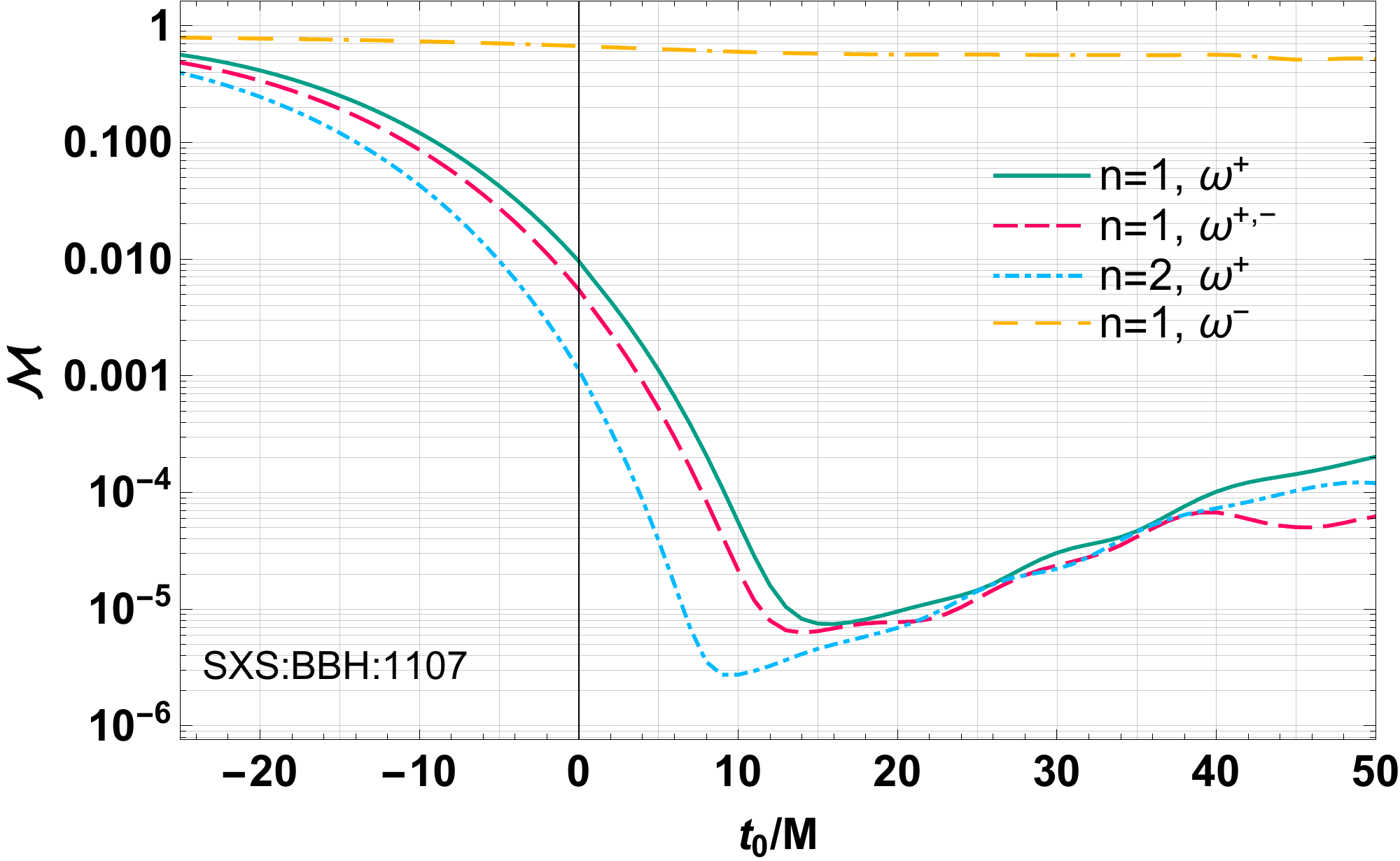}}
    \caption{Mismatch as in Fig.~\ref{fig:matchplot} for three different 2-tone models ($n=0,1$) and one 3-tone model ($n=0,1,2$) for the SXS:BBH:1143 and SXS:BBH:1107 NR  simulations. The solid green and dash dotted blue curves represent two separated RD models calibrated only with corotating modes ($\omega^+$), the red dashed line accounts for a 2-tone model calibrated with both the positive and negative frequency contributions ($\omega^{+,-}$), and the dashed dotted yellow curve is calibrated only with counterrotating modes ($\omega^-$). Notice that the mismatch of the model fit with positive and negative frequencies (red dashed) is comparable to the one obtained with $\omega^+$ only and larger than the one which includes only corotating modes up to the next-order overtone with $n=2$. For the model with $\omega^-$ only (dot-dashed yellow), we obtain a mismatch close to one.}
    \label{fig:counter_mismatch}
\end{figure*}

In Fig.~\ref{fig:counter_mismatch} we show the mismatch \eqref{eq:mismatch} for SXS:1143 (left panel) and SXS:1107 (right panel) and for three different 2-tone models with $n=0,1$ and one 3-tone model with $n=0,1,2$. The solid green and dash dotted blue curves correspond to only corotating modes, $ \omega^+\equiv \omega^+_{22n}$; the red dashed curve accounts for a 2-tone model fit with both the positive and negative frequency contributions; and the dashed dot yellow curve corresponds to only counterrotating modes, $\omega^- \equiv \omega^-_{22n}$. We find that the mismatch weakly depends on the inclusion of the counterrotating frequency. For the $\omega^{+,-}$ model with $n=1$ (red dashed curve), the mismatch at each starting time $t_0/M$ is similar to that obtained from the $n=1$, $\omega^{+}$ model (solid green), whereas is significantly smaller than the mismatch obtained for corotating modes only including the next-order contribution ($n=2$) overtone. As expected, for the RD model with $\omega^-$ only (yellow dot-dashed), we obtain a mismatch close to unity, showing a very poor agreement with the original NR waveforms if one includes only counterrotating modes. 

\begin{figure*}
 \subfloat{\includegraphics[width=0.48\textwidth]{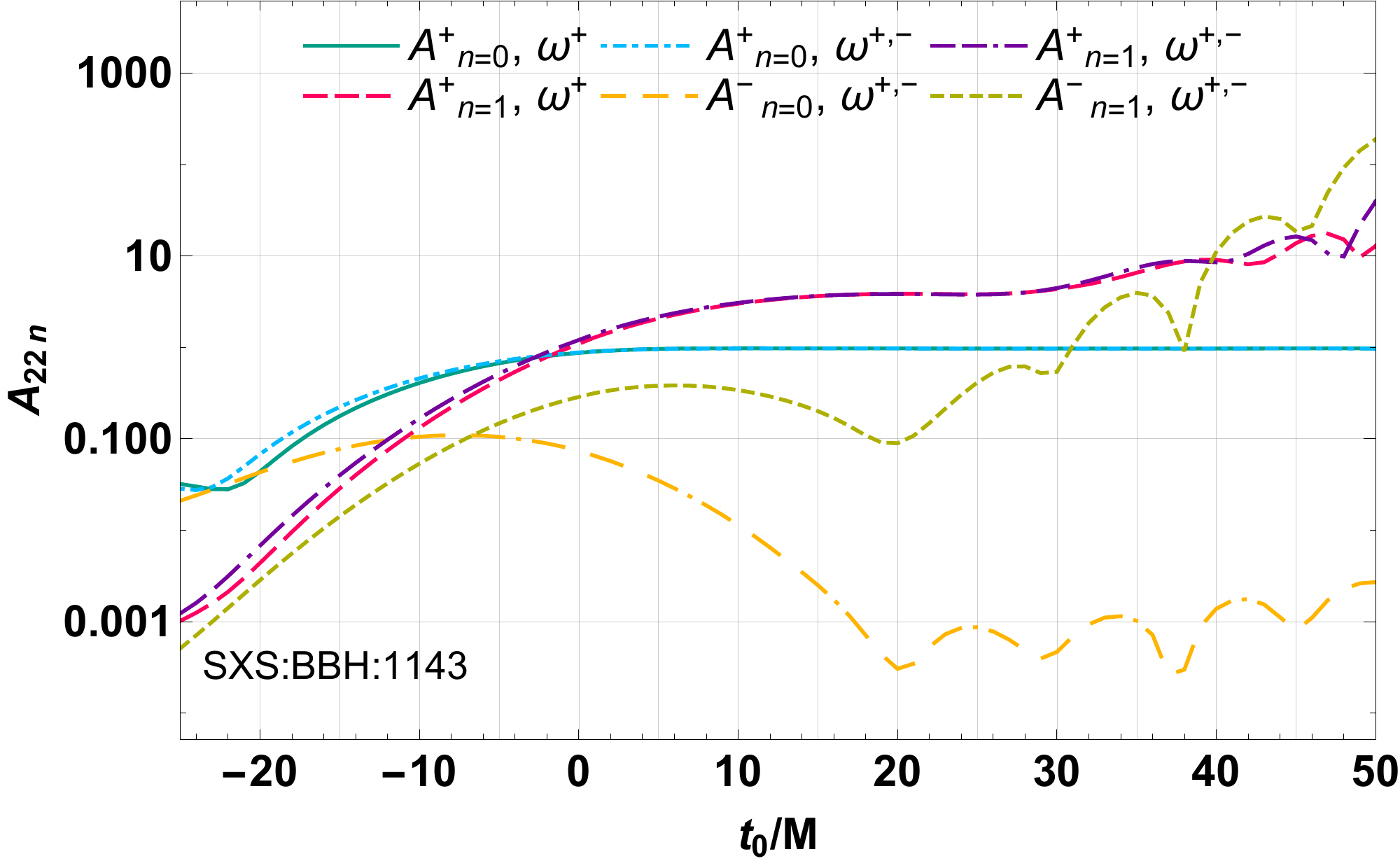}}
  \subfloat{\includegraphics[width=0.48\textwidth]{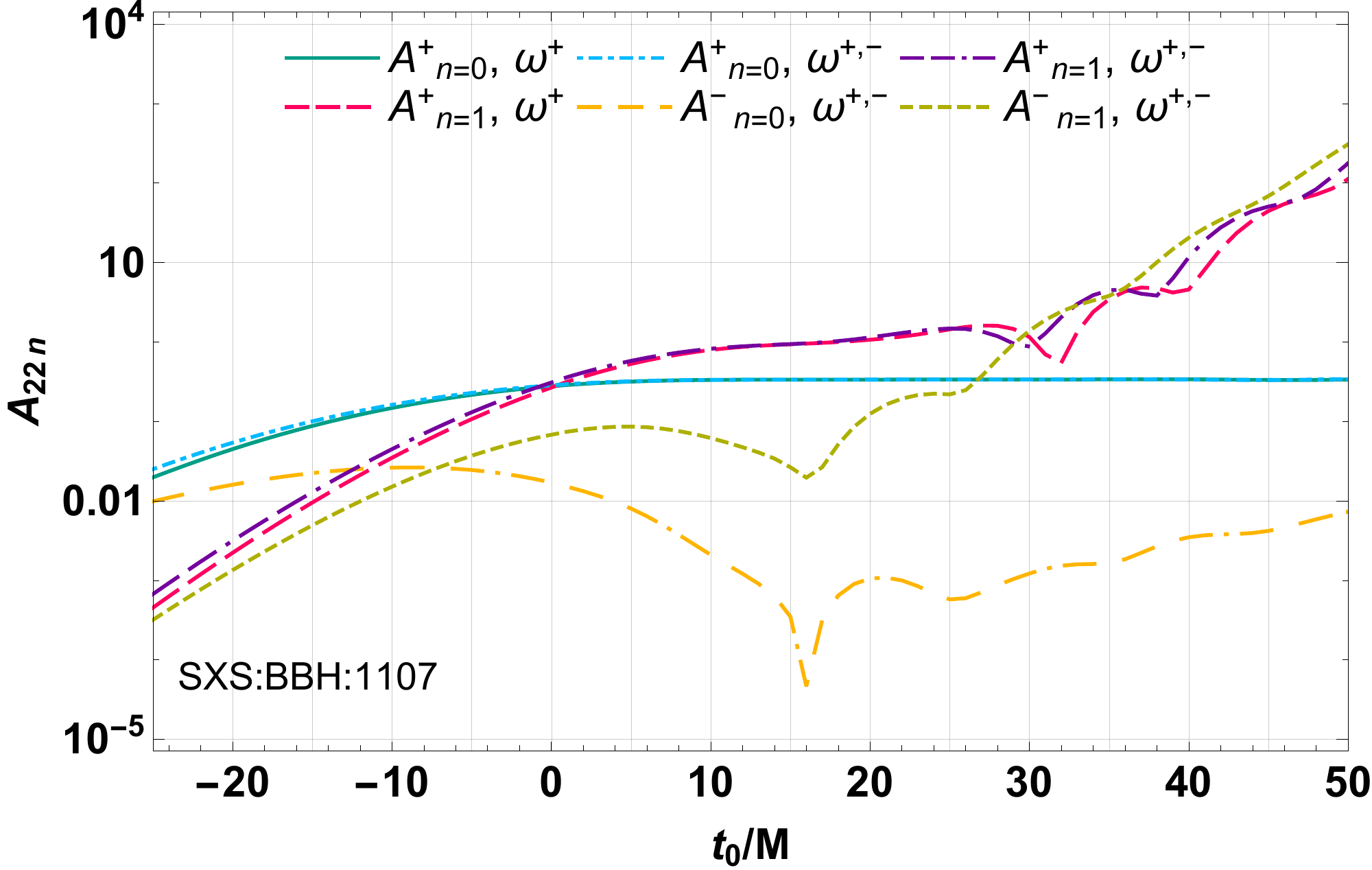}}
    \caption{Overtone amplitudes obtained for the two overtone models $n=1, (\omega^+)$ and $n=1,(\omega^{+,-})$ of Fig.~\ref{fig:counter_mismatch} for the SXS:BBH:1143 and SXS:BBH:1107 waveforms. The minus superscript refers to the amplitudes of the counterrotating modes, the $+$ superscript refers to models calibrated only with positive frequencies, and the $+,-$ superscript is for models calibrated with both the positive and negative frequencies. We observe that the counterrotating amplitudes are at least 1 order of magnitude smaller than the corresponding corotating ones for $0<t_0/M\lesssim 20$. Furthermore, the value of the amplitude of the corotating tones does not dependent significantly on the inclusion of counterrotating modes in the model.}
    \label{fig:counter_mismatch_amp}
\end{figure*}

These results are better understood by computing the relative amplitudes of the counterrotating modes $A^-_{lmn}$ obtained from the same fits. In Fig.~\ref{fig:counter_mismatch_amp} we show the fit amplitudes $A^{+,-}_{n}\equiv A^{+,-}_{22n}$ for the two $n=1$ models with $\omega^+$ and $\omega^{+,-}$. Note that at $0<t_0/M\lesssim 20$, the $A^{-}_n(\omega
^{+,-})$ amplitudes (green and yellow dashed curves) are about 1 order of magnitude smaller than the corresponding $A^{+}_n (\omega
^{+,-})$ amplitudes (blue and purple dashed curves). The values of the corotating frequency amplitudes $A^{+}_n (\omega
^{+,-})$ are not affected significantly by the addition of the counterrotating terms, as can be seen by comparing the solid green and red dashed curves (both corotating and counterrotating modes in the model) with the dot dashed blue and purple curves (only corotating modes in the model or $A^{+}_n (\omega^{+})$. For $t_0/M>20$ the $A_1^{+,-}$ amplitudes are no longer stable \cite{Bhagwat:2019dtm}.
\newpage
\section{Minimum SNR for detectability, resolvability, and measurability as a function of amplitudes and phases} \label{app:contour}

\begin{figure*}[ht!]
    \subfloat{\includegraphics[width=0.3\textwidth]{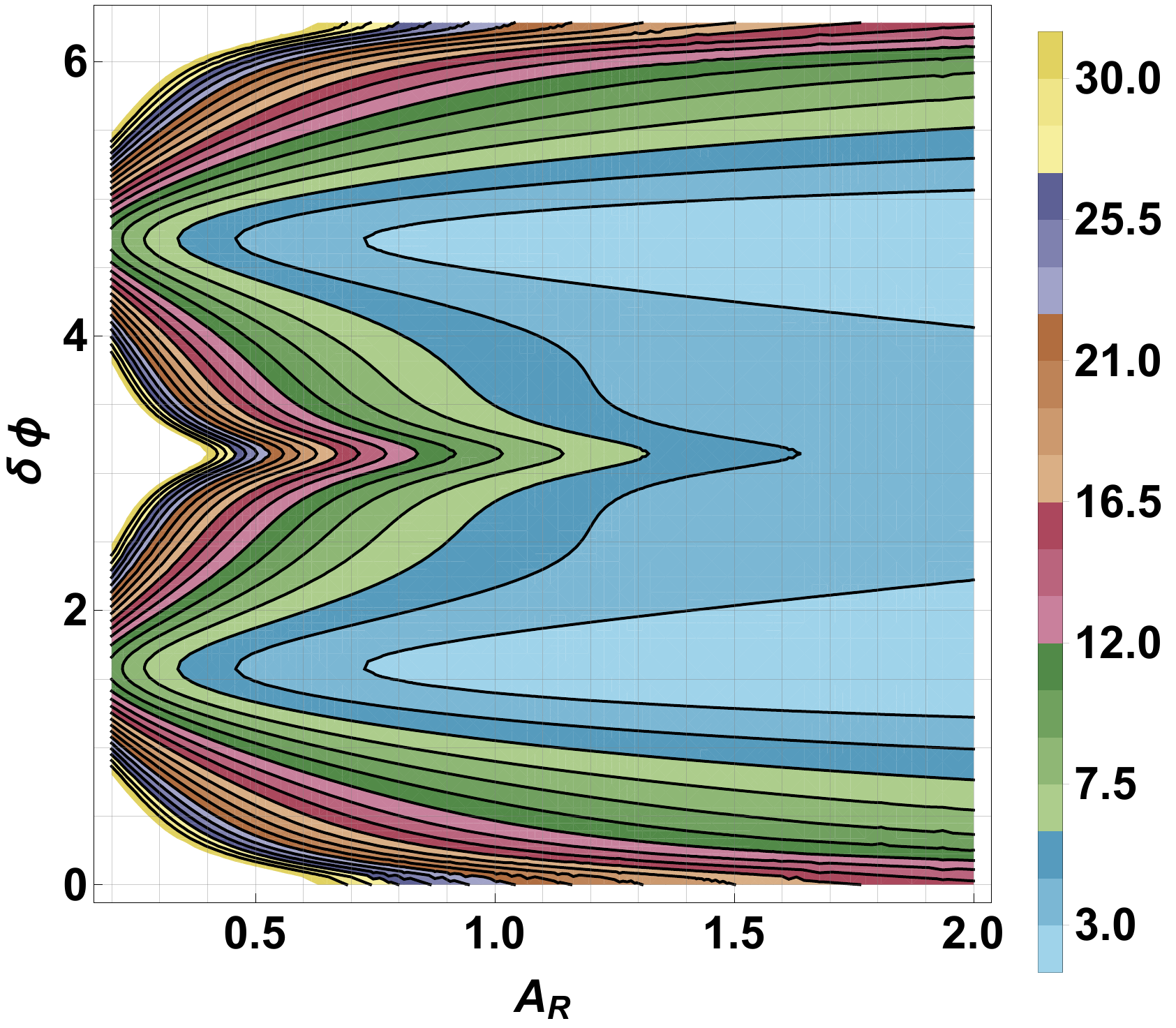}} 
     \subfloat{\includegraphics[width=0.3\textwidth]{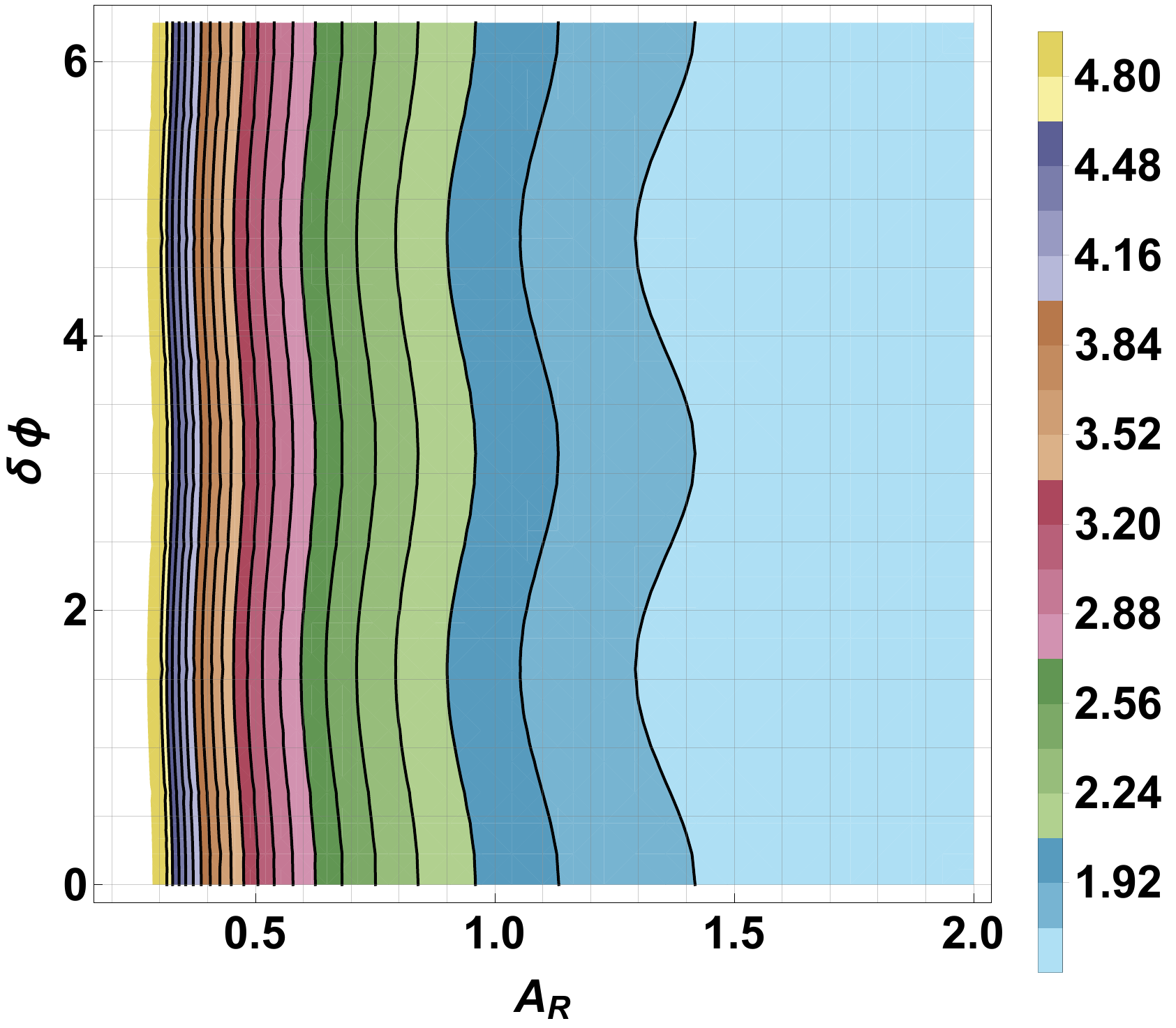}} 
     \subfloat{\includegraphics[width=0.3\textwidth]{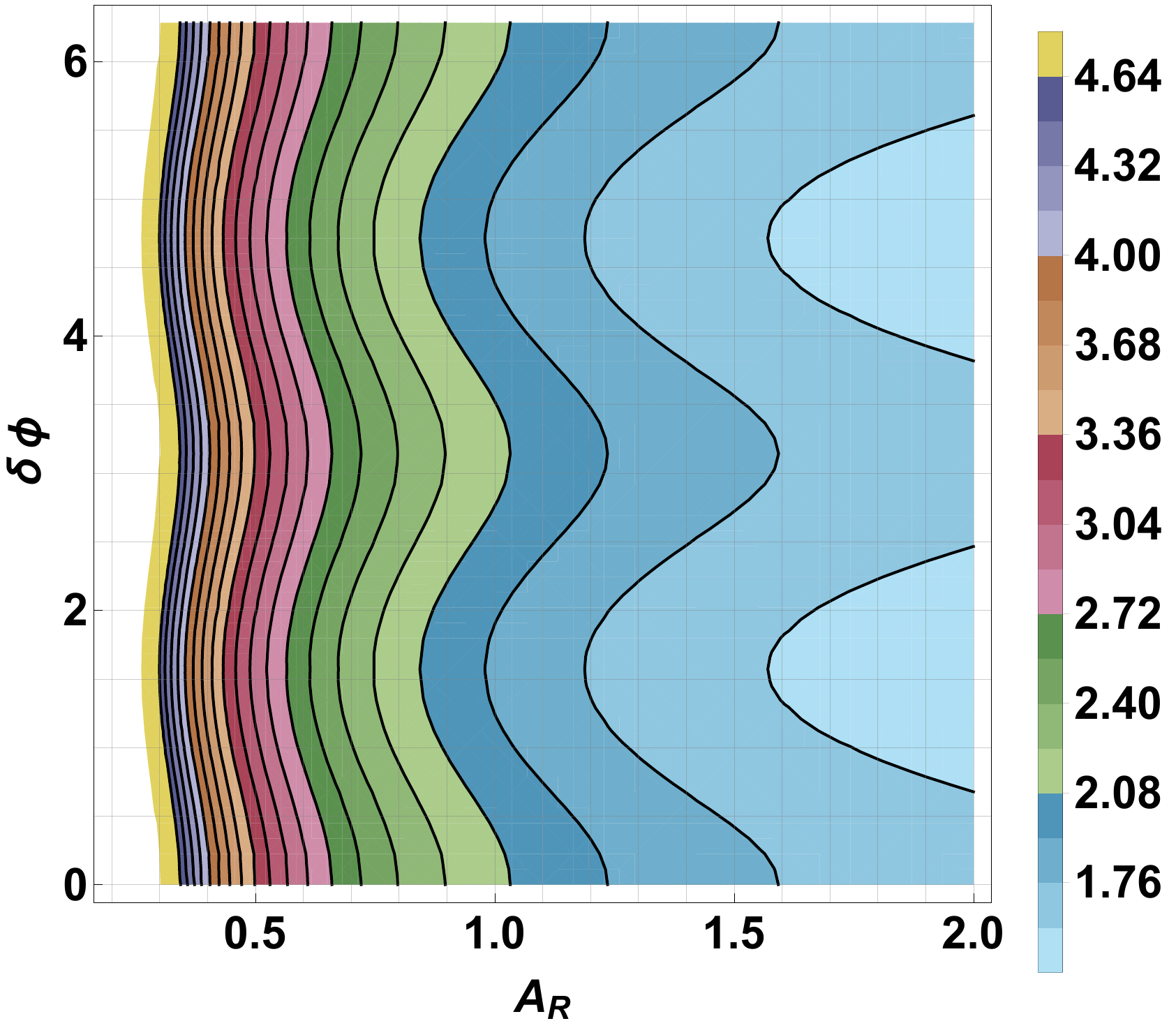}} \\
     \subfloat{\includegraphics[width=0.3\textwidth]{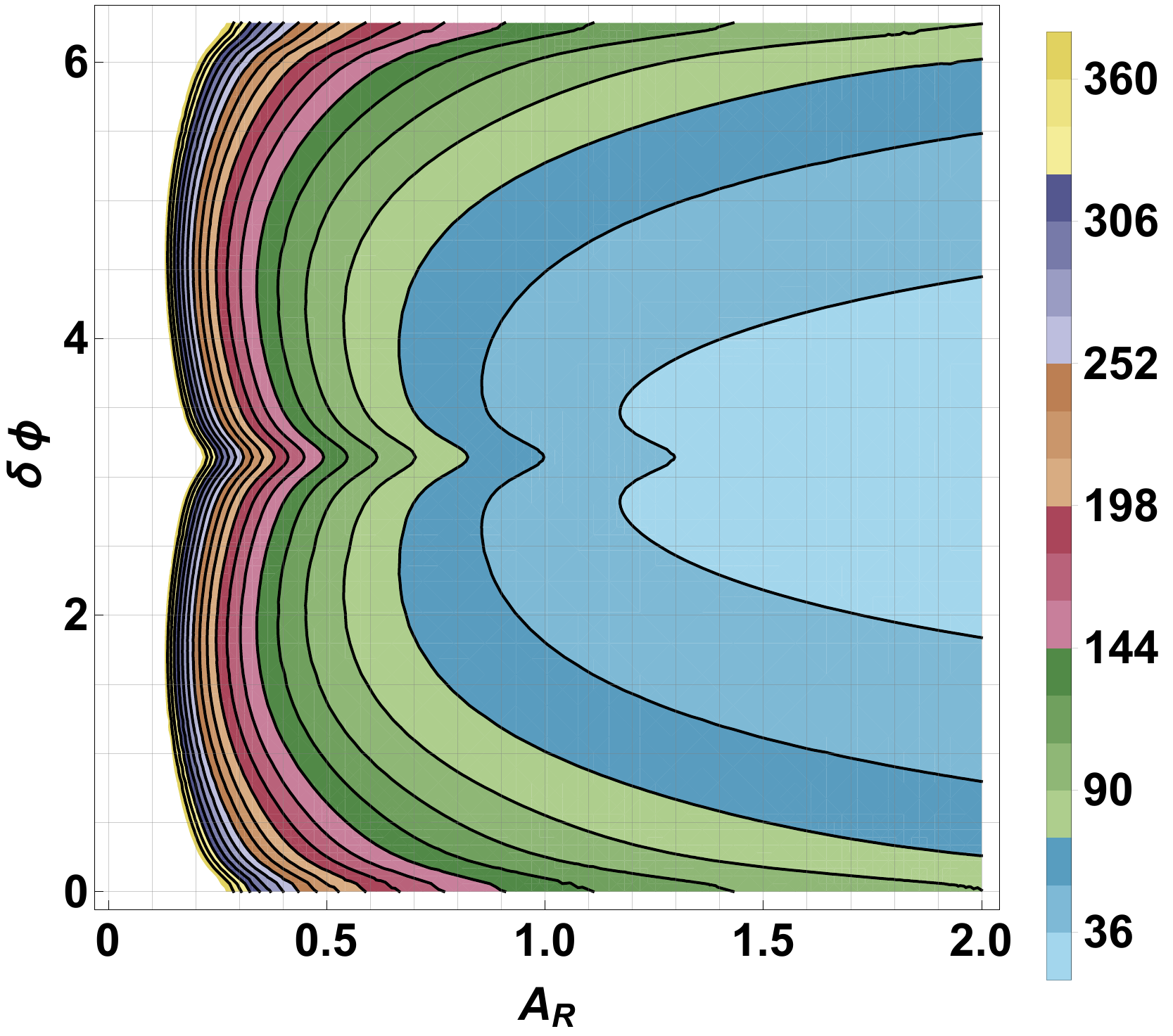}} 
     \subfloat{\includegraphics[width=0.3\textwidth]{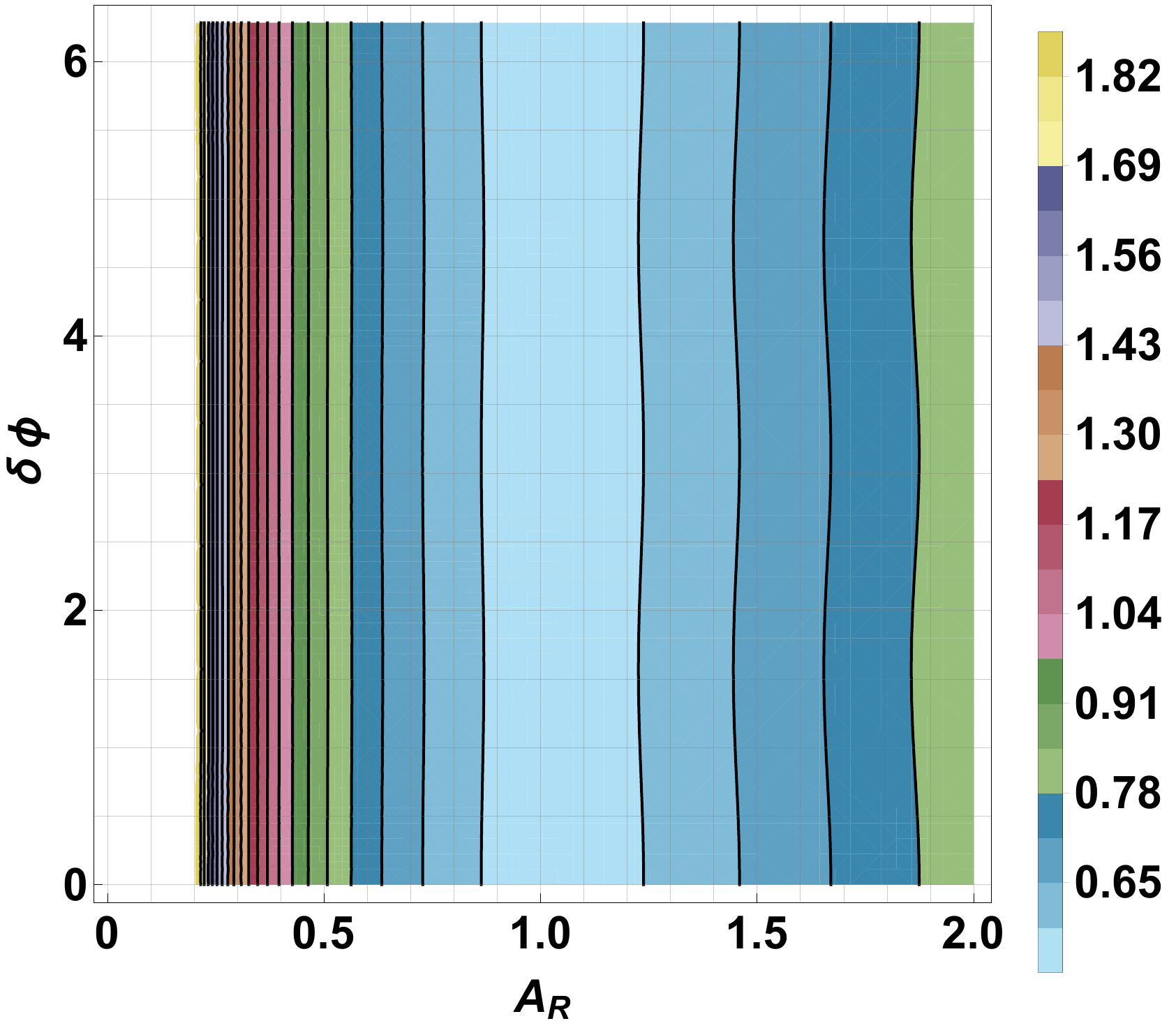}} 
    \subfloat{\includegraphics[width=0.3\textwidth]{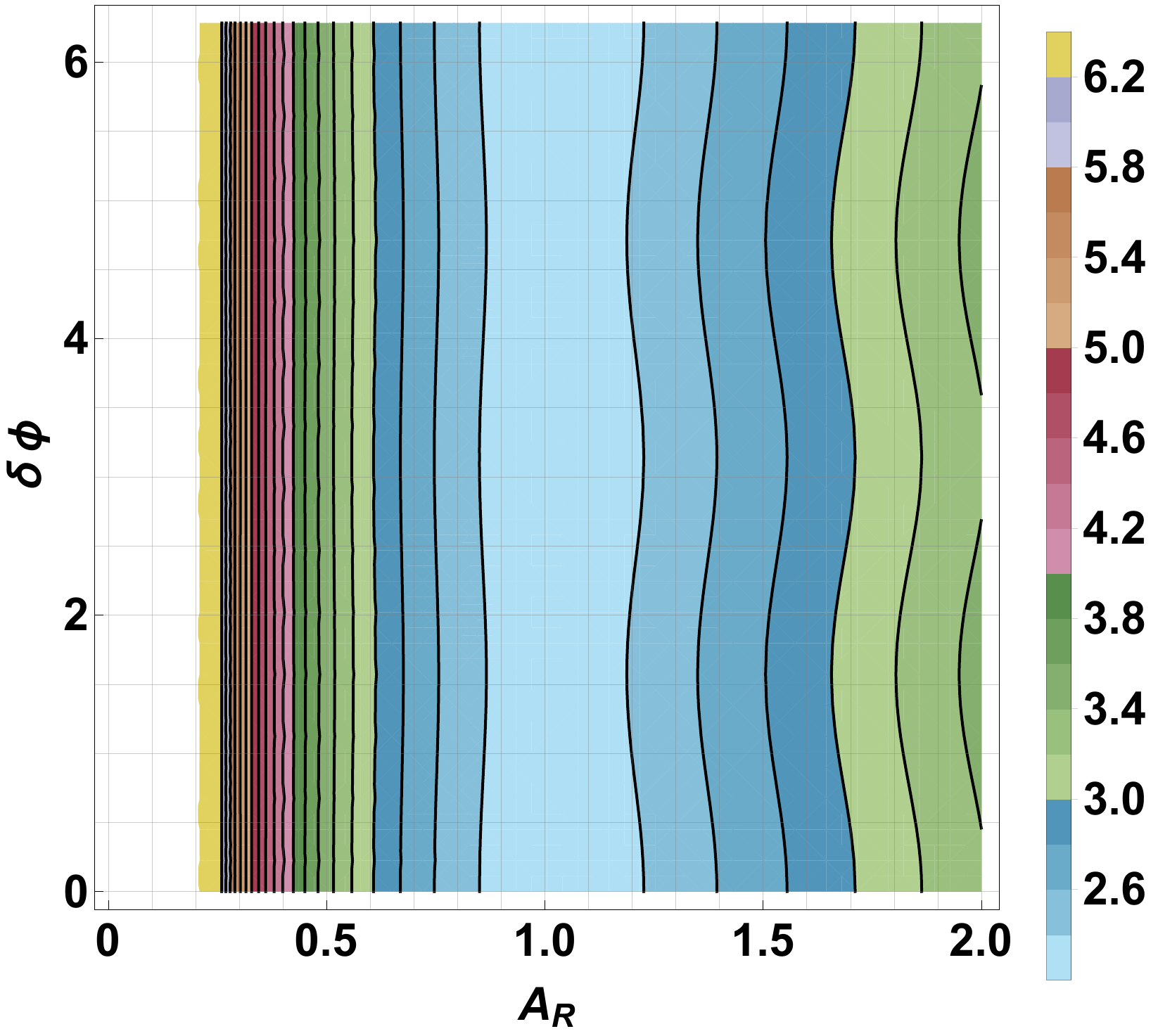}} \\
      \subfloat{\includegraphics[width=0.3\textwidth]{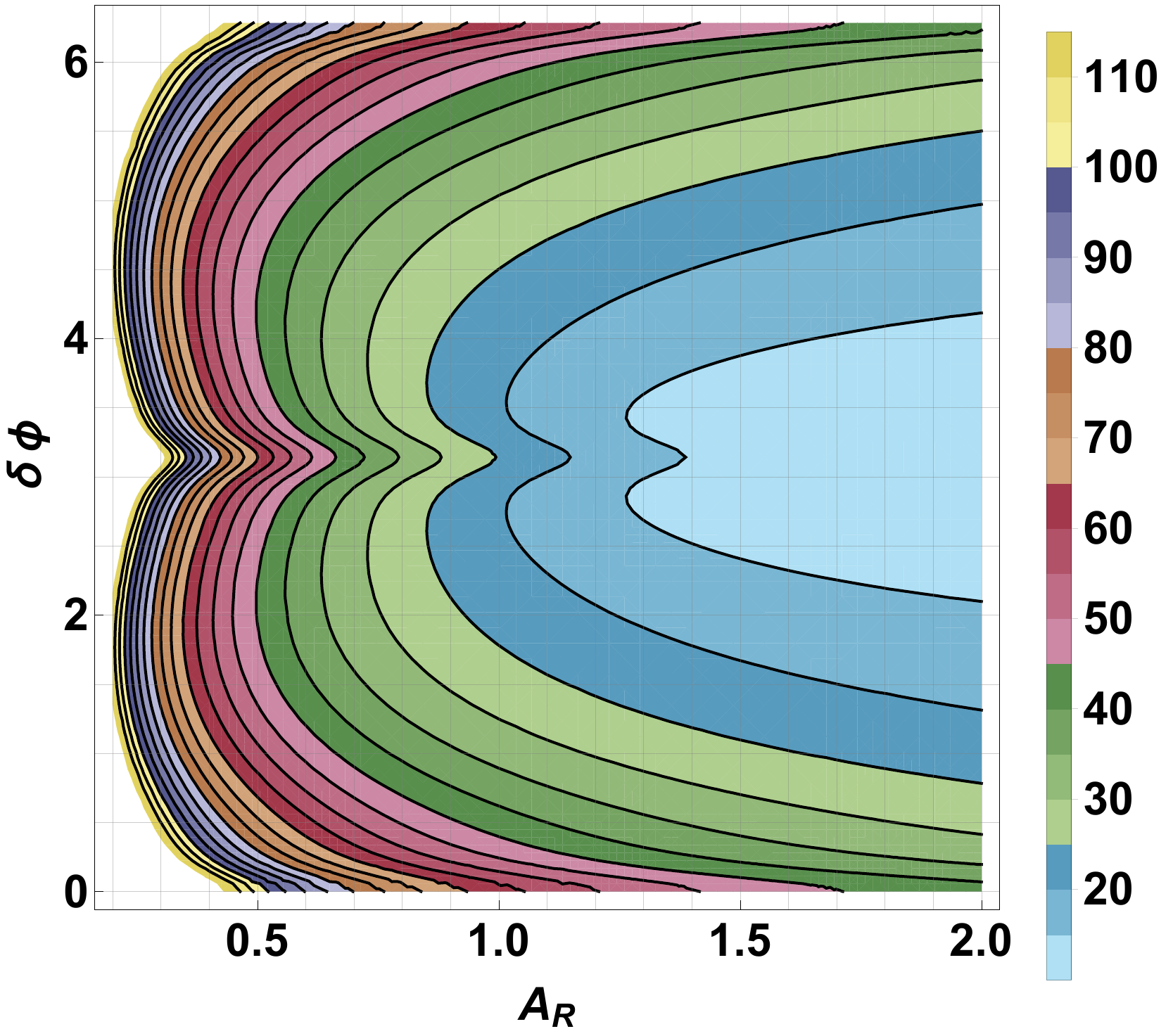}} 
     \subfloat{\includegraphics[width=0.3\textwidth]{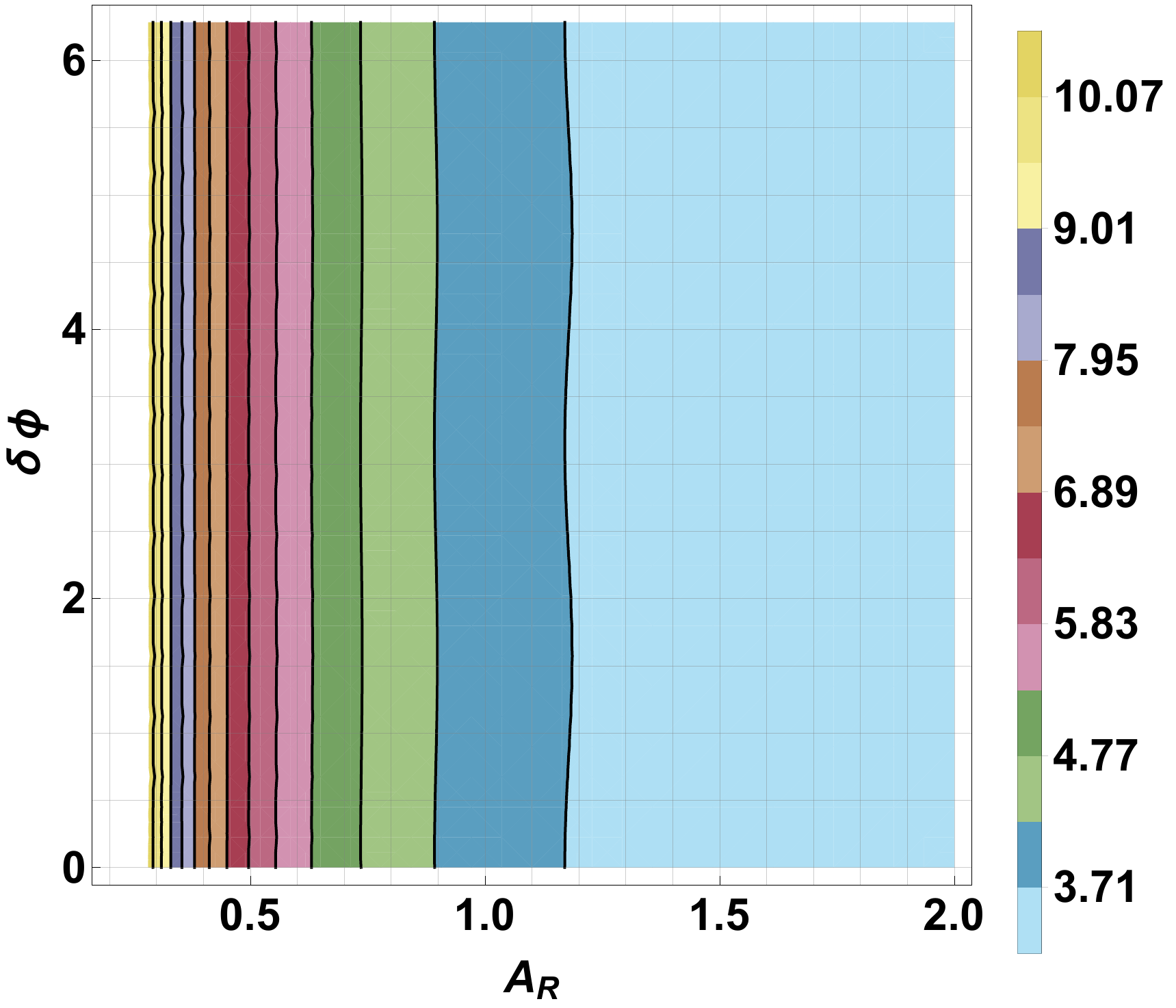}} 
    \subfloat{\includegraphics[width=0.3\textwidth]{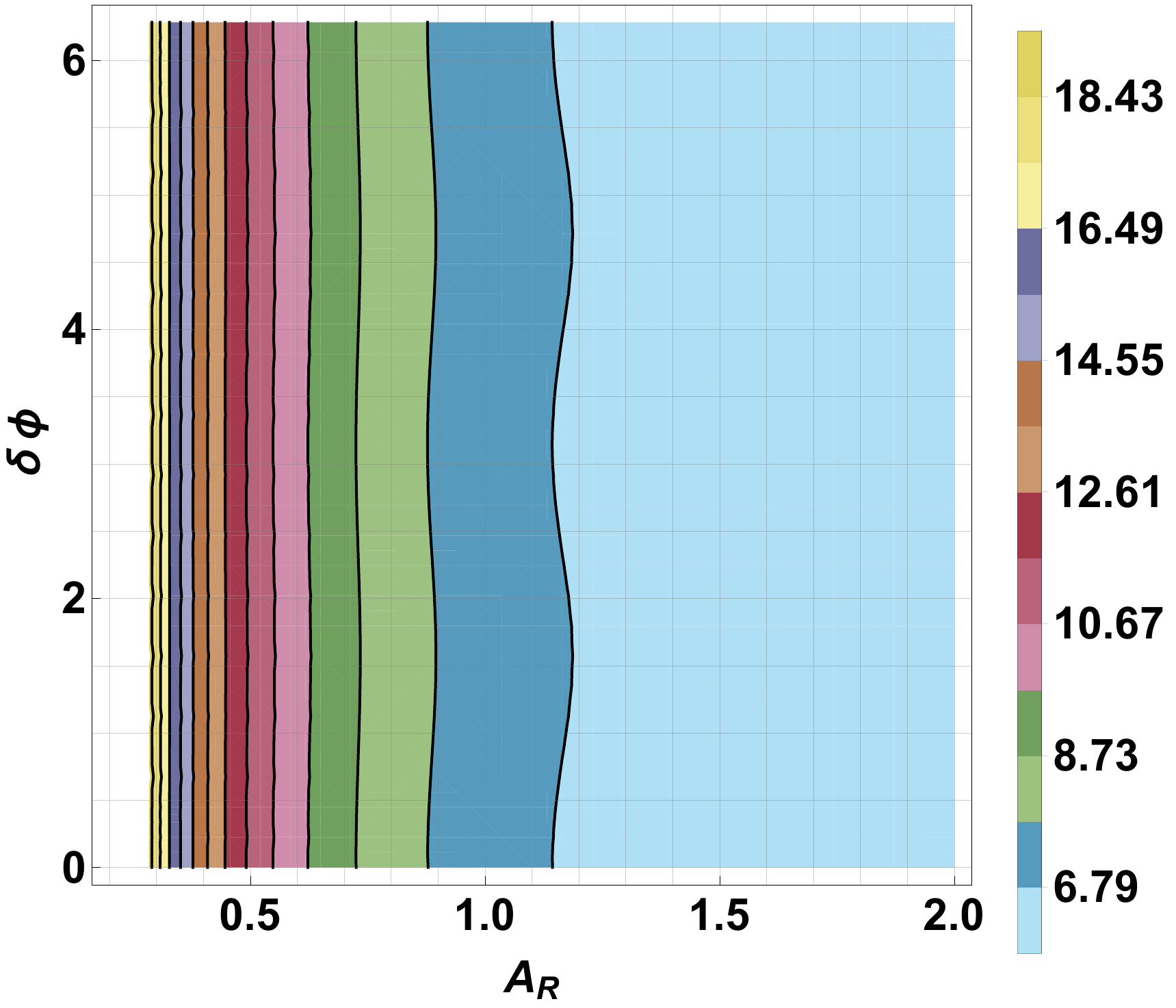}}
    \caption{Contour plot showing the minimum SNR required to satisfy each of the criteria discussed in Sec.~\ref{sec:crit} (detectability, resolvability, and measurability) for a two-mode/tone model as a function of the amplitude ratio $A_{R}$ and of the phase difference $\delta \phi$. The top, middle, and bottom panels correspond to the detectability, the resolvablity, and the measurablity of the subdominat mode frequency to a $5\%$ precision, respectively. The left, center, and right panels show the case of a secondary QNM with $l=m=2$, $n=1$ (left), $l=m=3$, $n=0$ (center), and $l=2$, $m=1$, $n=0$ (right), respectively. } 
    \label{fig:contour}
\end{figure*}

In Fig.~\ref{fig:contour} we show the minimum SNR required to satisfy each of the criteria discussed in Sec.~\ref{sec:crit} -~ namely detectability, resolvability, and measurability~- as a function of the amplitude ratio $A_R$ and of the phase difference $\Delta \phi$ of a secondary mode/tone.
This plot can be useful to immediately estimate the minimum SNR once $A_R$ and $\Delta\phi$ are known for a given system, also beyond those studied in this work (e.g., when including spinning progenitor binaries).
\clearpage
\bibliography{Reference}
\end{document}